\documentclass[pdflatex,sn-mathphys-num]{sn-jnl}

\usepackage{graphicx}%
\usepackage{multirow}%
\usepackage{amsmath,amssymb,amsfonts}%
\usepackage{amsthm}%
\usepackage{mathrsfs}%
\usepackage[title]{appendix}%
\usepackage{xcolor}%
\usepackage{textcomp}%
\usepackage{manyfoot}%
\usepackage{booktabs}%
\usepackage{algorithm}%
\usepackage{algorithmicx}%
\usepackage{algpseudocode}%
\usepackage{listings}%
\usepackage{array}
\usepackage[labelformat=simple]{subcaption}

\begin{document}

\title[Article Title]{Scalable Fault Vulnerability Analysis for ReRAM-Based Process-in-Memory CNN Accelerators
}
\author{\fnm{Aniseh} \sur{Dorostkar}}\email{a\_dorostkar@aut.ac.ir}

\author*[1]{\fnm{Hamed} \sur{Farbeh}}\email{farbeh@aut.ac.ir}

\author{\fnm{Hamid R.} \sur{Zarandi}}\email{h\_zarandi@aut.ac.ir}

\author{\fnm{Meysam} \sur{Sharifzadeh}}\email{msharifzadeh@aut.ac.ir}

\affil{\orgdiv{Department of Computer Engineering}, \orgname{Amirkabir University of Technology (Tehran Polytechnic), \city{Tehran}, \country{Iran}}}


\abstract{Resistive random-access memory (ReRAM)-based processing-in-memory (PIM) accelerators offer a highly efficient platform for parallel matrix-vector multiplications (VMM) in neural networks, leveraging analog computation, high density, low leakage, and non-volatility. However, their susceptibility to permanent faults, such as stuck-at-high (\textit{SaH}) and stuck-at-low (\textit{SaL}) resistance states, permanently corrupts the convolutional neural networks (CNNs) weights mapped onto ReRAM cell conductance values, thereby degrading inference accuracy and leading to system unreliability in safety-critical applications. Fault injection is a commonly used technique to assess the vulnerability of neural networks by systematically injecting faults into individual parameters and measuring the resulting accuracy degradation. Although this method is accurate, it is prohibitively time-consuming and computationally expensive, especially for large-scale CNNs. In this paper, we propose a fast, scalable, and analytical framework to quantify the fault vulnerability of convolutional neural networks deployed on ReRAM-based PIM accelerators. This method quantifies vulnerability at different granularities, including weights, kernels, and layers, without the need for exhaustive simulation, enabling the design of fault-tolerant accelerators for edge and safety-critical applications. The comparison between the results of the proposed vulnerability score (VS) metrics with performance metric of CNNs captured with fault injection simulations confirms the effectiveness of our proposed method. Unlike exhaustive fault injection, which requires days or weeks, whereas the proposed approach achieves a speedup of about $10^4\times-10^5\times$, enabling fast and accurate vulnerability analysis for large-scale CNNs within only a few minutes or hours.}




\maketitle

\section{Introduction}\label{Intro}
Deep neural networks (DNNs), and in particular convolutional neural networks (CNNs), have achieved high attention and great performance in classification, detection, and recognition tasks in various domains such as healthcare, autonomous driving, space applications, and other safety-critical systems~\cite{Ibrahim2020}. Their success is primarily due to their ability to extract complex features through deep, multi-layered architectures, which is based on computationally intensive vector-matrix multiplication (VMM). To achieve high accuracy, the complexity of CNNs increases, which in turn leads to significant growth in computational cost, memory requirements, and energy consumption.

Despite the widespread use of traditional computing platforms, including central processing units (CPUs), graphics processing units (GPUs), and field-programmable gate arrays (FPGAs), for convolutional neural networks (CNNs), these platforms increasingly face performance bottlenecks, particularly with respect to memory bandwidth and power consumption. Resistive random-access memory (ReRAM)-based Process-in-Memory (PiM) accelerators have emerged as a promising candidate that provide high parallelism and energy efficiency for large-scale CNNs inference. ReRAM devices, fabricated at the nanoscale, can be seamlessly integrated into crossbar arrays, providing an efficient solution for implementing large numbers of synaptic connections in neural networks~\cite{Guo2019, Pedretti2021} and graph processing~\cite{Ghasemi2022}. ReRAM crossbars can perform massively memory-intensive VMM operations in parallel and have been widely adopted in accelerators for CNN such as \textit{ISAAC}~\cite{Shafiee2016}, \textit{PRIME}~\cite{Chi2016}, and \textit{PipeLayer}~\cite{Song2017}.

ReRAM-based accelerators face critical reliability challenges due to the limitations of technology fabrication. Accordingly, they suffer from non-idealities such as process variations (PVs), limited write endurance, discrete and limited number of conductance states, sources of noise, and non-linearity of current-voltage characteristics~\cite{Lammie2020, Li2019, Chakraborty2020}. Transient and permanent faults, such as stuck-at-high (\textit{SaH}) and stuck-at-low (\textit{SaL}) faults, introduce significant errors in VMM computations, ultimately degrading CNN inference accuracy. As CNNs are increasingly deployed in safety-critical applications, ensuring reliable inference becomes essential, as even a single fault may propagate through the network and compromise the correctness of the entire system~\cite{Dorostkar2024}. Therefore, it is critical to investigate and analyze the vulnerability of CNN models deployed on ReRAM-based accelerators to transient and permanent faults, particularly \textit{SaH} and \textit{SaL} faults, at the level of weights, kernels, and layers. Such analysis is crucial for understanding the fault behavior of each kernel or layer and for ensuring the design of robust CNN models or the design of a fault-tolerant ReRAM accelerator suitable for safety-critical systems.

To investigate the reliability of CNNs, there is some fault injection and analytical model to analyze the vulnerability of CNNs to permanent or transient faults across a variety of hardware platforms, including CPUs, GPUs, FPGAs, Tensor Processing Units (TPUs), Application-Specific Integrated Circuits (ASICs)~\cite{Ibrahim2020, Gambardella2019, Zhang2019, Khoshavi2020, Taheri2024} and ReRAM-based PIM accelerators~\cite{Dorostkar2024, Sawal2024, Shin2023}.

Traditional fault injection provides an accurate way to assess the vulnerability of neural networks by systematically injecting faults and measuring the resulting accuracy drop or loss increase~\cite{Taheri2024, Schroedter2024}. However, this approach is computationally prohibitive for large-scale CNNs with millions of parameters. Several work have attempted to reduce the complexity of fault injection simulations~\cite{Weng2024, Chen2019}. In practice, most fault injection studies rely on random fault injection into a subset of layers, kernels, or parameters. 

However, conducting a comprehensive analysis that covers all weights, kernels, and layers requires simulating faults and re-executing inference for every parameter. Therefore, exhaustive analysis requires running inference and fault simulation for every parameter, kernel, and layer, which is impractical for large-scale and deep models. Therefore, alternative approaches are necessary to enable scalable and efficient vulnerability analysis. 

In CNNs, specific components, such as certain neurons, weights, kernels, and layers, exhibit higher computational significance, influencing overall model efficiency and accuracy. Consequently, faults occurring in different components have unequal impacts on the network's output. Despite ongoing advancements in neural network resilience, a significant gap remains in accurately identifying the most critical weights, biases, kernels, and layers whose faults lead to substantial accuracy degradation in PIM-based ReRAM CNN accelerators. 

In this work, we propose a comprehensive framework to analyze fault sensitivity for ReRAM-based PIM accelerators for CNNs. Our method introduces an element-wise sensitivity metric that estimates the vulnerability of individual weights and biases across different CNN layers under stuck-at faults (SAFs), leveraging both first- and second-order gradient approximation. Furthermore, the framework provides a fast, scalable, and efficient mechanism to rank the vulnerability of weights, kernels, and layers, which can be utilized in the design of fault-tolerant PIM-based ReRAM accelerators. The main contributions to this article are as follows: 
\begin{enumerate}
	\item We provide a scalable, fast, and efficient vulnerability analysis framework for SAFs occurring in ReRAM-based PIM accelerators targeting large scale CNNs.
	\item We introduce a sensitivity metric that combines first- and second-order gradient terms to capture both linear and curvature effects of weight perturbations caused by SAFs.
	\item We analyze the sensitivity of different CNNs at three levels of granularity, i.e., element-wise, kernel-wise, and layer-wise, enabling a hierarchical ranking of network components.
	\item We conduct a comprehensive analysis of CNN parameter sensitivity across diverse network topologies and scales, revealing that vulnerability to permanent faults depends on fault type and magnitude, network architecture, layer configuration, distance from the output, and the role of learning parameters such as weight or bias.
	\item We validate the proposed sensitivity metrics through systematic fault injection simulations, demonstrating their effectiveness in evaluating CNN performance under faults.
\end{enumerate}

The remainder of this article is organized as follows. Section~\ref{prelim} provides background information on neural networks, ReRAM-based PIM accelerators, reliability considerations, and fault models associated with ReRAM-based architectures. A comprehensive discussion of related work is presented in Section~\ref{rel-work}. Our proposed vulnerability analysis framework is detailed in Section~\ref{scal-sen}. Extensive analysis and evaluations using our scalable framework, along with validation results, are provided in Section~\ref{Exp-setup}. Finally, Section~\ref{concl} concludes the paper. 

\section{Preliminaries}\label{prelim}

\subsection{Convolutional Neural Networks}\label{CNN}

CNNs are composed of a sequence of convolutional (CONV) and pooling layers, along with optional layers such as batch normalization (BatchNorm) for feature extraction, followed by a small number of fully connected (FC) layers (typically fewer than three) for classification. The trainable parameters, namely, the kernels and biases in CONV layers, as well as the weights and biases in FC and BatchNorm layers, are adjusted during the training phase. The main computations in CNNs are dominated by multi-dimensional convolution and VMM, primarily carried out in the CONV and FC layers.

\begin{figure}[!t]
	\centering
	\subfloat[\label{Fig1-a}]{
		\includegraphics[width=0.33\linewidth]{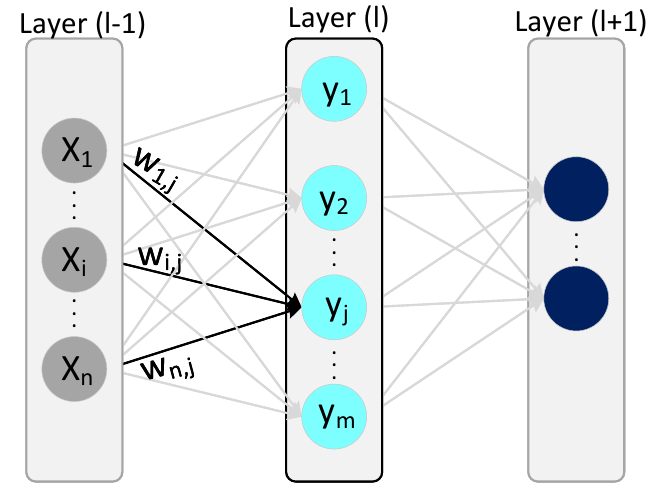}
	}
	\subfloat[\label{Fig1-b}]{
		\includegraphics[width=0.66\linewidth]{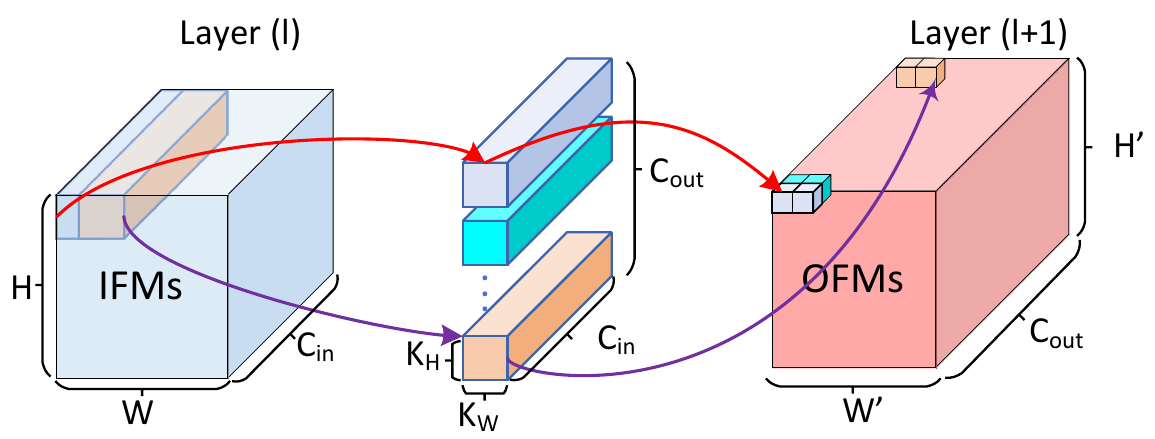}
	}
	\caption{The main computations in CNNs (a) vector-matrix multiplications in FC layer, and (b) Convolutional in CONV layer.}
	\label{Fig1}
\end{figure}

In FC layer, each neuron receives weighted input from all activations of the previous layer, equivalent to a VMM, added by a bias passed through an activation function $\sigma$, (commonly \textit{sigmoid}, \textit{ReLU}, or \textit{tanh}) to compute the neuron output (as shown in Fig.~\ref{Fig1-a}). The operation of an FC layer can be formulated as:

\begin{equation}
	y_{j} = \sigma \left( \sum_{i=1}^{n} x_{i} w_{ij} + b_{j} \right)
	\label{Eq1}
\end{equation}
where $y_{j}$ denotes the output of the $j$-th neuron, 
$x_{i}$ is the $i$-th input of the neuron (the output of the $i$-th neuron from the previous layer), 
$w_{ij}$ refers to the trainable (synaptic) weight associated with input $i$ and neuron $j$, 
and $b_{j}$ is the bias applied to neuron $j$. 

In CONV layers, instead of a simple weighted sum, a convolution is performed between each kernel and its associated input feature map (IFM), producing an output feature map (OFM) (as shown in Fig.~\ref{Fig1-b}). Given an IFM with $C_{\text{in}}$ channels, the $n$-th output feature map (OFM) at spatial location $(x,y)$ is computed by:
\begin{equation}
	\begin{aligned}
		\text{OFM}_n(x, y) = \sigma \sum_{i=0}^{C_{in}} 
		\Bigg( \sum_{dx=0}^{K_{W}} 
		\Bigg( \sum_{dy=0}^{K_{H}} 
		&  W_{n}(i,dx,dy) ^*IFM(i,x+dx,y+dy)
		\Bigg) + B_{n} 
		\Bigg),\\[1ex]
		&\phantom{=} 1 \leq n \leq C_{out}
	\end{aligned}
	\label{Eq2}
\end{equation}
where $W_{n}(i,dx,dy)$ denotes the $n$-th kernel applied to input channel $i$ at spatial location $(dx,dy)$, and $C_{\text{out}}$ is the number of kernels (output channels). $K_{W}$ and $K_{H}$ are the width and height of the kernels, respectively, and $B_{n}$ is the bias associated with the $n$-th kernel.

\subsection{ReRAM-based PIM Accelerators}

\begin{figure}[!t]
	\centering
	\subfloat[\label{Fig2-a}]{
		\includegraphics[width=0.53\linewidth]{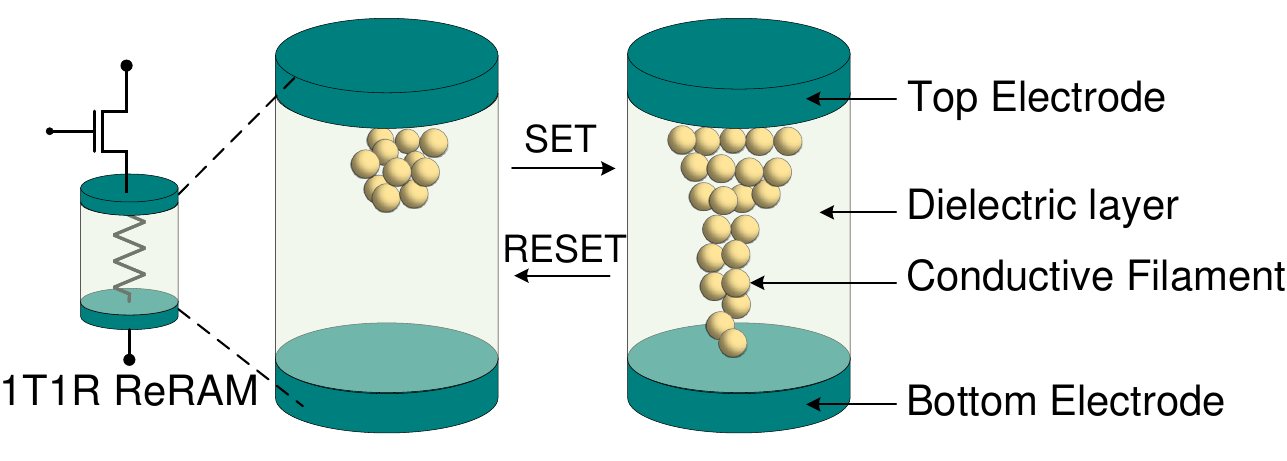}
	}
	\subfloat[\label{Fig2-b}]{
		\includegraphics[width=0.45\linewidth]{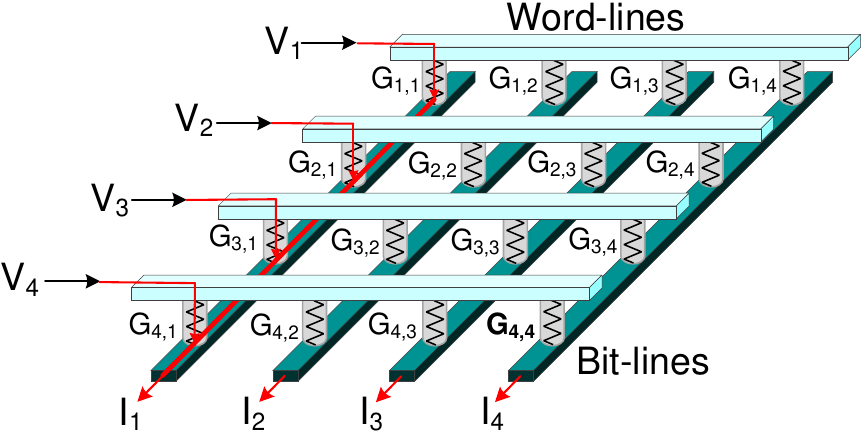}
	}
	\caption{ReRAM Structure, (a) 1T1R ReRAM cell structure, and (b) ReRAM crossbar array performing analog VMM operation.}
	\label{Fig2}
\end{figure}

ReRAM is a non-volatile memory technology that operates by switching the resistance state of its cells. A typical \textit{1T1R} cell consists of a resistive switching device based on a metal--insulator--metal (MIM) structure, integrated with an access transistor (shown in Fig~\ref{Fig2-a}). Depending on the applied voltage pulses, the device can switch between different resistive states, by the formation and rupture of conductive filaments within the insulating layer. Reading the cell requires only a small read voltage that does not disturb the current cell state.

A ReRAM crossbar array (shown in Fig.~\ref{Fig2-b}) is constructed by arranging multiple ReRAM cells at the intersections of perpendicular word-lines and bit-lines. The main operation of ReRAM-based neural network accelerators is VMMs, which are executed natively in the analog domain. In this process, input vectors are applied as voltages to the rows of the crossbar, while the stored conductance values represent the weight matrix. According to \textit{Ohm's law}, the product of input voltages and conductance values yields the corresponding multiplication results. By \textit{Kirchhoff's current law}, the currents along the bit-lines are naturally accumulated to generate the final vector-matrix product~\cite{Li2017}.

\begin{figure}[!t]
	\centering
	\includegraphics[width=0.99\textwidth]{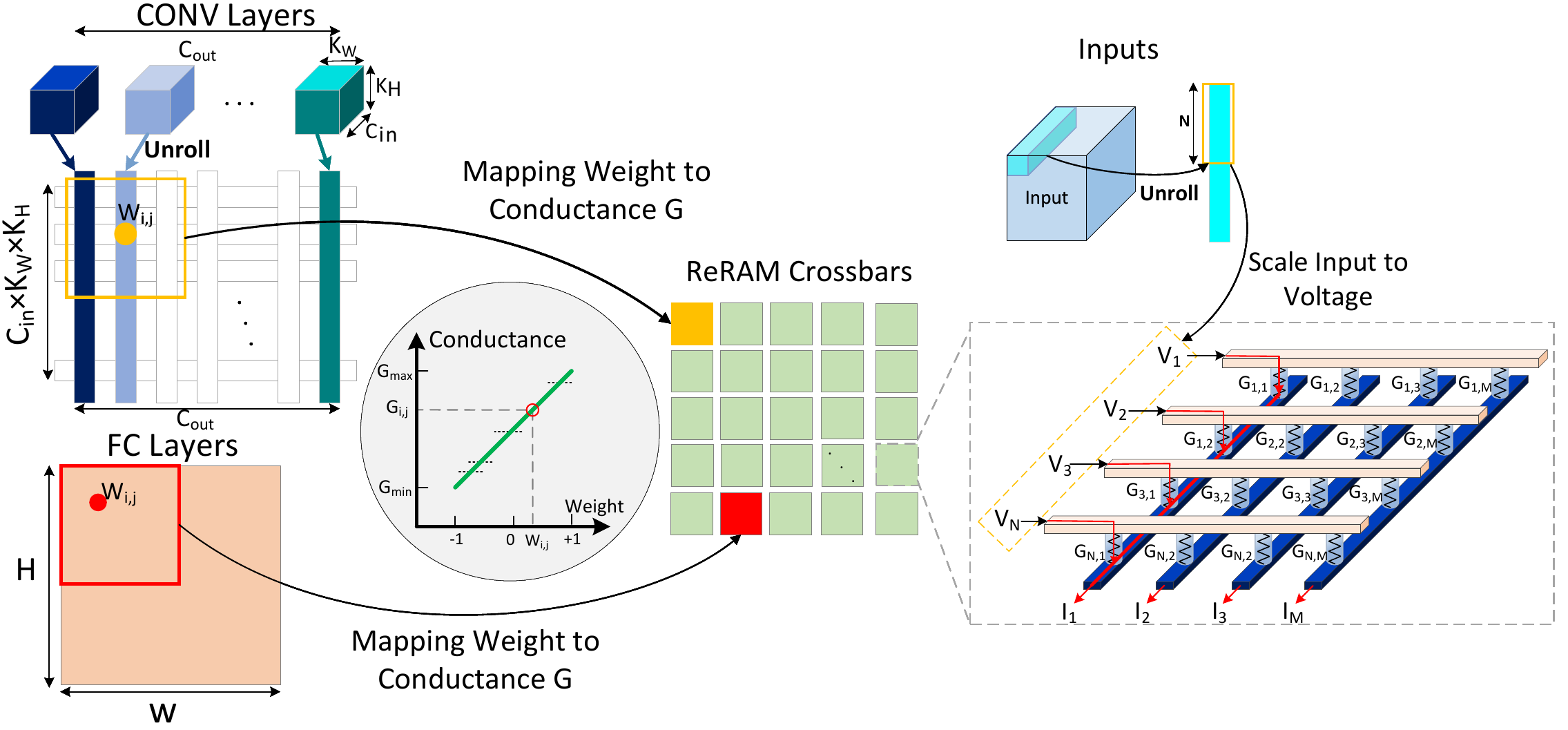}
	\caption{General process of mapping of CNN learning parameters to ReRAM crossbars.}
	\label{Fig3}
\end{figure}

Both fully connected (FC) and convolutional layers can be efficiently implemented in ReRAM crossbars, enabling scalable, high-density, and energy-efficient PIM for deep neural networks. The process of mapping CONV and FC layers is shown in Fig.~\ref{Fig3}. For executing matrix--vector multiplication in a FC layer, the input vector is mapped to input voltages $V$, while the weight matrix of dimension $(H, W)$ is mapped to the conductance values of the ReRAM crossbar.
 
To execute convolutional operations within ReRAM crossbars, the convolution kernels are first flattened and transformed into equivalent matrix representations. The resulting weight matrix of dimension $(C_{\text{in}}\times K_{W}\times K_{H}, C_{\text{out}})$, 
where each column corresponds to the weights of a kernel, is mapped to the conductance values of the crossbar array. Similarly, the IFM is unrolled into an input voltage matrix of size $(H\times W, C\times K\times K)$, which is applied to the rows of the crossbar. 

Since the number of parameters and the corresponding matrix dimensions differ across layers, the weight matrices may not match the size of a single crossbar. When the dimensions of the weight matrices (both FC and CONV layers) and inputs exceed the size of the crossbar, the matrices and inputs are partitioned into blocks matching the crossbar size and mapped across multiple crossbars. Consequently, depending on the CNN architecture, an appropriate mapping strategy is required to map weight matrices across a hierarchy of crossbars, tiles, and memory banks within the ReRAM-based accelerator.

Since ReRAM cells can only store non-negative conductance states, signed neural network weights are implemented using a dual-crossbar architecture. In this scheme, positive and negative weights are mapped to separate crossbars, while the corresponding cells in the dual crossbar are set to the minimum conductance value~\cite{Chi2016, Sun2021, Chen2021}.

\subsection{Fault Model and Reliability of ReRAM-based Accelerators for CNNs}

In ReRAM-based PIM accelerators, the learning parameters of neural networks are mapped to the conductance of ReRAM cells. Consequently, any variations or defects in ReRAM cells directly lead to variations in the neural network parameters, thereby affecting inference accuracy and system reliability.

The reliability of ReRAM accelerators is influenced by data-dependent and data-independent factors. Data-dependent factors include the mapping of weights and biases to conductance values of ReRAM cells, scaling of inputs into voltages, process-variations (cell-to-cell and cycle-to-cycle variations), resistance drift, quantization errors, and device faults such as stuck-at-high (\textit{SaH}) and stuck-at-low (\textit{SaL}). In contrast, data-independent factors are related to peripheral circuitry and fabrication technology, including access transistors or selectors, ADC/DAC resolution, sneak-path currents, and parasitic resistances.

ReRAM faults are commonly categorized into transient and permanent types. Transient faults are temporary and reversible, typically caused by external disturbances such as high-energy particles, temperature fluctuations, resistance drift, or read/write disturbance~\cite{Ibrahim2020, Xia2019, Cheshmikhani2020, Hajisadeghi2024}. These faults alter cell resistance temporarily without causing physical damage and ReRAM cells can still function. Permanent faults, however, are persistent and result from irreversible physical damage caused by limitations in fabrication (e.g., oxide thickness variations, forming-stage defects, or reduced feature sizes~\cite{Yeo2019, Ching-Yi2015}) or limited write endurance~\cite{Beckmann2016, Grossi2019}\}. The forming process of ReRAM cells is particularly sensitive, as high-voltage stress is applied to establish conductive filaments in devices. Defects such as over-forming or short defects~\cite{Ching-Yi2015} often cause the resistance of cell to becomes irreversibly stuck-at high-resistance (\textit{SaH}), stuck-at low-resistance (\textit{SaL}), or more generally, stuck-at value (\textit{SaV}) states. Experimental studies report that approximately 9.04\% of ReRAM cells exhibit \textit{SaH} and 1.75\% exhibit \textit{SaL}~\cite{Ching-Yi2015}. 

Even after successful fabrication and initial testing, additional faults can occur during operation due to limited endurance, especially after repeated write-0 or write-1 pulses~\cite{Grossi2019, Mittal2018}\} or wear-out, which cause damaging the device or transistor. 

\section{Related Work}\label{rel-work}

Faults occurring during neural network execution can lead to errors in the output. Prior studies have focused on the software- and hardware-based investigation and analysis of various types of faults and errors associated with these networks. Importantly, faults originating from different network layers or parameters have been shown to have varying impacts on the network's output accuracy and overall performances.

Software-based fault injection tools typically evaluate neural network reliability without considering the underlying hardware. For example, TensorFi~\cite{Chen2020} and PyTorchFi~\cite{Mahmoud2020} are open-source fault injection tools built on TensorFlow and PyTorch, respectively. These frameworks enable the injection of both transient and permanent faults at the algorithmic-level. TensorFi injects the faults at the data path of neural networks, while PyTorchFi modifies neuron activations and weights at specific locations. However, their functionality is limited to random data modifications, and they are unable to provide a detailed vulnerability analysis across CNN models or hardware platforms. In contrast, GPU-oriented tools such as SASSIFI~\cite{Hari2017} and NVBitFI~\cite{Tsai2021} perform fault injection at the instruction-level and can also be applied to CNN workloads.

A simulation is done in~\cite{Neggaz2020} that injects random faults into the data path of several CNN architectures, including \textit{AlexNet}, centroid-aware feature recalibration network (CafeNet), network-in-network (NiN), and ConvNet, across both floating-point and fixed-point data types with varying bit widths.

Furthermore, to improve efficiency and identify critical parameters in neural networks, several extended FI frameworks have been proposed. BinFI~\cite{Chen2019}, is an extension of TensorFI, focusing on detecting critical bits in machine learning applications whose corruption by transient hardware faults may lead to safety violations. Similarly, LLTFI~\cite{Agarwal2022} enables transient fault injection at the LLVM intermediate representation (IR) level, supporting machine learning workloads implemented in TensorFlow and PyTorch. DeepVigor~\cite{Ahmadilivani2023} is a resilience analysis framework for DNNs that introduces the concept of a neuron vulnerability range, which characterizes how bit-level faults can alter neuron outputs and potentially lead to misclassification in the network.

From another aspect, several hardware-based fault-injection tools have been proposed for different DNN models running on different accelerator platforms (GPU, TPU, FPGA, and ASIC). Transistor-level fault simulations have so far been limited to individual neurons in spiking neural networks (SNNs)~\cite{El-Sayed2019} or to small-scale networks~\cite{Temam2012}, typically executed on specialized spiking-based hardware designed for small and specific SNN or DNN models. Transistor-level simulations are prohibitively expensive in terms of time and resources, making them impractical for large networks. Therefore, fault injection for large-scale neural networks necessitates higher levels of abstraction, which reduce computational costs while still enabling meaningful reliability analysis. Such abstract fault models not only make large-scale simulations feasible but also enable more reliability analysis that is independent of specific hardware implementation details.

For this purpose, fault injection has been investigated across different aspects. In~\cite{Vatajelu2019, El-Sayed2020}, a behavioral modeling of faults had been done in neurons of spiking neural networks (SNNs). There is some work that model permanent and transient bit-flip faults in the data path and memory of CNNs~\cite{Neggaz2020,Li2017un, Reagen2018}. At the hardware level, studies have examined stuck-at faults at gate-level in TPUs~\cite{Zhang2019}, the faults in resolution of activation of QNN in FPGA~\cite{Gambardella2019}, and faults in conductance of ReRAM crossbar cells~\cite{Dorostkar2024, Zhao2020}.

ReConFI~\cite{Dorostkar2024} is a fault injection framework that performs fault injection at both software- and hardware-level to characterize the impact of permanent faults on PIM-based ReRAM accelerators that run CNNs. ReConFI analyzes the impact of weight variations and \textit{SaH} and \textit{SaL} fault models on different layers and locations of CNN learning parameters.

To evaluate the impact of faults on different network parameters, researchers have defined a set of sensitivity metrics. A brute-force fault impact mapping algorithm was proposed in~\cite{Schroedter2024} to evaluate the effect of SAFs in memristive crossbars. The algorithm iteratively injects SAFs into weights and biases and measures the corresponding accuracy degradation. The weights causing the most significant drop in accuracy are identified as critical weights. To demonstrate its effectiveness, the algorithm was applied to a two-layer perceptron network trained on a reduced version of the MNIST dataset (digits 0-3) with an input image size of $8\times8$.

Fiji-FIN~\cite{Khoshavi2020} considers the accuracy degradation as the impact of transient faults on the quantized neural network (QNN) performance on an FPGA. For each single fault injection, the inference phase for the dataset is executed to calculate the accuracy. SAFFIRA~\cite{Taheri2024} proposed a software-based hardware-aware fault injection method for systolic based DNN accelerators, using the Uniform Recurrent Equations (URE) system. SAFFIRA introduces a faulty distance to assess the reliability metric.

To reduce the time of fault injection simulation, APPRAISER~\cite{Taheri2023} applies approximate computing error to improve the cost of analyzing the transient faults in DNNs emulated on an FPGA. FKeras~\cite{Weng2024} provides bit-level sensitivity metrics to evaluate the fault tolerance of edge neural networks. It reduces the simulation complexity by ranking NN weights according to their vulnerability and guiding fault injection campaigns.

The proposed fault injection approaches typically require repeatedly simulating a DNN with randomly selected faulty parameters or activations, and then measuring the impact on the outputs to derive reliability metrics. However, the accuracy of these metrics strongly depends on the number of simulations performed. Conducting exhaustive FI, where every possible fault in all learning parameters is considered, results in an enormous search space that becomes infeasible for modern DNNs with millions of parameters and activations. To overcome this, we propose a fast and scalable method that assesses the sensitivity of learning parameters of CNNs to faults. This method uses sensitivity of weights and fault distribution of ReRAM crossbars to exhaustively analyze the vulnerability of large-scale CNNs at different granularities of weights, kernels, and layers. 

\section{Scalable Fault Vulnerability Analysis of DNNs}\label{scal-sen}

\begin{figure}[!t]
	\centering
	\includegraphics[width=1\textwidth]{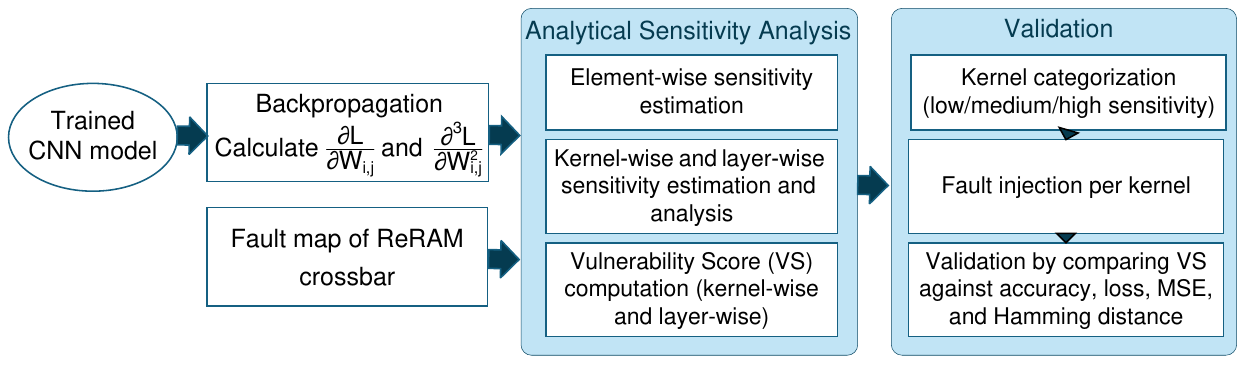}
	\caption{General process of mapping of CNN learning parameters to ReRAM crossbars.}
	\label{Fig4}
\end{figure}

Traditional fault injection schemes offer accurate vulnerability analysis but are prohibitively expensive for exhaustively analyzing the fault vulnerability of deep CNNs with millions of parameters. To address this limitation, we introduce a scalable gradient-based sensitivity analysis framework that estimates the impact of \textit{SaH} and \textit{SaL} faults. The main idea is that the partial derivative of the loss function L with respect to a weight $W_(i,j)$ indicates how a small perturbation in that weight would affect the overall loss. The overall workflow of the proposed method is illustrated in Fig.~\ref{Fig4}. For a trained CNN model, sensitivity is computed for each weight using backpropagation gradients and faulty weight modeling, and then aggregated hierarchically at the weights, kernels, and layer levels, enabling a multi-granularity exploration of CNN vulnerability. Furthermore, to capture realistic hardware behavior, we define a \textit{vulnerability score} (VS) that combines sensitivity values with the actual distribution of \textit{SaH} and \textit{SaL} faults in ReRAM devices. The sensitivity and VS are further validated by fault injection on some small and medium CNN models and shown correlation with accuracy, loss, MSE, and Hamming distance. Unlike exhaustive fault injection, the proposed approach is scalable to large networks such as \textit{InceptionV3}, while also offering practical benefits by identifying critical kernels and layers for fault-aware mapping and selective redundancy allocation. 

To comprehensively evaluate the fault vulnerability of CNNs, sensitivity is analyzed at three different granularity levels:
\begin{itemize}
	\item Element-wise: Determines the vulnerability of each individual learning parameter, including weights/kernels and biases of CONVs, BatchNorm, and FC layers. It is useful to identify critical weights whose faults can significantly degrade overall network performance.
	\item Kernel-wise: Since convolutional kernels operate as functional units, aggregating sensitivities over their weights allows us to identify the most vulnerable kernels. This view highlights critical kernels that can strongly impact feature extraction and is also useful for kernel mapping in hardware accelerators such as columns of ReRAM crossbars.
	\item Layer-wise: At a higher abstraction, the average sensitivity of all kernels within a CONV layer or the average of elements in FC and BatchNorms captures how faults affect different stages of the CNN. This is particularly useful for layer mapping or redundancy strategies, where resources are often allocated at the layer level.
\end{itemize}

The framework provides a hierarchical understanding of CNN vulnerability. Element-wise analysis reveals the most sensitive parameters, kernel-wise analysis identifies vulnerable functional filters, and layer-wise analysis enables comparison across the network hierarchy. Together, they offer both fine-grained and system-level perspectives on fault tolerance.

First, the sensitivity of each weight (element-wise) is computed based on the first and second order gradients of the loss function $L$ with respect to that weight and the difference between its actual and faulty value under \textit{SaH} or \textit{SaL} faults. 
If, in the presence of a fault, the weight value of $W_{i,j}$ is changed to a faulty value $W_{i,j}^{\text{Faulty}}$ 
($\Delta W_{i,j} = W_{i,j} - W_{i,j}^{\text{Faulty}}$), the impact on the loss function (element-wise sensitivity) can be approximated as:
\begin{equation}
	S_{i,j}^{f} = 
	\Big| g_{i,j} \cdot \Delta W_{i,j}^{\text{Faulty},f} 
	+ \frac{1}{2} h_{i,j} \cdot \big(\Delta W_{i,j}^{\text{Faulty},f}\big)^2 \Big|,
	\quad f \in \{ \text{SaH}, \text{SaL} \}
	\label{Eq3}
\end{equation}
Where $g_{i,j} = \frac{\partial L}{\partial W_{i,j}}$ is the first-order gradient term, representing the partial derivative of the loss function with respect to the weight (and is used to calculate first-order sensitivity). $h_{i,j} = \frac{\partial^2 L}{\partial W_{i,j}^2}$ is the second-order gradient term (diagonal Hessian entry, per-weight curvature), representing how fast or slow the gradient itself changes when weights change and is used to calculate second-order sensitivity. 

In the fault model considered in this study, the faulty weight value $W_{i,j}^{\text{Faulty},f}$ is set to either $w_{\min}$ or $w_{\max}$, corresponding to \textit{SaH} and \textit{SaL} faults, respectively, and $w_{\min}$ and $w_{\max}$ represent the minimum and maximum weight values, respectively. 

When faults cause small deviations in the weight values ($\Delta W$ is small), the quadratic term become negligible, and the first-order approximation provides a reasonable estimate of the resulting loss variation. However, SAFs may cause large deviations (stuck weights at $w_{\min}$ or $w_{\max}$), and the first-order sensitivity alone may underestimate the effect of faults. To better capture the effect of such perturbations, we incorporate the second-order term of the Taylor expansion, which reflects the local curvature of the loss surface through the Hessian. Consequently, both the gradient term (first-order sensitivity) and the Hessian term (second-order sensitivity) are considered to approximate the loss variation caused by SAFs. While higher-order terms could further refine the approximation, they incur significantly higher computational cost and limit scalability for large CNN models. In practice, retaining the first- and second-order terms provides an effective trade-off between analytical accuracy and computational efficiency for vulnerability estimation.

For a trained neural network operating in the inference phase, the test dataset is first propagated once to compute the loss of the network. Subsequently, backpropagation is applied to extract the first- and second-order partial derivatives with respect to all weights. 

Using these gradients together with the fault model, the sensitivity metrics are calculated for each weight, kernel, and layer under \textit{SaH} and \textit{SaL} faults based on Eq.~\eqref{Eq4}. 
$S_{l,k,c,i,j}^{\text{SaH}}$ and $S_{l,k,c,i,j}^{\text{SaL}}$ show the sensitivity of weight $w$ at position $(c,i,j)$ in kernel $k$ of CONV layer $l$ to \textit{SaH} and \textit{SaL} faults, respectively. 
$S_{l,i,j}^{\text{SaH}}$ and $S_{l,i,j}^{\text{SaL}}$ determine the sensitivity of weights at position $(i,j)$ in FC layer $l$. 
$S_{l,i}^{\text{SaH}}$ and $S_{l,i}^{\text{SaL}}$ show the sensitivity of elements in layer $l$ belonging to the weights of BatchNorm and the biases of all layers. 

To calculate the vulnerability of each convolutional kernel $K_{l,k}$ with size $(C_{\text{in}}, k_1, k_2)$, 
element-wise sensitivities are summed over all weights belonging to that kernel. 
The kernel-wise sensitivity is obtained by averaging over the weights of the kernel, as given in Eq.~\eqref{Eq4}:

\begin{equation}
	S_{l,k}^{\text{kernel-wise},f} = 
	\frac{1}{|K_{l,k}|} 
	\sum_{c=1}^{C_{\text{in}}} 
	\sum_{i=1}^{k_1} 
	\sum_{j=1}^{k_2} 
	S_{l,k,c,i,j}^{f}, 
	\quad f \in \{\text{SaH}, \text{SaL}\}
	\label{Eq4}
\end{equation}
Where $|K_{l,k}|$ is the number of elements in the kernel $k$ of layer $l$. 

At the layer-wise level, sensitivities are aggregated over all kernels or elements within the layer. 
The layer-wise sensitivity ($S_{l}^{\text{layer-wise},f}$) is obtained by averaging sensitivity over all elements of that layer. 

Furthermore, to account for realistic fault behavior in ReRAM devices, we incorporate the percentages of \textit{SaH} and \textit{SaL} faults into our analysis and define a fault distribution-aware vulnerability score (VS). 
This metric enables multi-fault and hardware-relevant estimation of CNN vulnerability. 

The kernel-wise VS is calculated by:
\begin{equation}
	VS_{l,k}^{\text{kernel-wise}} = 
	F_{\text{SaH}} \cdot S_{l,k}^{\text{kernel-wise,SaH}} + 
	F_{\text{SaL}} \cdot S_{l,k}^{\text{kernel-wise,SaL}}
	\label{Eq5}
\end{equation}
Where $F_{\text{SaH}}$ and $F_{\text{SaL}}$ are the percentages of \textit{SaH} and \textit{SaL} faults in each kernel and each layer of the CNN model. 

More accurately, by knowing the fault distribution and fault map of the ReRAM crossbar, the vulnerability factor can be calculated by:
\begin{equation}
	VS_{l,k}^{\text{kernel-wise},f} = 
	\sum_{\substack{c,i,j \\ \text{Cell}_{c,i,j} \text{ is faulty (SaH or SaL)}}} 
	S_{l,k,c,i,j}^{f}, 
	\quad f \in \{\text{SaH}, \text{SaL}\}
	\label{Eq6}
\end{equation}

Similarly, the layer-wise VS ($VS_{l}^{\text{layer-wise},f}$) 
can be calculated as the sum of sensitivities of all faulty elements in each layer, 
where $f \in \{\text{SaH}, \text{SaL}\}$.

\subsection{Validation of the Proposed Sensitivity Estimation Method}

The accuracy of the proposed analytical method is validated through fault injection simulation on three CNN architectures: \textit{SimpleNet}, \textit{LeNet5}, and \textit{ResNet18}. Based on the gradient-based sensitivity values, weights and kernels in each layer are categorized into three levels of vulnerability, low-, medium-, and high-sensitivity level). \textit{SaH} and \textit{SaL} faults are then injected at different rates into the kernels of each category, and the corresponding accuracy, loss, mean squared error (MSE), and prediction Hamming distance are measured for the faulty networks. The results are compared against the predicted sensitivities and the computed VS.

This validation confirms a strong correlation between the analytical sensitivity predictions and the observed degradation obtained from fault injection, highlighting the accuracy of the proposed framework. While we validate our method on relatively small and medium-scale CNNs, its scalability and low-cost computation make it applicable for vulnerability analysis even to large and complex CNNs, such as \textit{InceptionV3}. 

Moreover, the results of our analysis and observation can be applied to prioritizing the most critical kernels and layers identified through sensitivity analysis for fault-tolerant methods such as fault-aware mapping and selective redundancy allocation, thereby improving the reliability of ReRAM-based CNN accelerators.

\section{Experimental Setup}\label{Exp-setup}

\subsection{Simulation Platform and Neural Network Models}

We demonstrate how the proposed method can efficiently analyze a NNs fault vulnerability by performing simulations on six well-known models including \textit{SimpleNet}~\cite{SimpleNet}, \textit{LeNet5}~\cite{Lecun1998}, \textit{AlexNet}~\cite{Krizhevsky2012}, \textit{ResNet18}~\cite{He2016}, \textit{VGG19}~\cite{Simonyan2015}, and \textit{InceptionV3}~\cite{Szegedy2016}. \textit{SimpleNet} and \textit{LeNet5} were trained on MNIST~\cite{Lecun1998} dataset, which consists of handwritten digits, while \textit{AlexNet}, \textit{ResNet18}, \textit{VGG19}, and \textit{InceptionV3} were trained on \textit{CIFAR10}~\cite{krizhevsky2009learning}, a collection of small, color images. Each dataset includes 60,000 training and 10,000 test images.

\begin{figure}[!t]
	\centering
	\includegraphics[width=\textwidth]{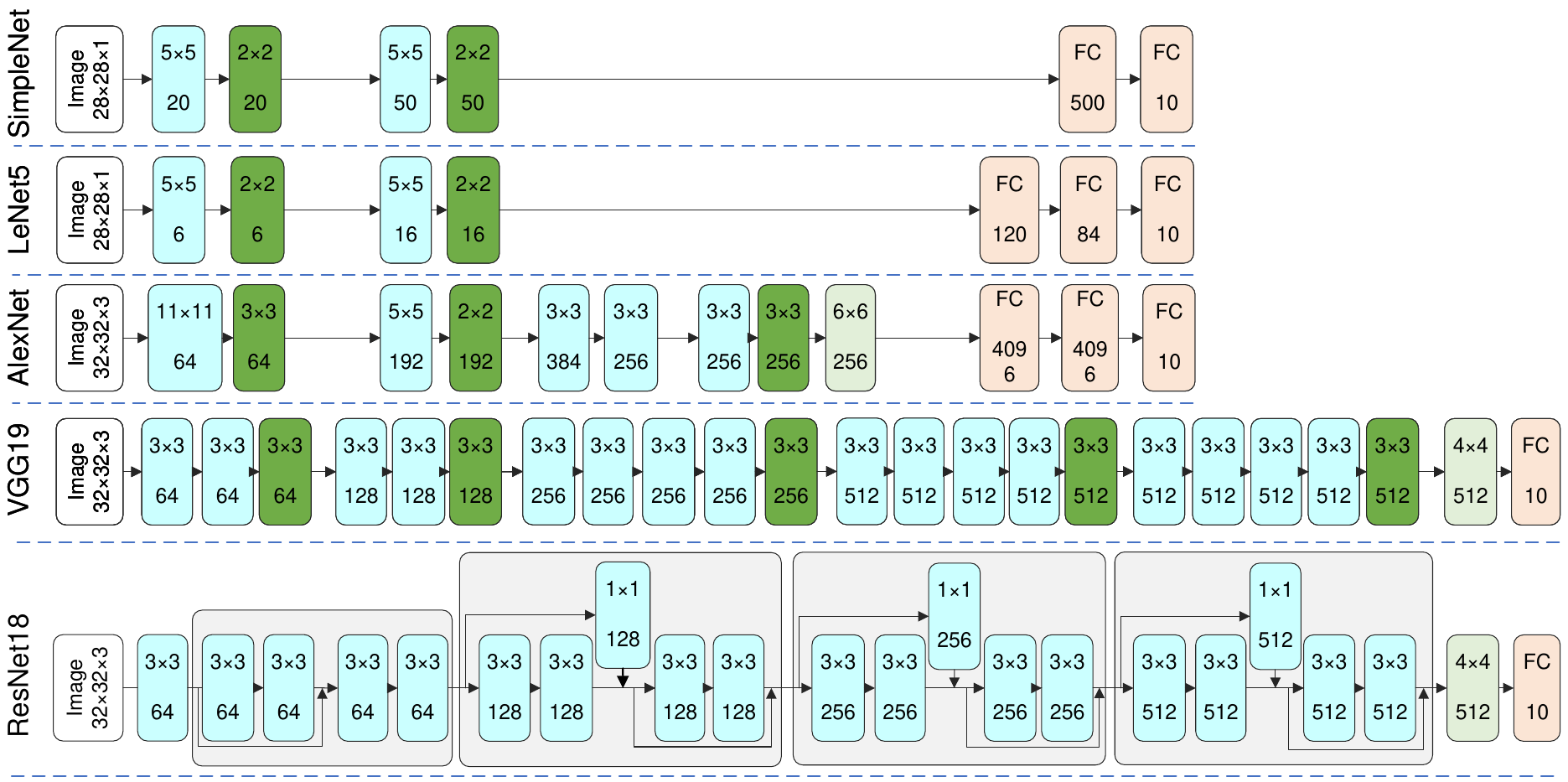}
	\includegraphics[width=\textwidth]{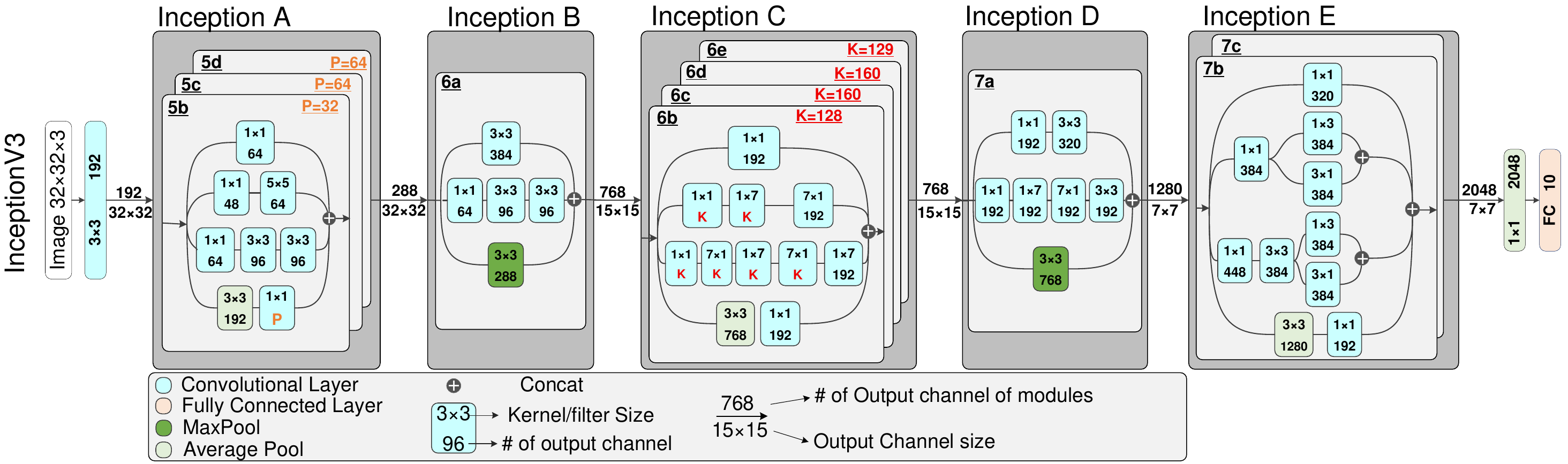}
	\caption{Architecture of CNN models under study.}
	\label{Fig5}
\end{figure}

Table~\ref{Table1} summarizes the specification and Fig.~\ref{Fig5} shows the architecture of CNN models evaluated in this study. \textit{SimpleNet} and \textit{LeNet5} each consist of two convolutional layers, while \textit{SimpleNet} includes two FC layers and \textit{LeNet5} contains three. In contrast, \textit{AlexNet} is composed of five CONV layers with varying kernel sizes followed by three FC layers. \textit{ResNet18} has an initial $3\times3$ CONV, followed by four sequential stages (layer 1 to 4) composed of BasicBlocks, each block contains two $3\times3$ CONV layers, and residual shortcuts and finally, ends with a FC (linear) classifier. In total, \textit{ResNet18} has 20 CONV layers, each followed by BatchNorms. Three of them are $1\times1$ CONVs used in four shortcut connections. In shortcuts, the input of a block is added to its output. When the number of input channels or size of inputs and outputs differ, a $1\times1$ CONV is applied to the inputs to match the output dimension (known as shortcut projection). \textit{VGG19} has 16 CONV $3\times3$ layers followed by BatchNorms, with more than 20 million parameters. \textit{InceptionV3} comprises 83 CONV layers, each followed by BatchNorms, with over 21 million learnable parameters. A key innovation of this model lies in its inception modules, each module composed of sequential blocks, where multiple parallel paths with different kernel sizes ($1\times1$, $3\times3$, $5\times5$, $7\times7$, and pooling) are applied to the same input and concatenated at the output.

We have developed our fault injection and sensitivity analyzer in Python using the PyTorch framework. The MemTorch library is used to model ReRAM-based DNNs and evaluate performance metrics of faulty and non-faulty CNNs. All ReRAM devices were simulated using the VTEAM model~\cite{Kvatinsky2015}.

\begin{table}[!t]
	\centering
	\caption{Specifications of the CNN Models Under Study.}
	\label{Table1}
	\begin{tabular*}{\textwidth}{
			@{\extracolsep{\fill}}
			>{\raggedright\arraybackslash}p{1.4cm}
			>{\raggedright\arraybackslash}p{1cm}
			>{\centering\arraybackslash}p{1.4cm}
			>{\centering\arraybackslash}p{2cm}
			>{\centering\arraybackslash}p{1.4cm}
			>{\centering\arraybackslash}p{2cm}
			>{\centering\arraybackslash}p{1.4cm}
		}
		\toprule
		CNNs & Dataset & Baseline Accuracy & Number of CONV Layers & Number of FC layers & Total Parameters & Total Size (MB) \\
		\midrule
		LeNet5      & MNIST   & 99.26\% & 2  & 3 & 44,426     & 0.30  \\
		SimpleNet   & MNIST   & 99.42\% & 2  & 2 & 431,080    & 1.76  \\
		ResNet18    & Cifar10 & 89.42\% & 20 & 1 & 11,173,962 & 53.89 \\
		VGG19       & Cifar10 & 88.40\% & 16 & 1 & 20,040,522 & 83.64 \\
		AlexNet     & Cifar10 & 86.82\% & 5  & 3 & 57,044,810 & 226.55\\
		InceptionV3 & Cifar10 & 93.35\% & 83 & 1 & 21,638,954 & 82.55 \\
		\bottomrule
	\end{tabular*}
\end{table}

Gradients and second-order derivatives of the Cross-Entropy loss function were computed with respect to the network weights using fully trained models in evaluation mode. The analysis was conducted on the entire test set of each dataset (comprising 10,000 samples for both MNIST and CIFAR-10) using a batch size of 128. To maintain computational efficiency, the second-order derivatives were approximated using the diagonal elements of the Hessian matrix, calculated via the PyTorch Autograd engine. Finally, weight sensitivity was averaged over all batches to obtain the final estimates.
 
To validate the proposed method, four complementary metrics are employed to quantify the impact of faults on CNN performance:
\begin{itemize}
	\item Top-1 accuracy: Reflects the overall performance of the model, defined as the percentage of test samples for which the predicted class matches the true class.
	\item Loss (Cross-Entropy Loss): measures the discrepancy between the predicted probability distribution and the true label distribution of the model. Lower loss values indicate that the model not only predicts the correct class but also does so with high confidence.
	\item Mean Squared Error (MSE): Measures the numerical deviation between the output vectors of the fault-free and faulty networks. Unlike accuracy, it is sensitive to changes in confidence scores, even when the final predicted class remains unchanged. The MSE is computed as the average squared difference between the corresponding elements of the two output vectors.
	\item Hamming Distance: Provides a direct comparison between the predictions of fault-free and the faulty networks. For each input sample, it checks whether the predicted class of the faulty network differs from that of the fault-free network.
\end{itemize}

While the relationship between accuracy and loss is not strictly linear, in practice, a lower loss generally corresponds to a higher accuracy. Together, these four metrics provide the impact of faults on CNN behavior.

\subsection{Sensitivity Evaluation}

By exploring the sensitivity parameter for different granularities of CNNs, the importance and priority of parameters can be determined for fault-tolerant and resilience approaches. For this purpose, the sensitivity is explored and analyzed for different granularity including weights, kernels, and layers of various CNNs.
 
\begin{figure}
	\centering
	\begin{subfigure}{0.33\linewidth}
		\centering
		\includegraphics[width=\linewidth]{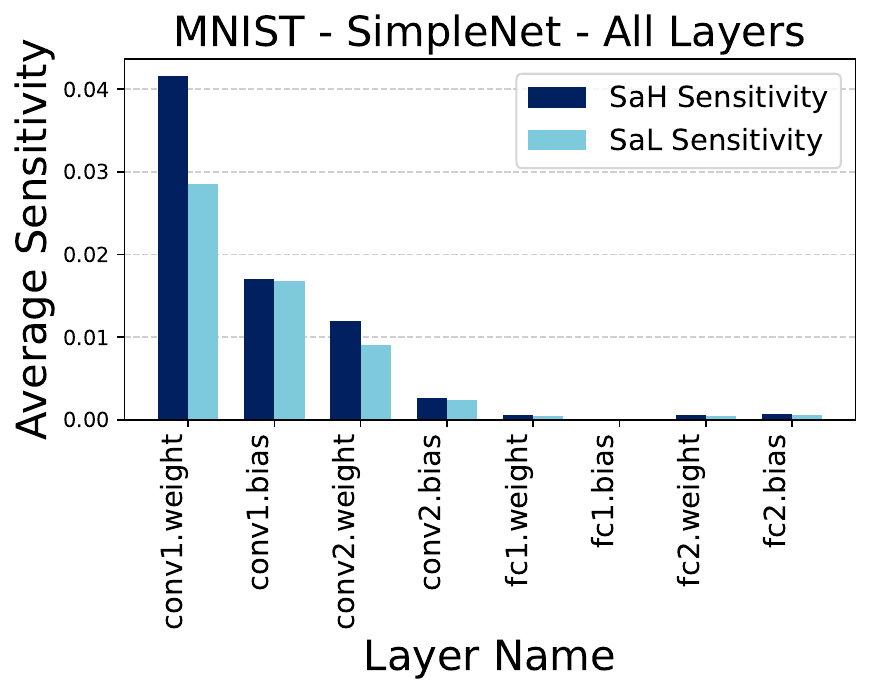} 
		\caption{}
		\label{fig6-a}
	\end{subfigure}\hfill
	\begin{subfigure}{0.335\linewidth}
		\centering
		\includegraphics[width=\linewidth]{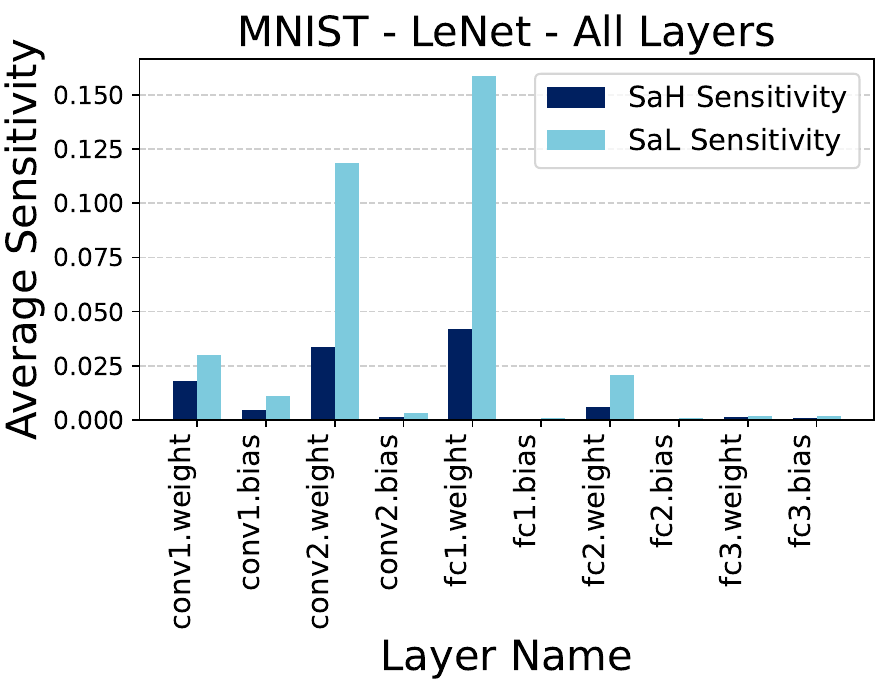}
		\caption{}
		\label{fig6-b}
	\end{subfigure}\hfill
	\begin{subfigure}{0.33\linewidth}
		\centering
		\includegraphics[width=\linewidth]{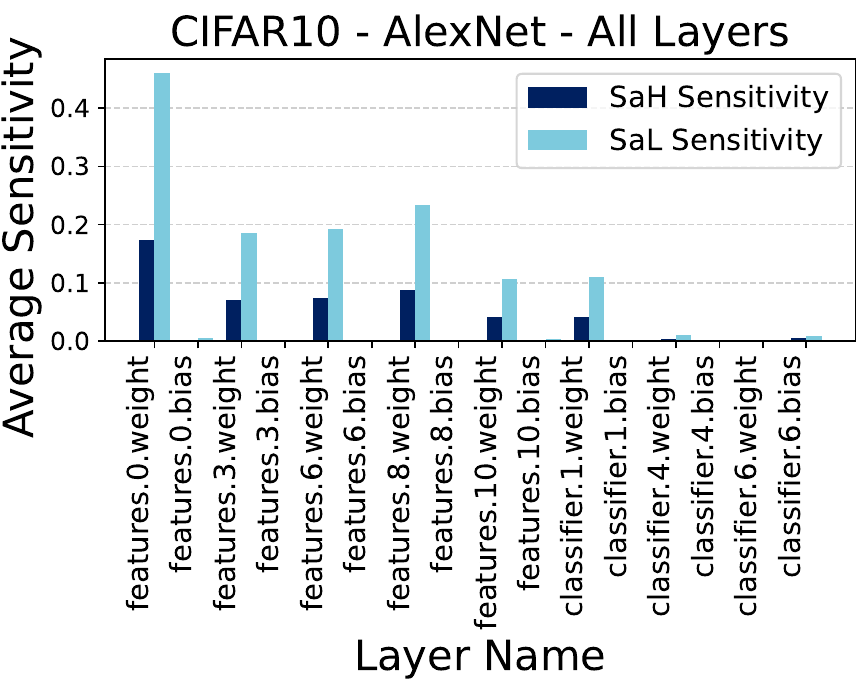}
		\caption{}
		\label{fig6-c}
	\end{subfigure}\hfill
	\caption{The layer-wise sensitivity of (a) \textit{SimpleNet}, (b) \textit{LeNet5}, and (c) \textit{AlexNet}. }
	\label{Fig6}
\end{figure}

The \textit{SaH} and \textit{SaL} sensitivity of each layer in \textit{SimpleNet}, \textit{LeNet5}, and \textit{AlexNet} are shown in Fig.~\ref{Fig6}. As shown, in \textit{LeNet5} and \textit{AlexNet}, the \textit{SaL} sensitivity of the weights is significantly higher than the \textit{SaH} sensitivity (on average, $2.45\times$ and $2.11\times$, respectively), indicating that the \textit{SaL} faults contribute more to CNN error compared to \textit{SaH} faults. In contrast, in \textit{SimpleNet}, the \textit{SaH} sensitivity is $1.21\times$ higher than the \textit{SaL} sensitivity, although in most layers their values are relatively close. In general, the early and middle layers are more vulnerable to faults, whereas the deeper layers are relatively robust.
Because, faults occurring in early CNN layers propagate through multiple subsequent layers and are repeatedly transformed by nonlinear operations, leading to cumulative distortion in feature extraction.
However, in the final layers (the classifiers), the sensitivity increases again.
Unlike earlier layers, errors in these layers are not compensated by subsequent transformations and directly alter the output logits. 
In \textit{AlexNet}, which includes three classifiers, the outer classifiers (those closer to the outputs) show high fault sensitivity. In earlier layers, the bias parameters have lower sensitivity than the weight parameters. For example, in \textit{LeNet5} and \textit{AlexNet}, the sensitivity of biases is negligible. However, the bias parameters of the last layer have higher sensitivity than the weights, as any change in these parameters directly alters the final output of CNNs. 

\begin{figure}
	\centering
	\begin{subfigure}{0.335\linewidth}
		\centering
		\includegraphics[width=\linewidth]{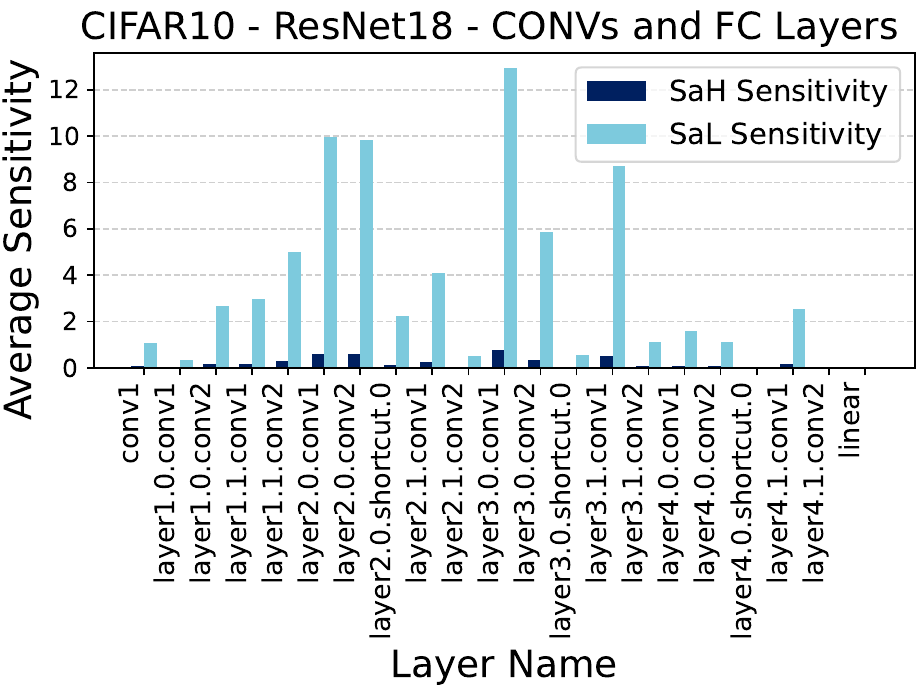} 
		\caption{}
		\label{fig7:a}
	\end{subfigure}\hfill
	\begin{subfigure}{0.32\linewidth}
		\centering
		\includegraphics[width=\linewidth]{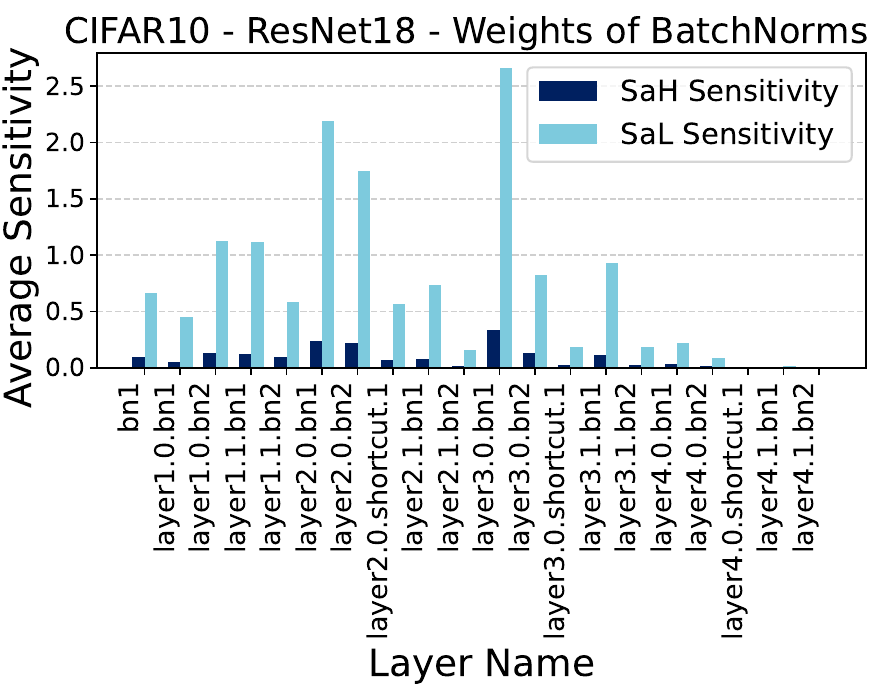}
		\caption{}
		\label{fig7:b}
	\end{subfigure}\hfill
	\begin{subfigure}{0.335\linewidth}
		\centering
		\includegraphics[width=\linewidth]{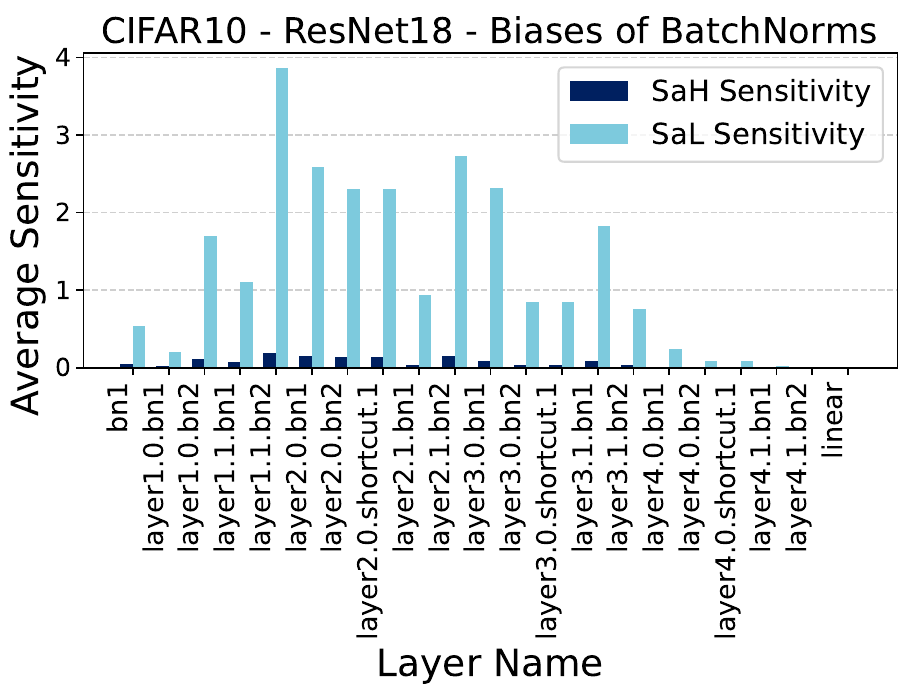}
		\caption{}
		\label{fig7:c}
	\end{subfigure}\hfill
	\caption{The layer-wise sensitivity of \textit{ResNet18} (a) weights of CONVs and FCs, (b) weights of BatchNorm, and (c) biases of BatchNorm and Conv $1\times1$}
	\label{Fig7}
\end{figure}

\begin{figure}
	\centering
	\begin{subfigure}{0.44\linewidth}
		\centering
		\includegraphics[width=\linewidth]{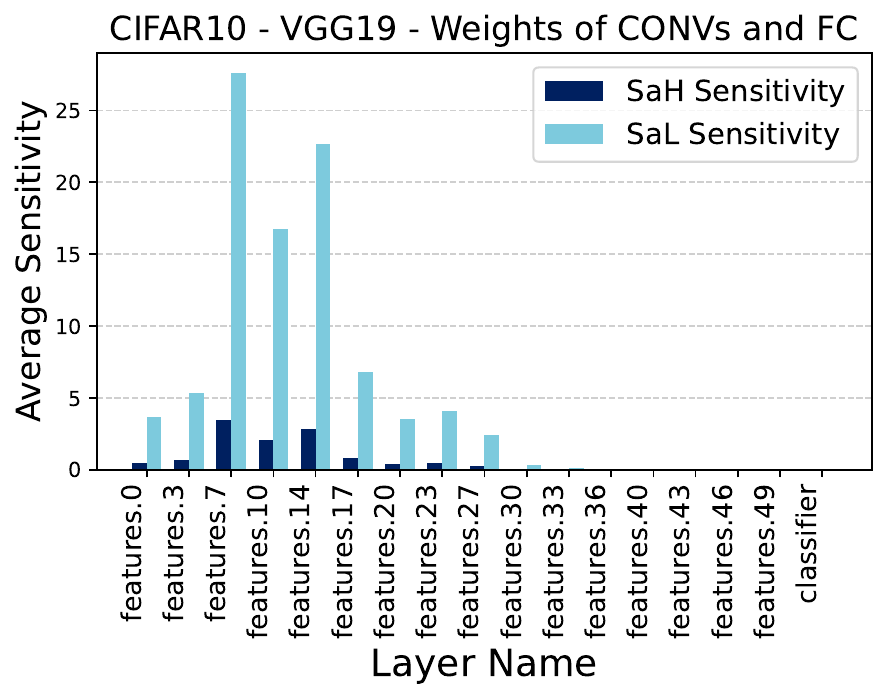} 
		\caption{}
		\label{fig8:a}
	\end{subfigure}\hfill
	\begin{subfigure}{0.44\linewidth}
		\centering
		\includegraphics[width=\linewidth]{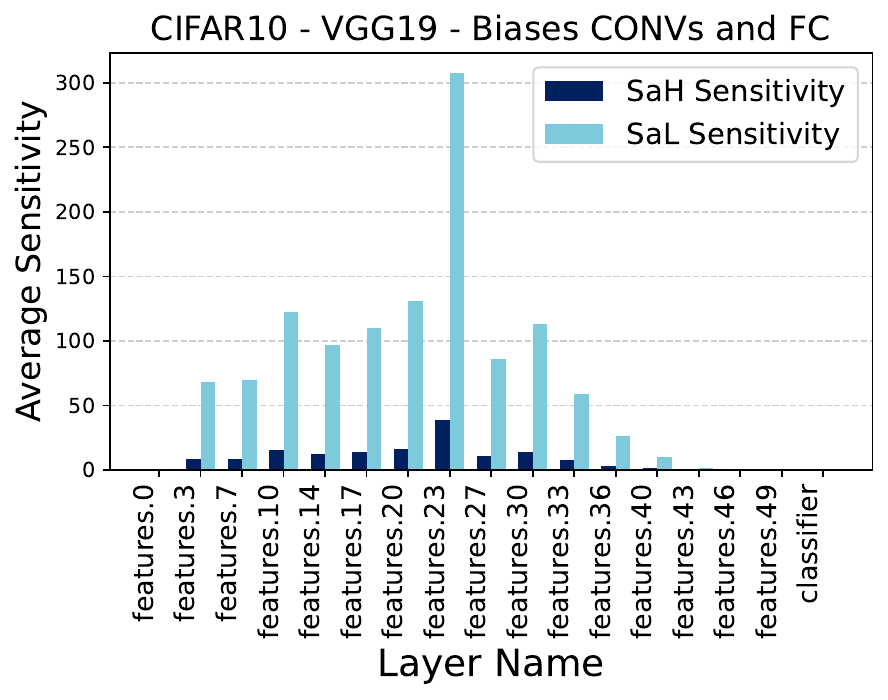}
		\caption{}
		\label{fig8:b}
	\end{subfigure}
	
	\begin{subfigure}{0.44\linewidth}
		\centering
		\includegraphics[width=\linewidth]{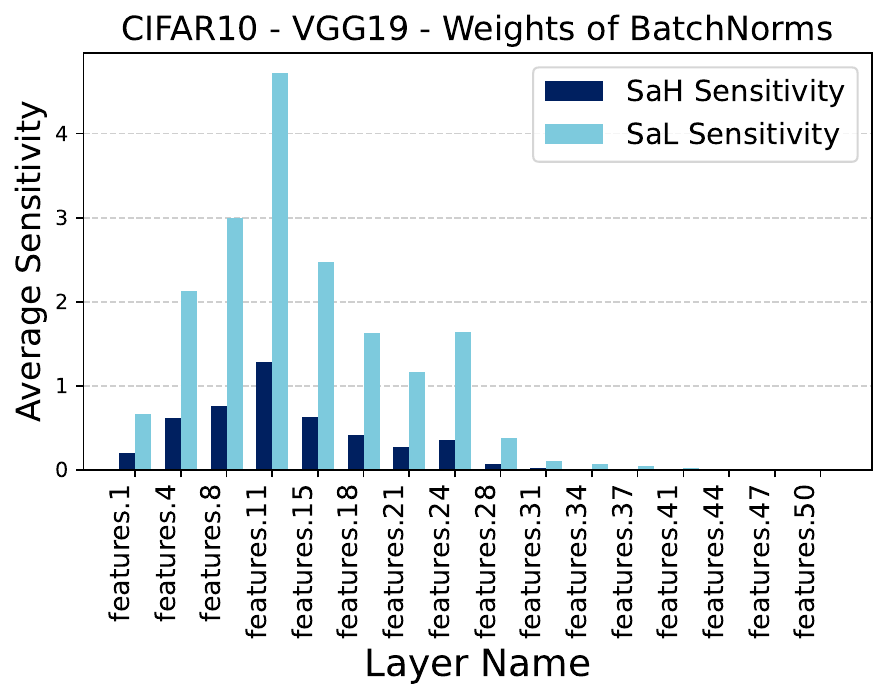}
		\caption{}
		\label{fig8:c}
	\end{subfigure}\hfill
		\begin{subfigure}{0.44\linewidth}
		\centering
		\includegraphics[width=\linewidth]{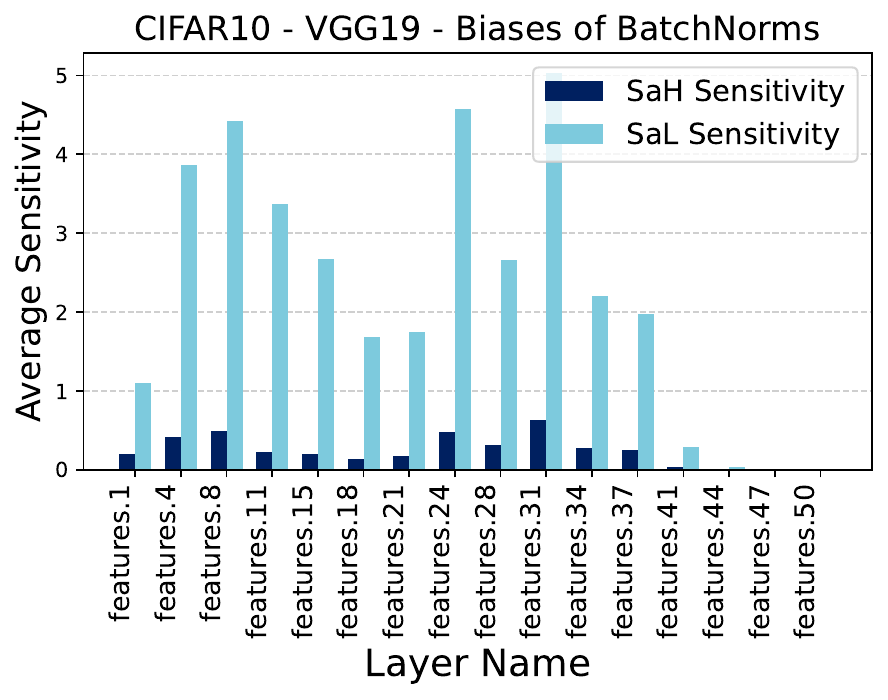}
		\caption{}
		\label{fig8:d}
	\end{subfigure}
	\caption{The layer-wise sensitivity of \textit{VGG19} (a) weights of CONVs and FCs, (b) biases of CONVs and FCs, (c) weights of BatchNorm, and (d) biases of BatchNorm.}
	\label{Fig8}
\end{figure}

The \textit{SaH} and \textit{SaL} sensitivity of weights and biases in CONV and FC layers, as well as the weights and biases of BatchNorm layers in \textit{ResNet18} and \textit{VGG19}, are shown in Figs.~\ref{Fig7} and~\ref{Fig8}. As shown, the overall \textit{SaL} sensitivity is $12\times$, on average, higher than \textit{SaH} sensitivity in \textit{ResNet18} and about $6.6\times$, on average, higher in \textit{VGG19}. Moreover, in \textit{ResNet18}, the sensitivity of weights and particularly the biases of CONV is more critical compared to that of the BatchNorm and FC layers. 
Residual connections in \textit{ResNet18} affect fault propagation by creating multiple signal paths, which may either reduce or increase the impact of a fault depending on the path in which it occurs.
Among the CONV weights, \textit{layer2.0.conv1}, \textit{layer2.0.conv2}, and \textit{layer3.0.conv1} exhibit relatively high sensitivity. For example, the \textit{SaH} and \textit{SaL} sensitivity values of \textit{layer3.0.conv1} are about $0.8$ and $12.9$, respectively, while for both \textit{layer2.0.conv1} and \textit{layer2.0.conv2} the corresponding values are about $0.61$ and $9.9$, respectively. In \textit{VGG19}, the biases of CONVs have the highest sensitivity, whereas the BatchNorms are comparatively less sensitive to faults. The sensitivity of the 9-th to 16-th layers (\textit{features.27} to \textit{features.49} of CONV and \textit{features.28} to \textit{features.50} in BatchNorms) is significantly lower than that of the preceding layers. 

\begin{figure}[!t]
	\centering
	\subfloat[\label{Fig9-a}]{
		\includegraphics[width=0.49\linewidth]{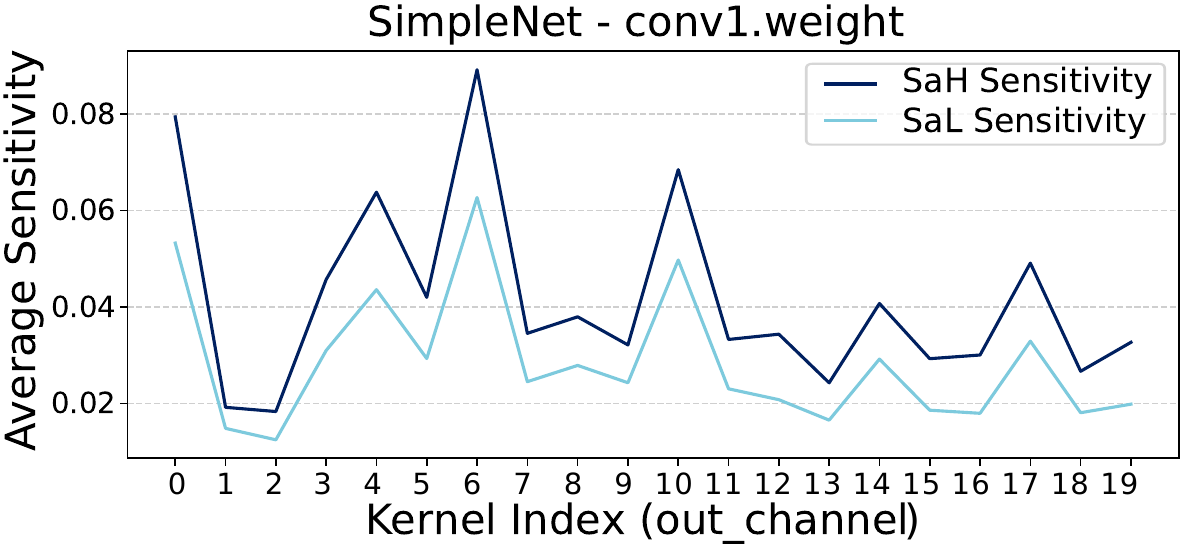} 
		\includegraphics[width=0.49\linewidth]{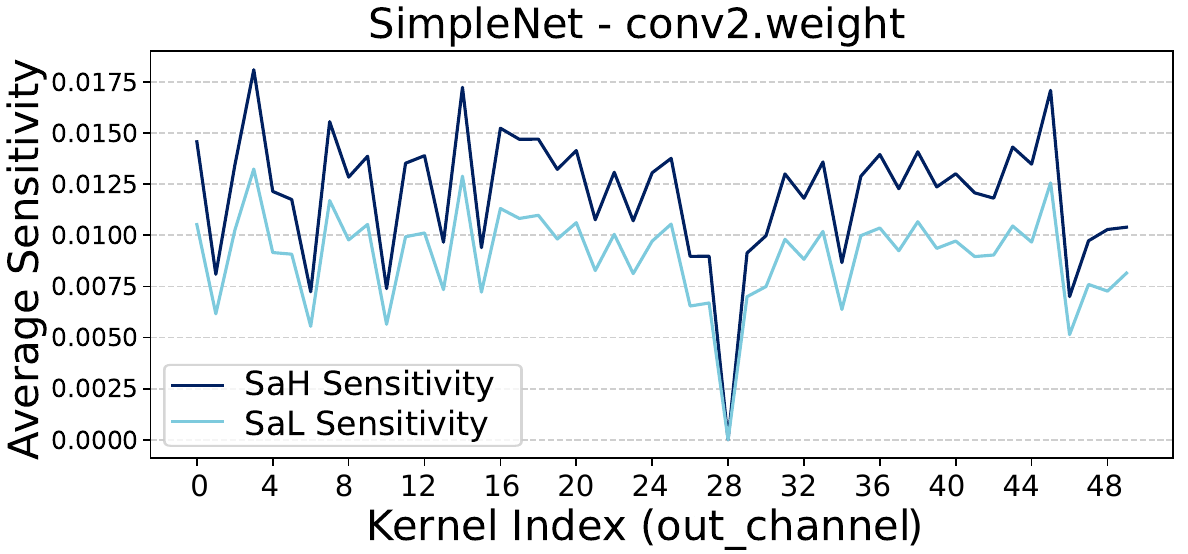}
	}
	
	\subfloat[\label{Fig9-b}]{
		\includegraphics[width=0.49\linewidth]{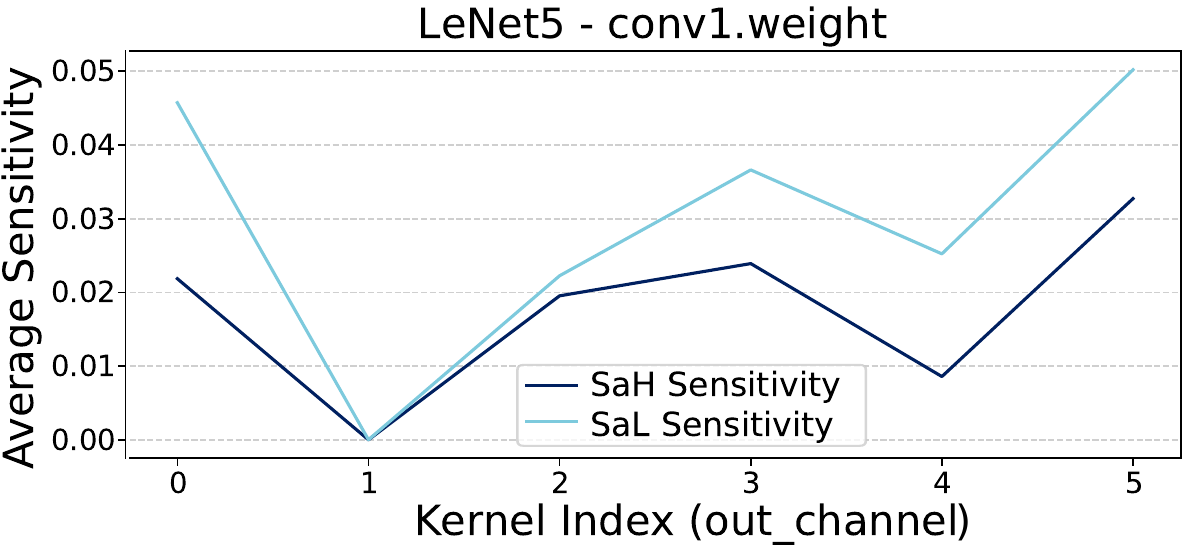}
		\includegraphics[width=0.49\linewidth]{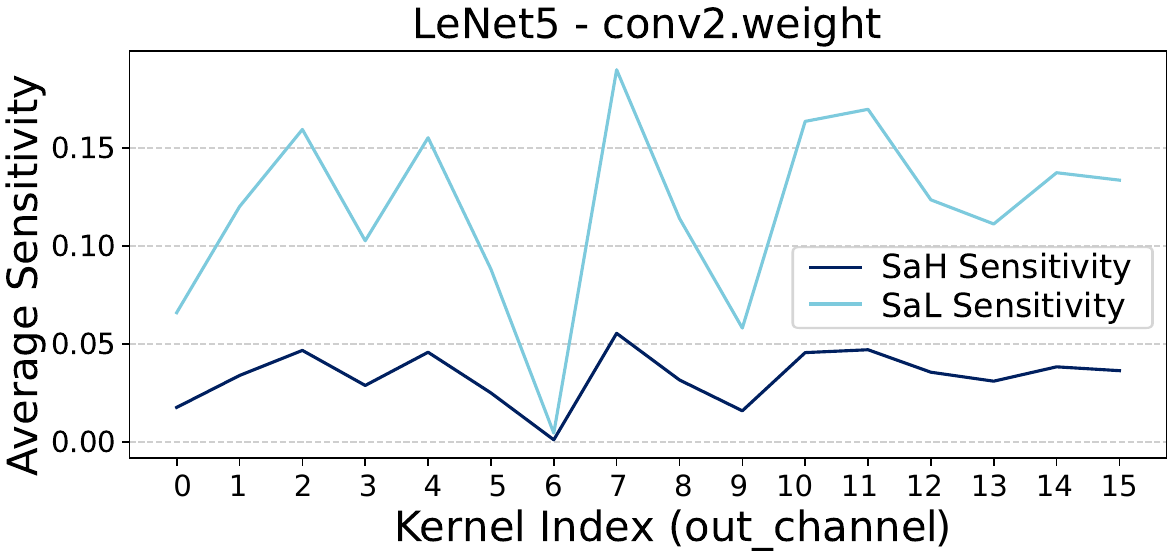}
	}
	\caption{The sensitivity of different kernels of CONV layers in (a) \textit{SimpleNet}, (b) \textit{LeNet5}.}
	\label{Fig9}
\end{figure}

\begin{figure}[!t]
	\centering
	\subfloat{
		\includegraphics[width=0.49\linewidth]{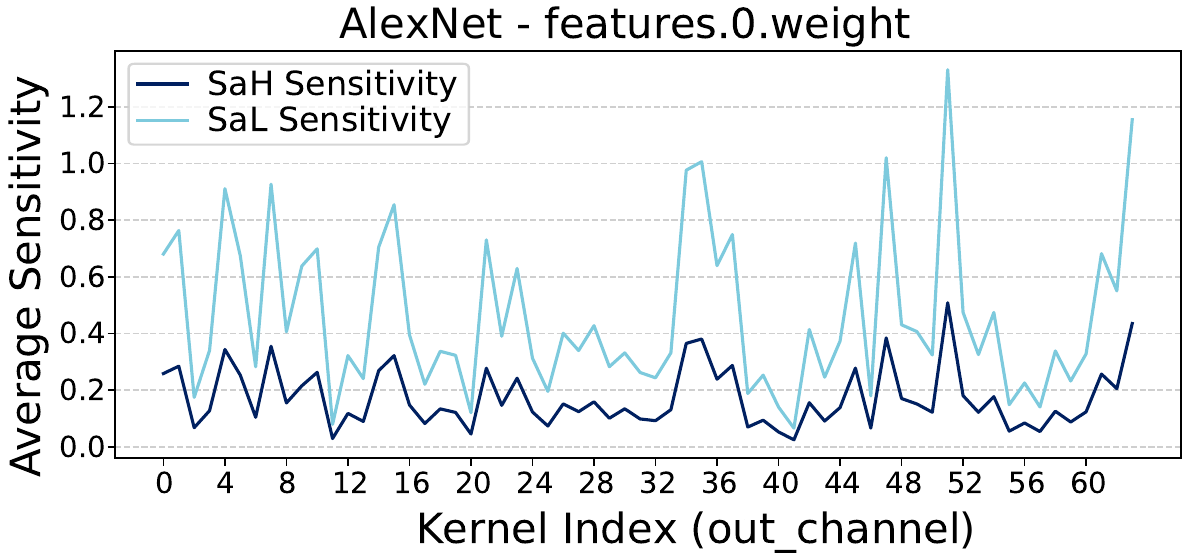} 
		\includegraphics[width=0.49\linewidth]{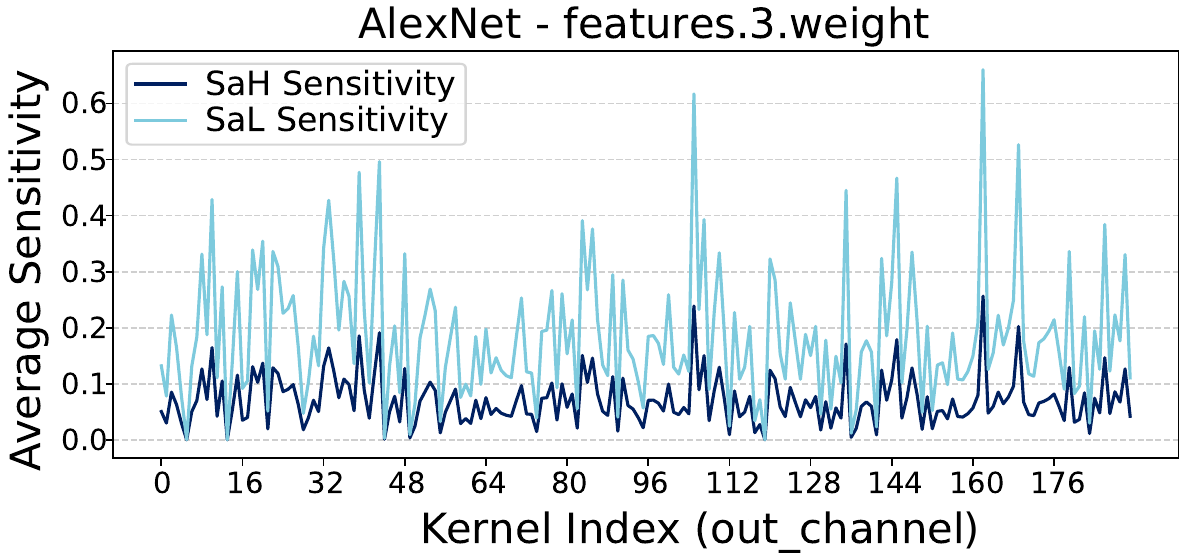}
	}
	
	\subfloat{
		\includegraphics[width=0.49\linewidth]{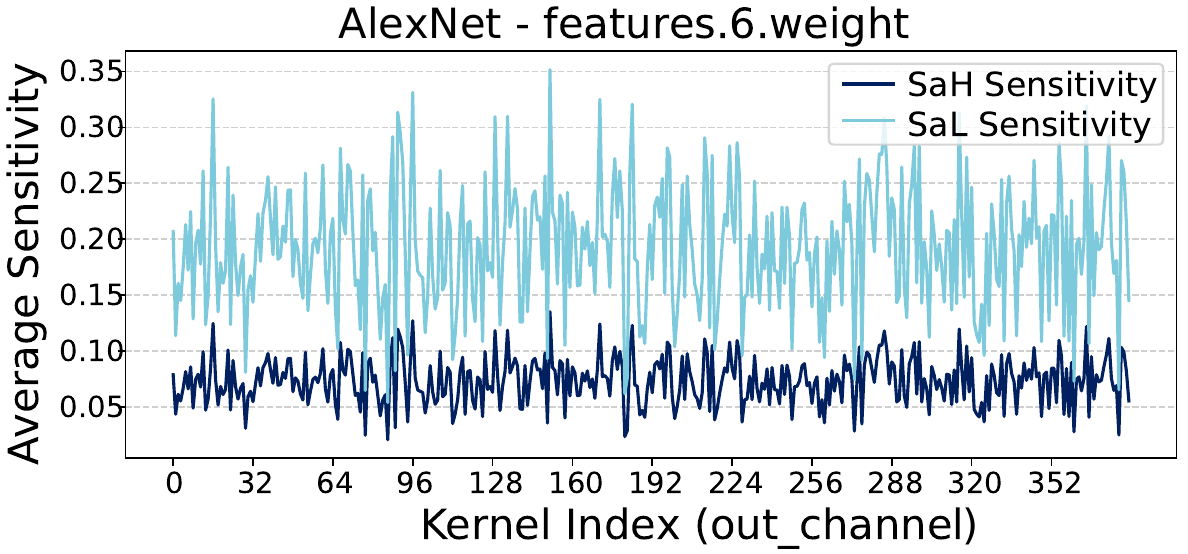}
		\includegraphics[width=0.49\linewidth]{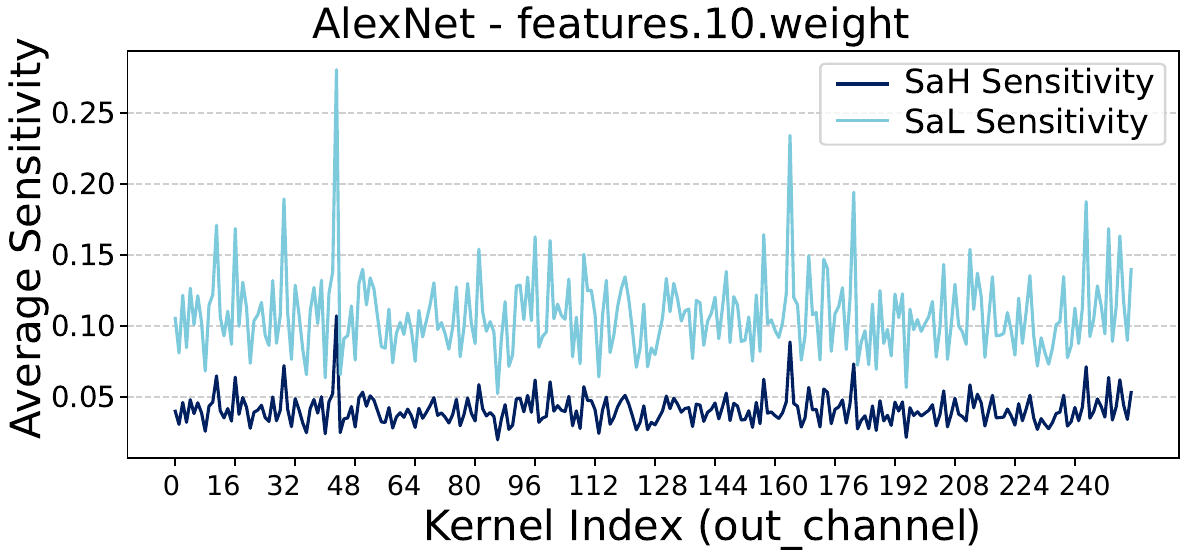}
	}
	\caption{The sensitivity of different kernels of CONV layers in \textit{AlexNet}.}
	\label{Fig10}
\end{figure}

\begin{figure}[!t]
	\centering
	\subfloat{
		\includegraphics[width=0.49\linewidth]{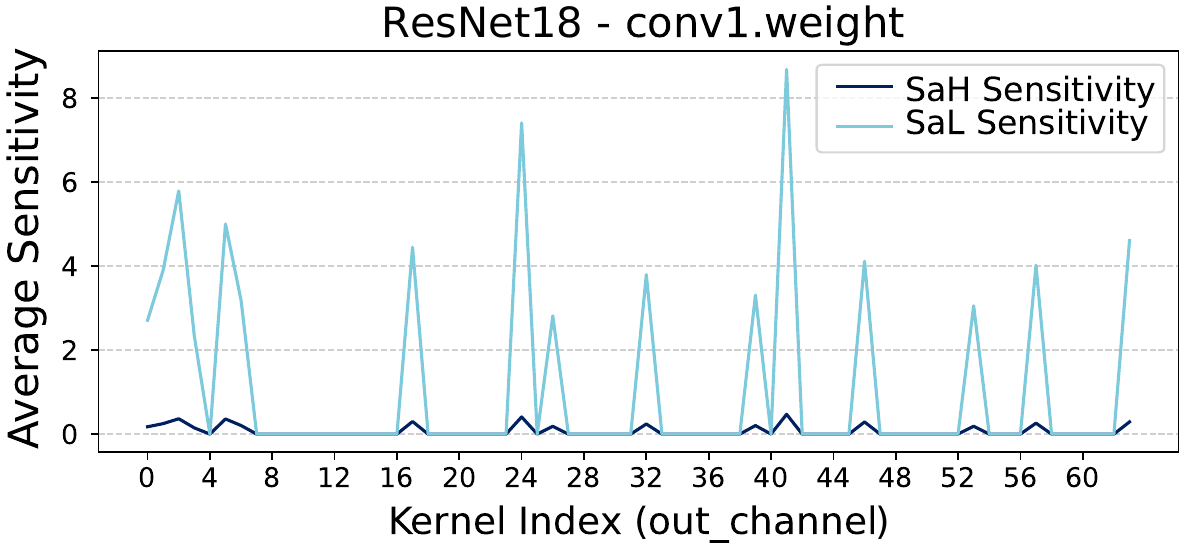} 
		\includegraphics[width=0.49\linewidth]{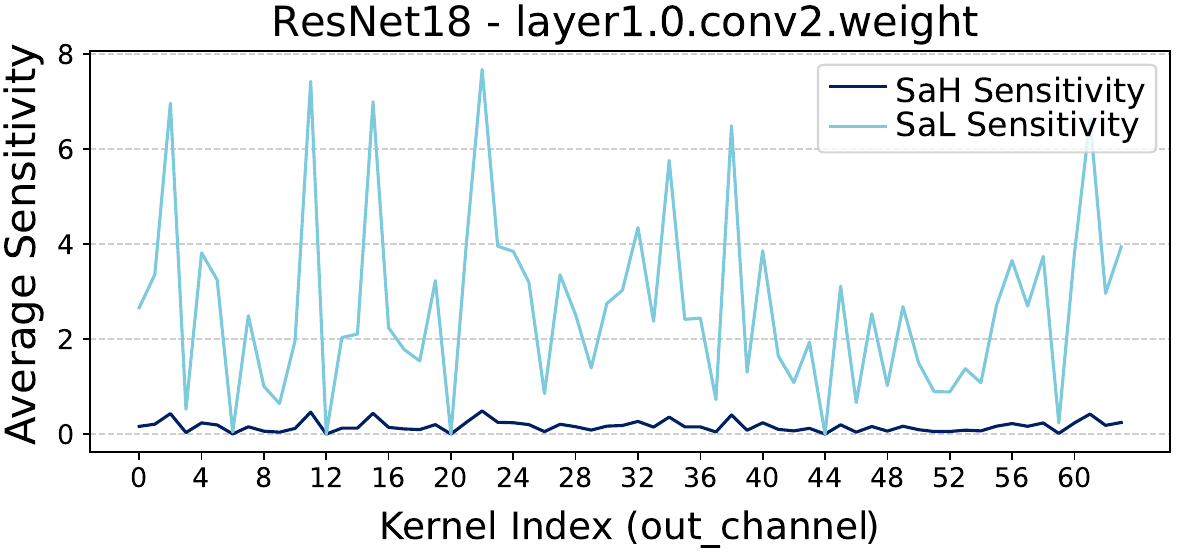}
	}
	
	\subfloat{
		\includegraphics[width=0.49\linewidth]{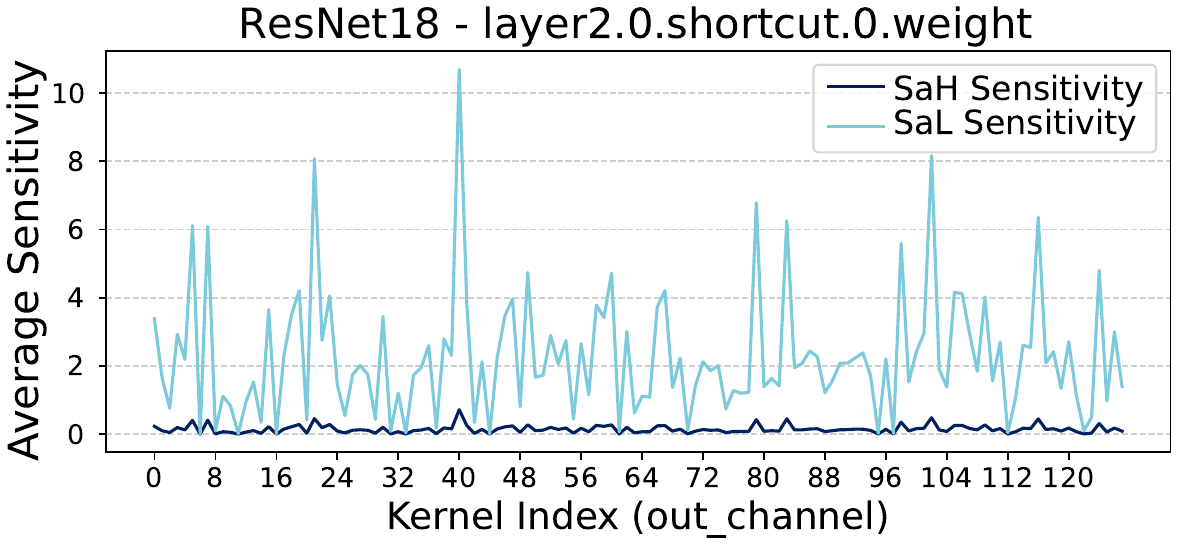}
		\includegraphics[width=0.49\linewidth]{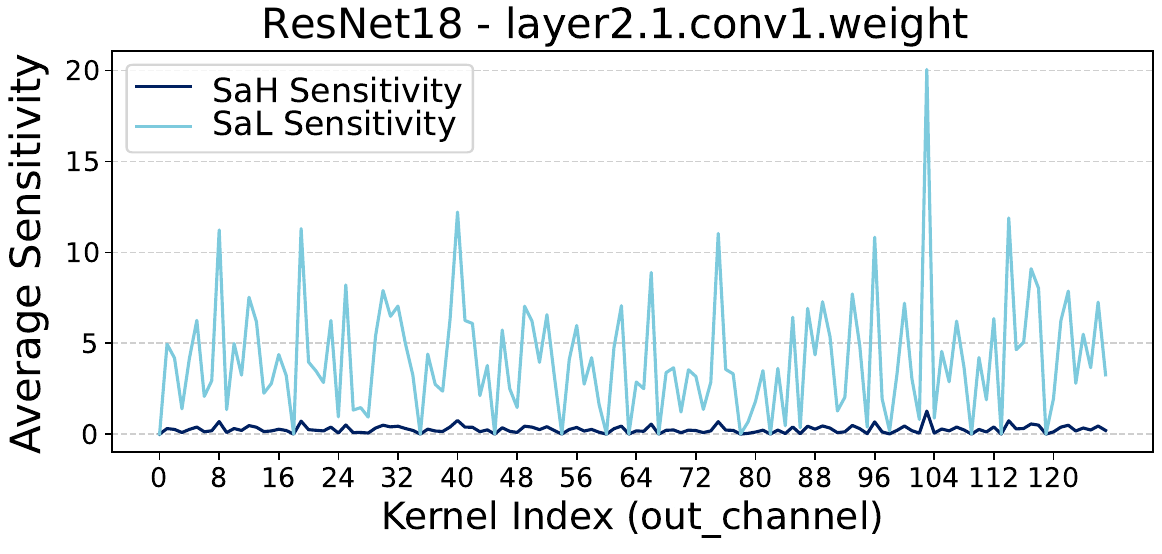}
	}
	
		\subfloat{
		\includegraphics[width=0.49\linewidth]{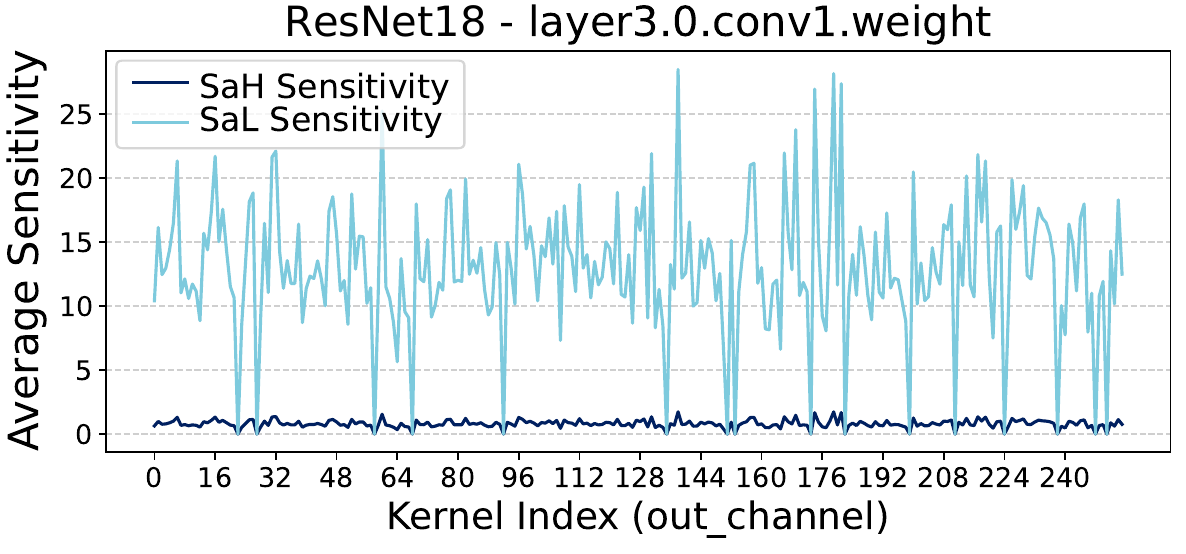}
		\includegraphics[width=0.49\linewidth]{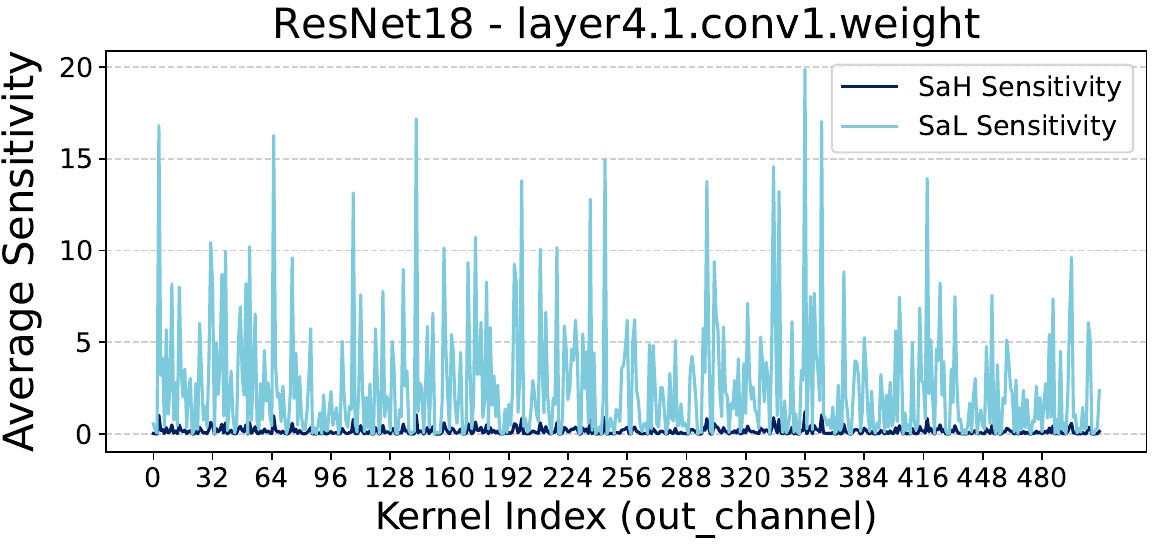}
	}
	\caption{The kernel-wise sensitivity analyzing of CONV layers in \textit{ResNet18}.}
	\label{Fig11}
\end{figure}

\begin{figure}[!t]
	\centering
	\subfloat{
		\includegraphics[width=0.49\linewidth]{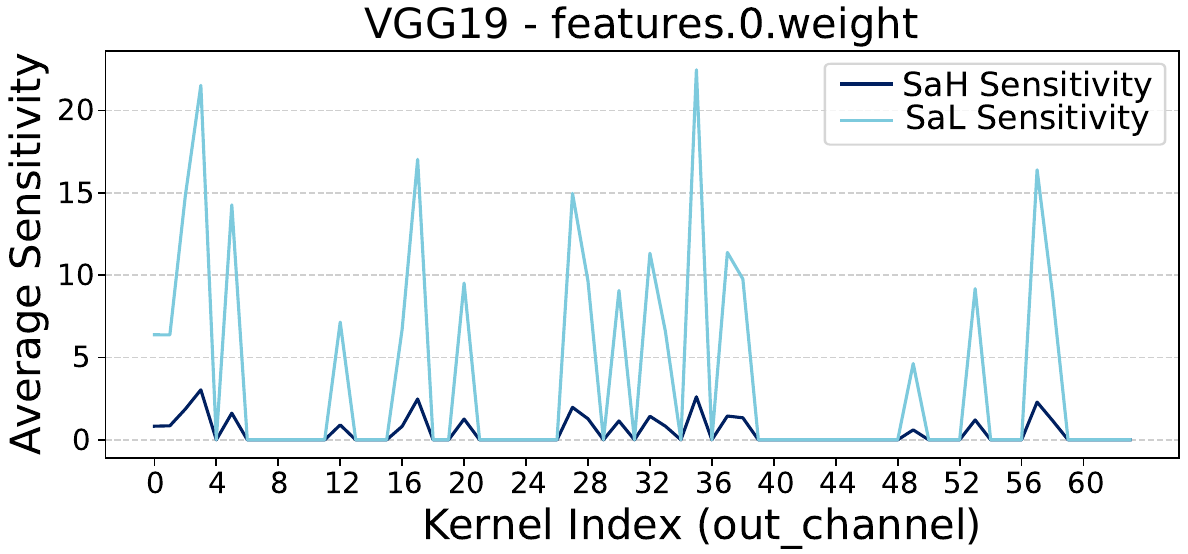} 
		\includegraphics[width=0.49\linewidth]{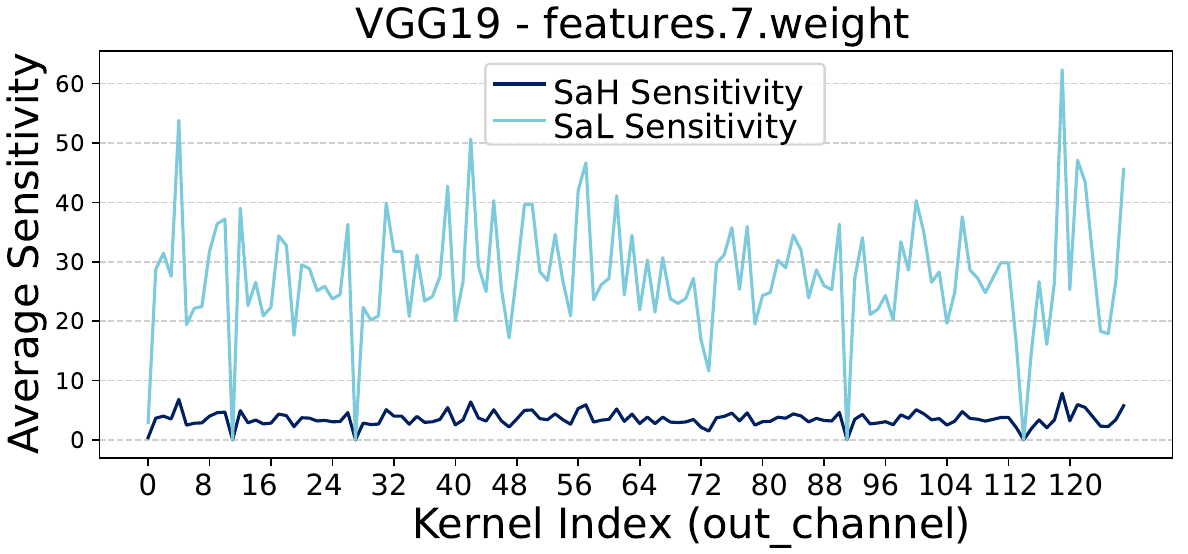}
	}
	
	\subfloat{
		\includegraphics[width=0.49\linewidth]{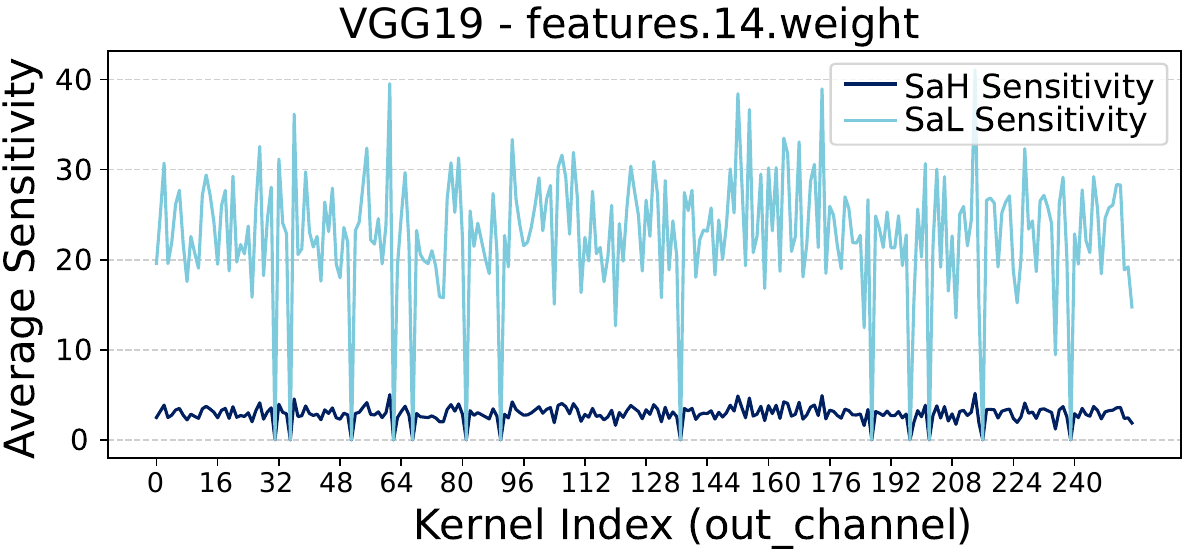}
		\includegraphics[width=0.49\linewidth]{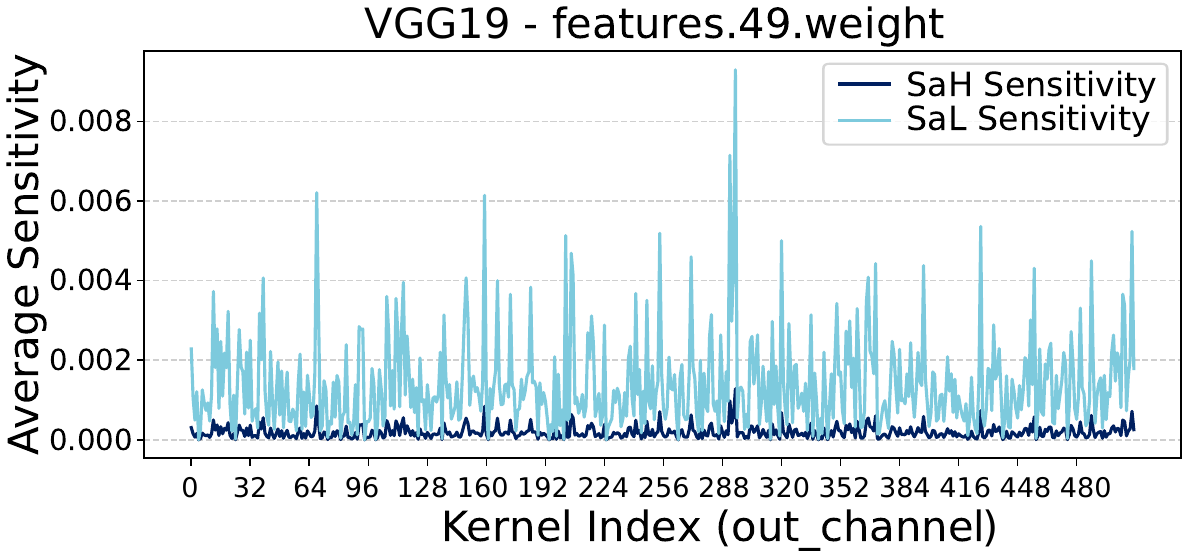}
	}
	\caption{The kernel-wise sensitivity analysis of CONV layers in \textit{VGG19}.}
	\label{Fig12}
\end{figure}

Different weights exhibit diverse levels of fault sensitivity, and this variation also extends to convolutional kernels, which also show distinct sensitivity levels across varying kernels of the same layers. This varying sensitivity reflects the non-uniform vulnerability of different layers. The kernel-wise \textit{SaH} and \textit{SaL} sensitivities of some selected CONV layers across different models are illustrated in Figs.~\ref{Fig9}~to~\ref{Fig12}. By kernel-wise analysis, highly sensitive kernels can be identified and prioritized in fault-tolerant techniques. Kernels that have high sensitivity values are more critical than others; consequently, faults in these kernels have a greater impact on accuracy. There is no clear or regular pattern in the order of kernels and sensitivity values.
Some kernels capture dominant features, while others extract less critical patterns. Consequently, faults in certain kernels lead to substantial accuracy degradation, whereas others remain non-critical.
For example, in \textit{conv1.weight} of \textit{SimpleNet} (Fig.~\ref{Fig9-a}), the kernel with index six has the highest sensitivity, whereas, kernels with index one and two have the lowest sensitivity. In \textit{ResNet18} (Fig.~\ref{Fig11}), most kernels in \textit{Conv1.weight} have negligible sensitivity values. These kernels are known as non-critical kernels. On the other hand, only 6 out of 64 kernels in \textit{layer1.0.conv2.weight} have low sensitivity values. 

\begin{figure}[!t]
	\centering
	\includegraphics[width=1\linewidth]{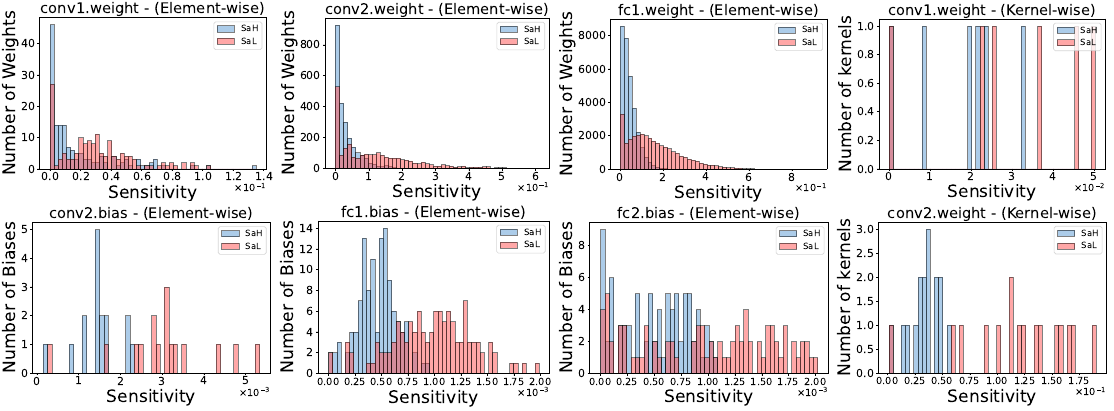}
	\caption{Element-wise and kernel-wise sensitivity histogram of \textit{LeNet5}.}
	\label{Fig13}
\end{figure}

\begin{figure}[!t]
	\centering
	\includegraphics[width=1\linewidth]{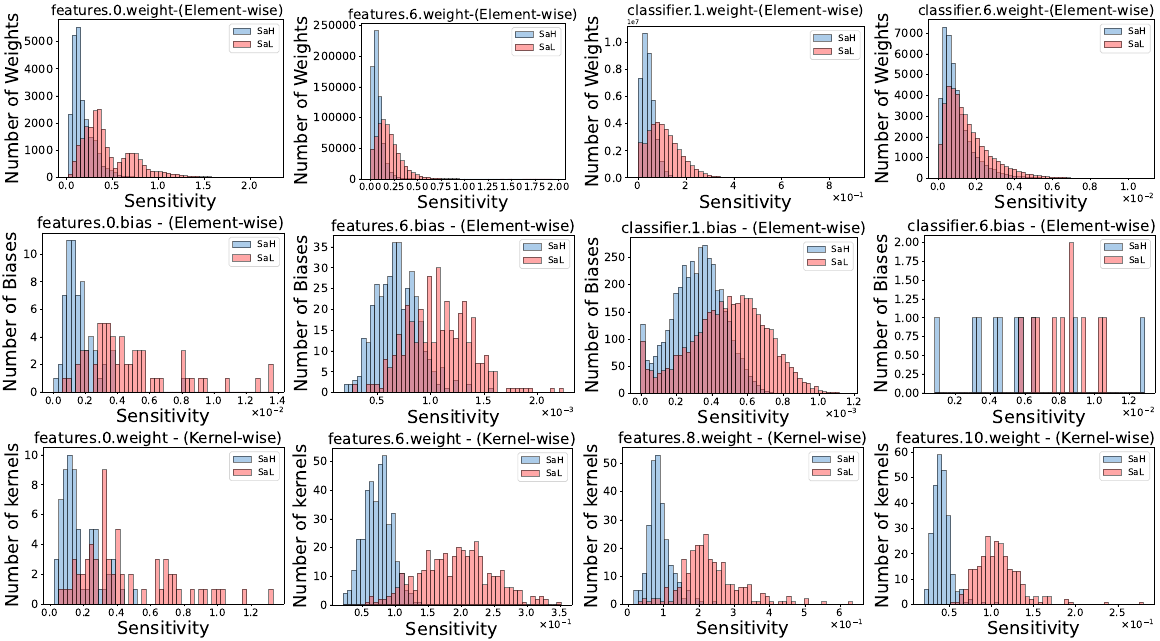}
	\caption{Element-wise and kernel-wise sensitivity histogram of \textit{AlexNet}.}
	\label{Fig14}
\end{figure}

\begin{figure}[!t]
	\centering
	\includegraphics[width=1\linewidth]{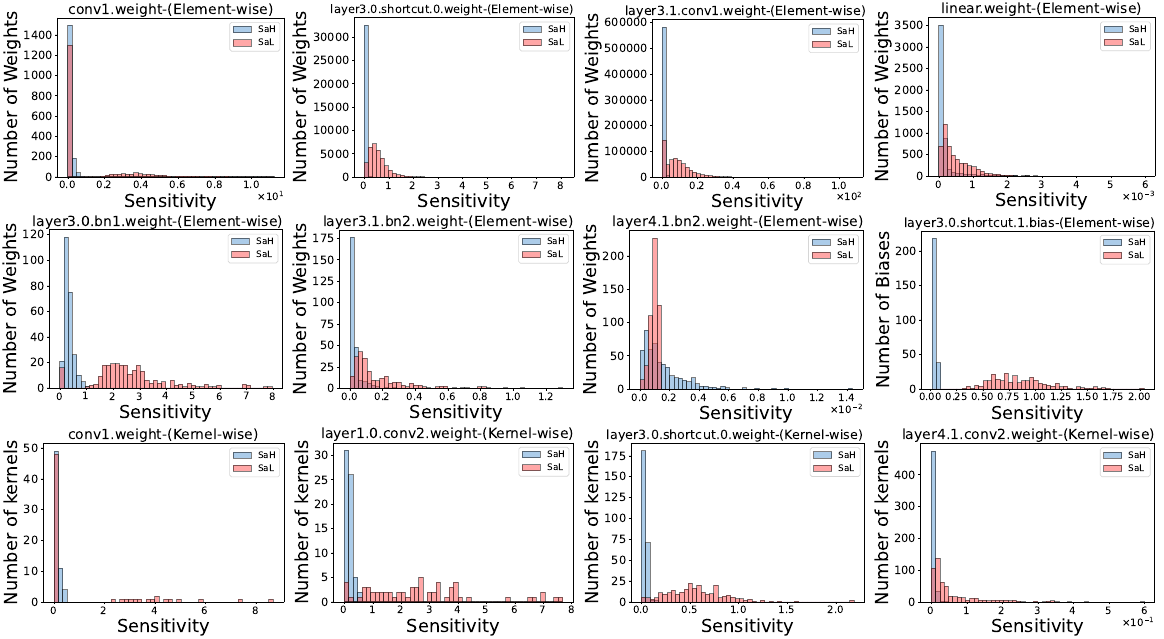}
	\caption{Element-wise and kernel-wise sensitivity histogram of \textit{ResNet18}.}
	\label{Fig15}
\end{figure}

The element-wise and kernel-wise sensitivity histogram of some layers are shown in Figs.~\ref{Fig13}~to~\ref{Fig15}. In the element-wise histograms, the number of weights at different \textit{SaH} and \textit{SaL} sensitivity values are depicted, while in the kernel-wise histogram, the number of kernels across this sensitivity values are shown for each layer. As shown in these figures, most of weights in the CONV layers exhibits low sensitivity values (close to zero). This observation indicates that faults in these weights have little impact on the accuracy of the CNNs. In contrast, only a few elements have high sensitivity. Nevertheless, these small numbers of weights with significantly high sensitivity are critical weights, as faults occurring in them drastically degrade the performance of CNNs. In fact, these few weights with high sensitivity values are potential weak points of the CNNs. For example, in \textit{classifier1.bias} of \textit{AlexNet} shown in Fig.~\ref{Fig14}, most weights have medium sensitivity values. Similarly, the majority of kernels also have medium sensitivity values in \textit{features6} and \textit{feautures8}. 

Moreover, the sensitivity distribution is not identical for \textit{SaH} and \textit{SaL} faults. The number of sensitive elements and kernels differs between the two fault types. For example, in \textit{AlexNet}, a larger proportion of weights and kernels are more vulnerable to \textit{SaL} faults compared to \textit{SaH}, indicating a higher vulnerability of the network to \textit{SaL} faults. As shown in Fig.~\ref{Fig15}, in \textit{layer4.1.bn2.weights} of \textit{ResNet18} (weights of BatchNorms), a greater number of elements have high \textit{SaH} sensitivity. Conversely, the number of elements with high \textit{SaL} sensitivity is more than element with high \textit{SAH} sensitivity in \textit{layer3.0.bn1.weights} and \textit{layer3.0.shortcut0.bias}.

\begin{figure}[!t]
	\centering
	\subfloat[\label{Fig16-a}]{
		\includegraphics[width=0.33\linewidth]{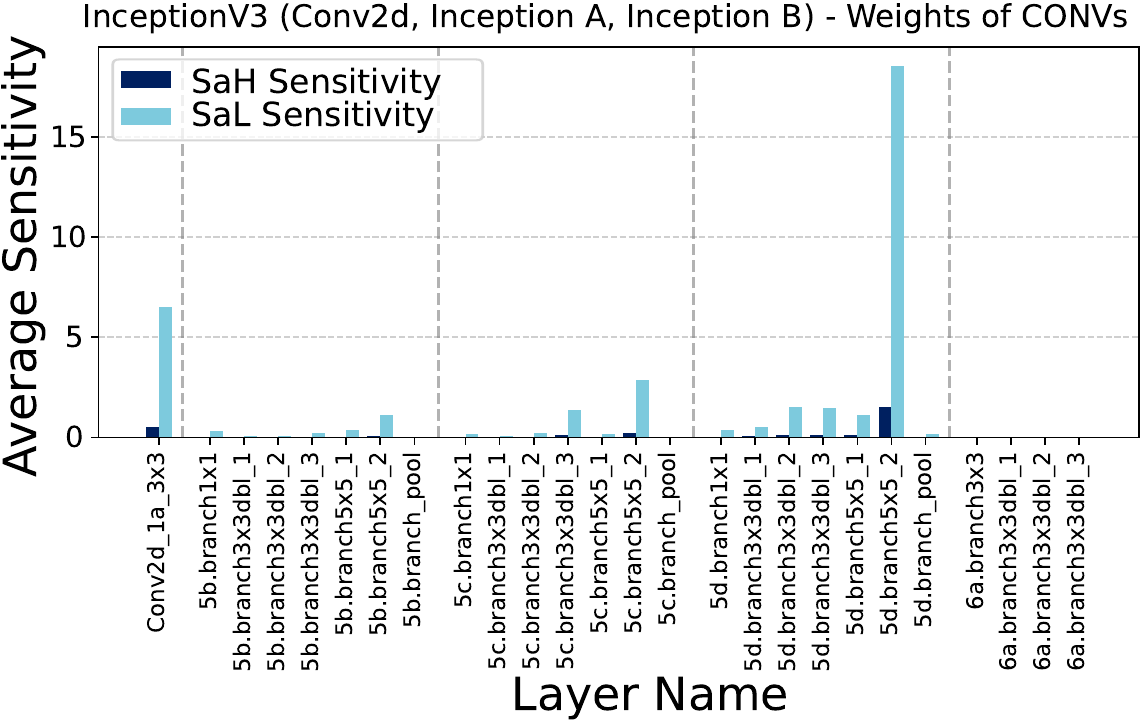} 
		\includegraphics[width=0.34\linewidth]{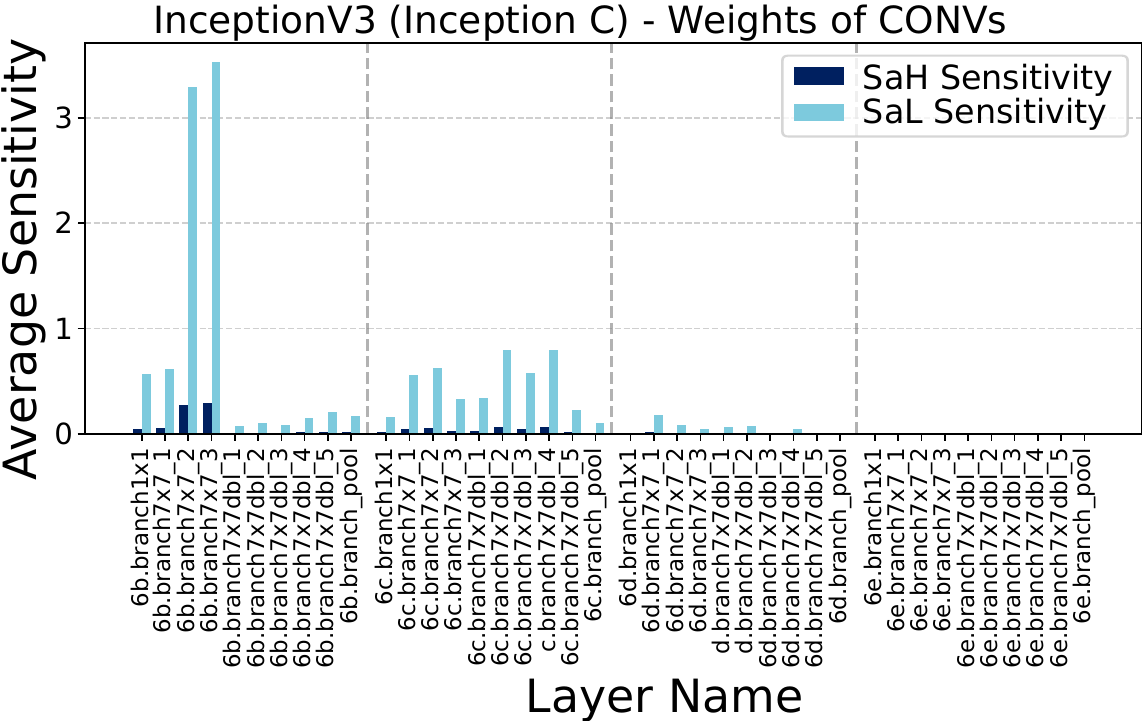}
		\includegraphics[width=0.33\linewidth]{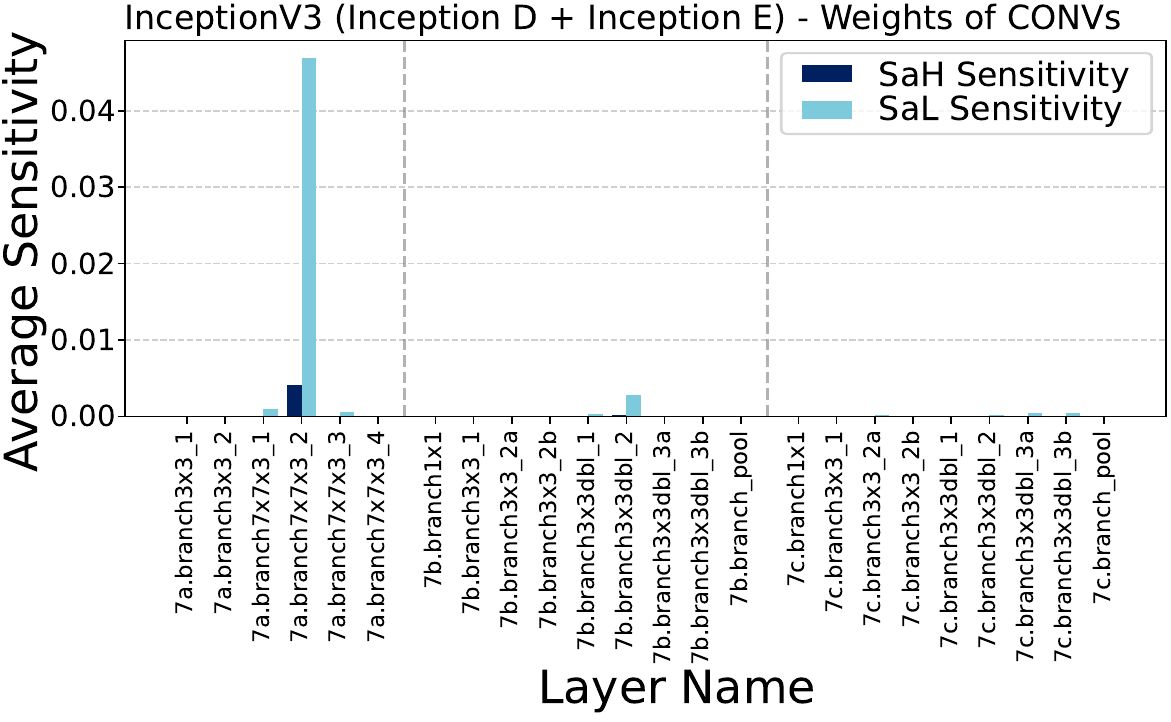}
	}
	
	\subfloat[\label{Fig16-b}]{
		\includegraphics[width=0.33\linewidth]{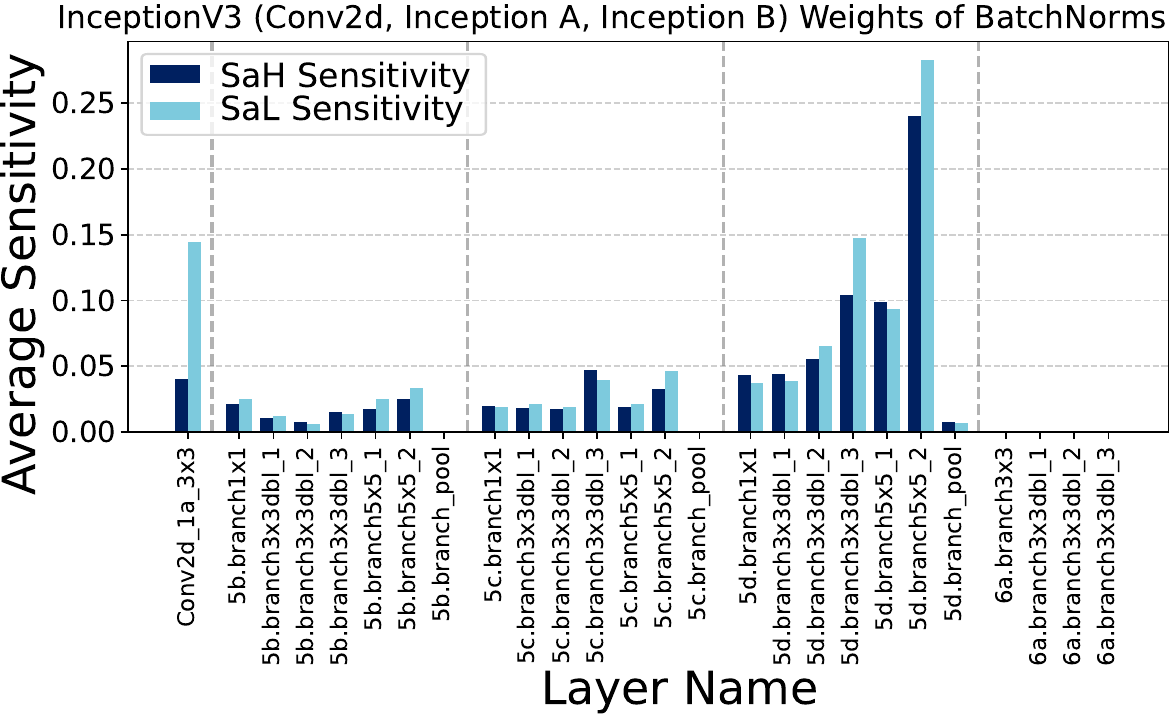}
		\includegraphics[width=0.34\linewidth]{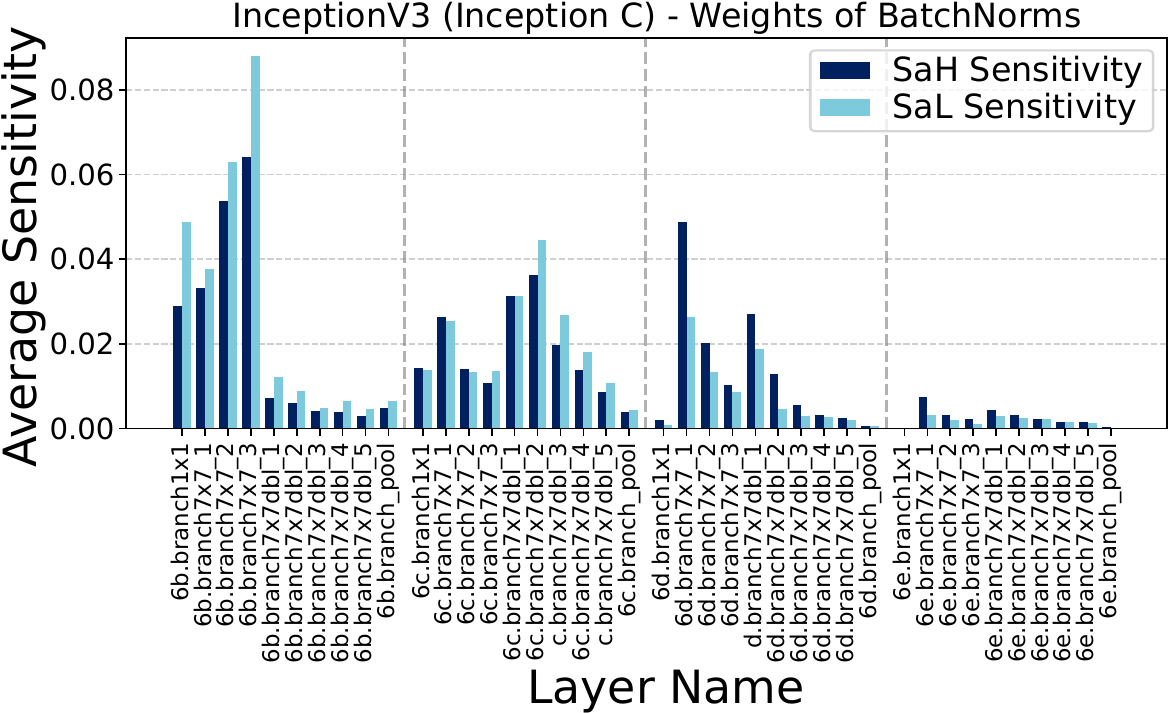}
		\includegraphics[width=0.33\linewidth]{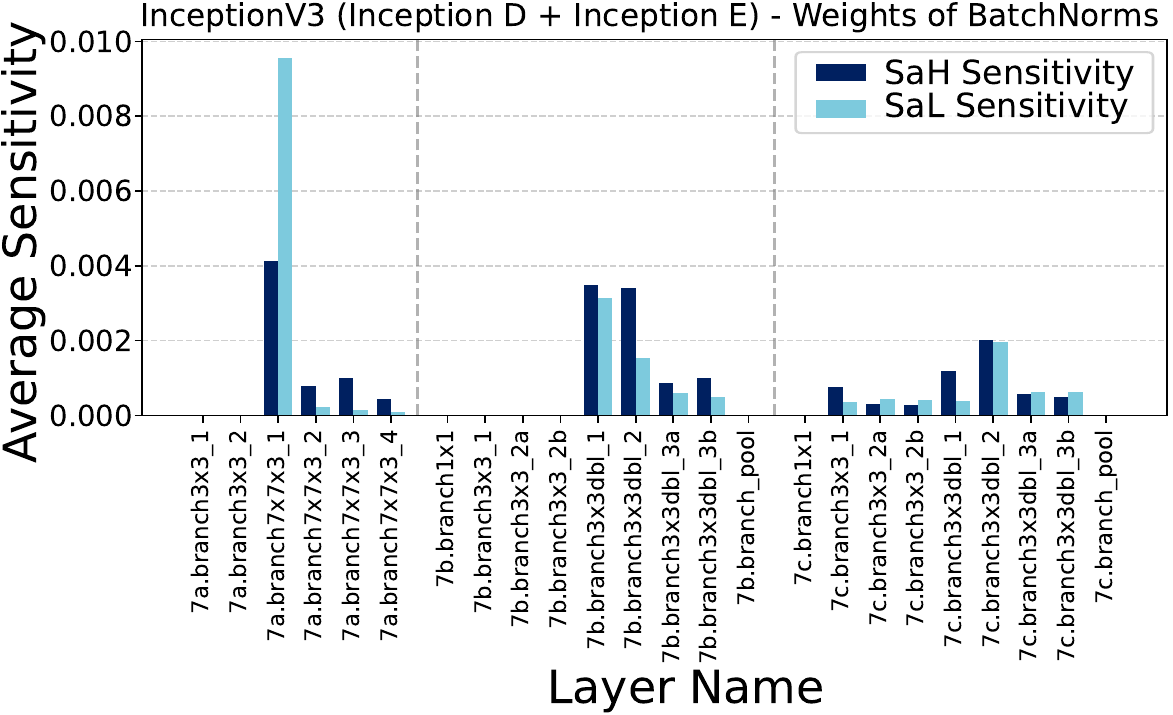}
	}
	
	\subfloat[\label{Fig16-c}]{
		\includegraphics[width=0.33\linewidth]{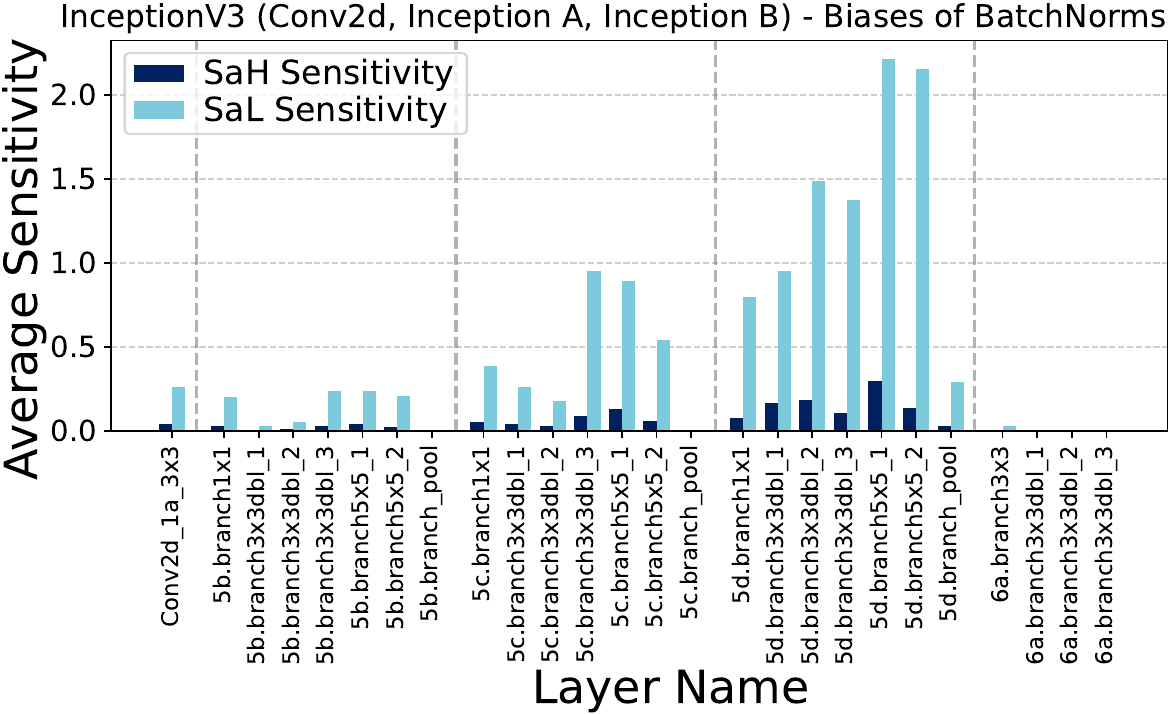}
		\includegraphics[width=0.34\linewidth]{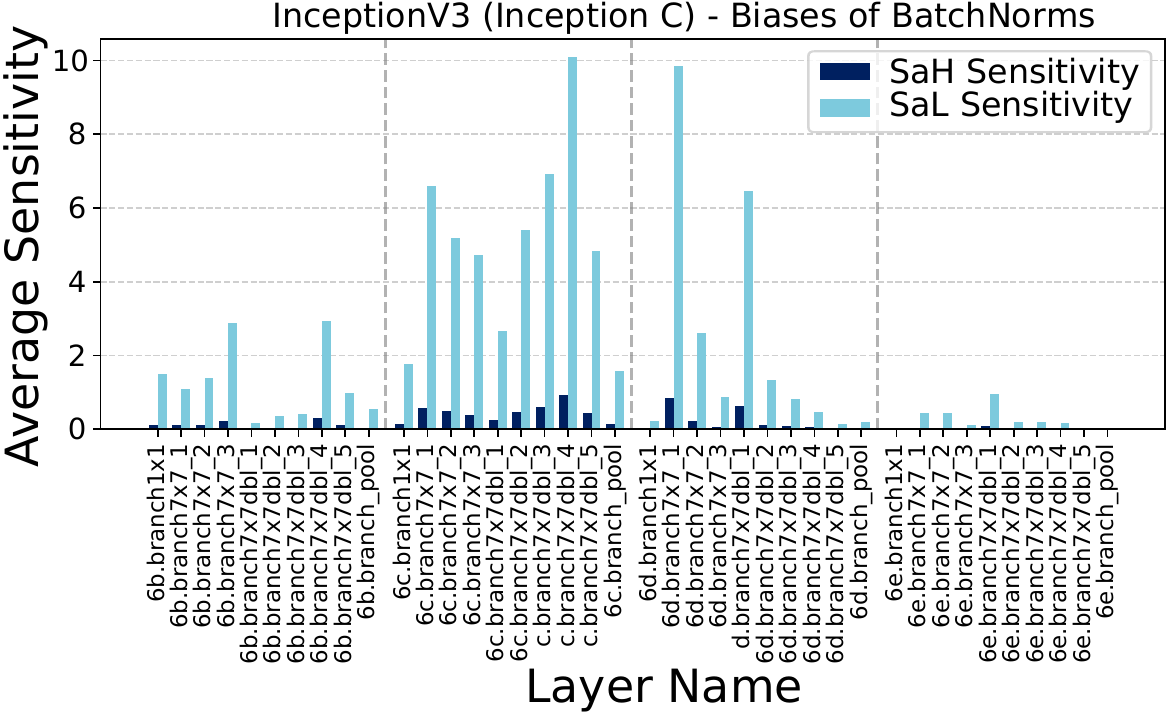}
		\includegraphics[width=0.33\linewidth]{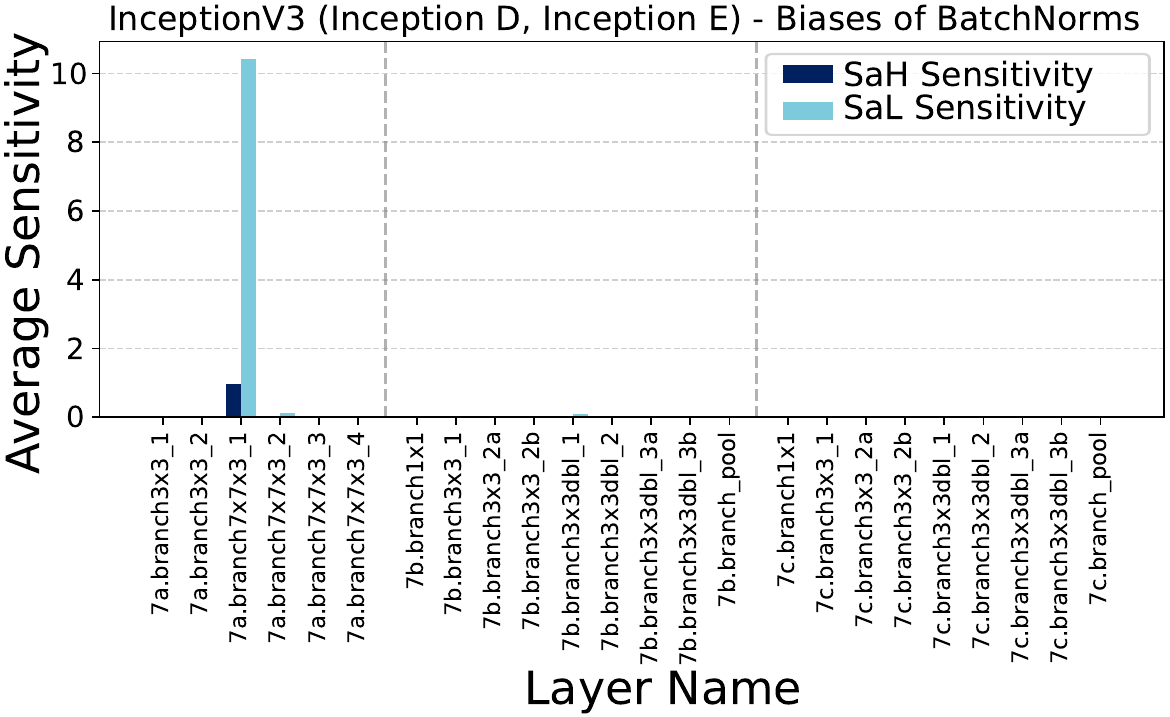}
	}
	\caption{The layer-wise sensitivity of various layers in \textit{InceptionV3} to \textit{SaH} and \textit{SaL} faults, (a) weights of CONVs, (b) weights of BatchNorm, and (c) biases of BatchNorm.}
	\label{Fig16}
\end{figure}

A comprehensive layer-wise sensitivity analysis was conducted for the CONV weights as well as the weights and biases of BatchNorm layers in \textit{InceptionV3}, and the results are illustrated in Fig.~\ref{Fig16}. Due to the large number of layers, results are presented in separate charts for the initial CONV layers, \textit{Inception A} and \textit{Inception B} modules, \textit{Inception C} module, and \textit{Inception D} and \textit{E} modules. Within each chart, layers of individual blocks are separated by vertical dashed lines. It should be noted that the sensitivity scales differ across the charts to show the variations in low values.

As shown, different blocks and modules exhibit varying degrees of sensitivity to \textit{SaH} and \textit{SaL} faults. On average, the \textit{SaL} sensitivity values of CONV weights are approximately $3.5\times-12.3\times$ higher than their \textit{SaH} sensitivity values. A similar trend is observed in BatchNorm biases, although not all BatchNorm weights follow this pattern. Specifically, among the 90 BatchNorm layers, 37 exhibit higher \textit{SaH} sensitivity than \textit{SaL} sensitivity, (by $1.4\times$, on average). For instance, nearly all BatchNorm layers in blocks \textit{6d} (except \textit{6d.branch7x7dbl\_4}) and \textit{6e} (except \textit{6e.branch1x1}) show higher sensitivity to \textit{SaH} faults. This variation is particularly observed in the third path of \textit{Inception E} (the 5th to 8th layers in both blocks \textit{7b} and \textit{7c}).

As shown, layers in the \textit{Inception B} module demonstrate notably low sensitivity, indicating that faults in these parameters do not significantly propagate through the network and have a low impact on the final output.
For CONV layers, the initial layer (\textit{Conv2d\_1a\_3x3}) and \textit{5d\_branch5x5\_2} exhibit the highest sensitivity. As the depth of layers increases, the sensitivity of CONV layers decreases. This behavior is explained by the fact that faults in early layers propagate through multiple parallel paths toward the network output.
However, due to the parallel structure of \textit{Inception B} blocks, sensitivity also varies across paths.
Because, the \textit{Inception} modules generate features through several parallel branches that are later concatenated, a fault in a single branch affects only part of the final feature representation, while the remaining branches still contribute correct features, mitigating the overall impact of the fault.
For example, the first and third paths of blocks in \textit{Inception A} have higher sensitivity values than other paths. Within \textit{Inception A}, later blocks (for example, \textit{5d}) show greater sensitivity than earlier ones (for example, \textit{5b}). In contrast, in \textit{Inception C}, the opposite trend is observed, with earlier blocks being more sensitive. Regarding BatchNorm biases, outer layers generally exhibit lower sensitivity values, while block \textit{6c} displays significantly higher values compared to other blocks.

\begin{figure}[!t]
	\centering
	\subfloat{
		\includegraphics[width=0.5\linewidth]{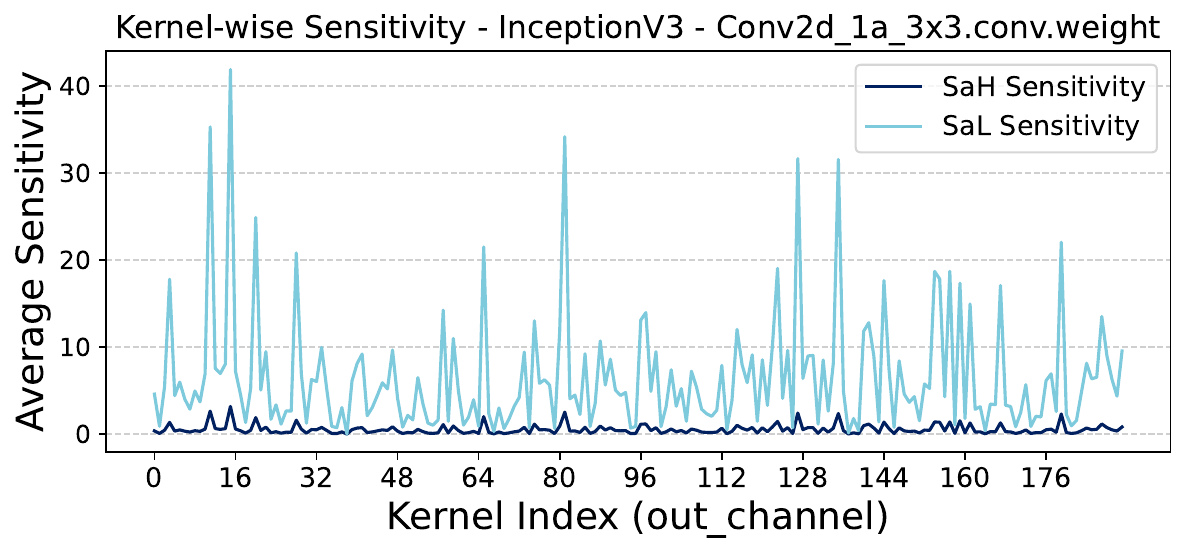} 
		\includegraphics[width=0.5\linewidth]{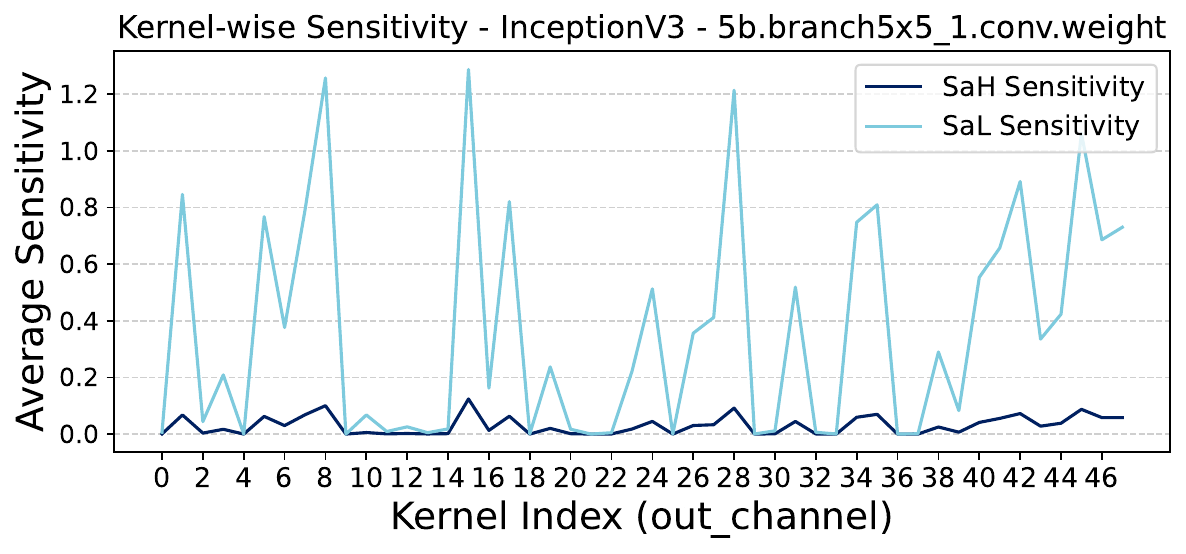}
	}
	
	\subfloat{
		\includegraphics[width=0.5\linewidth]{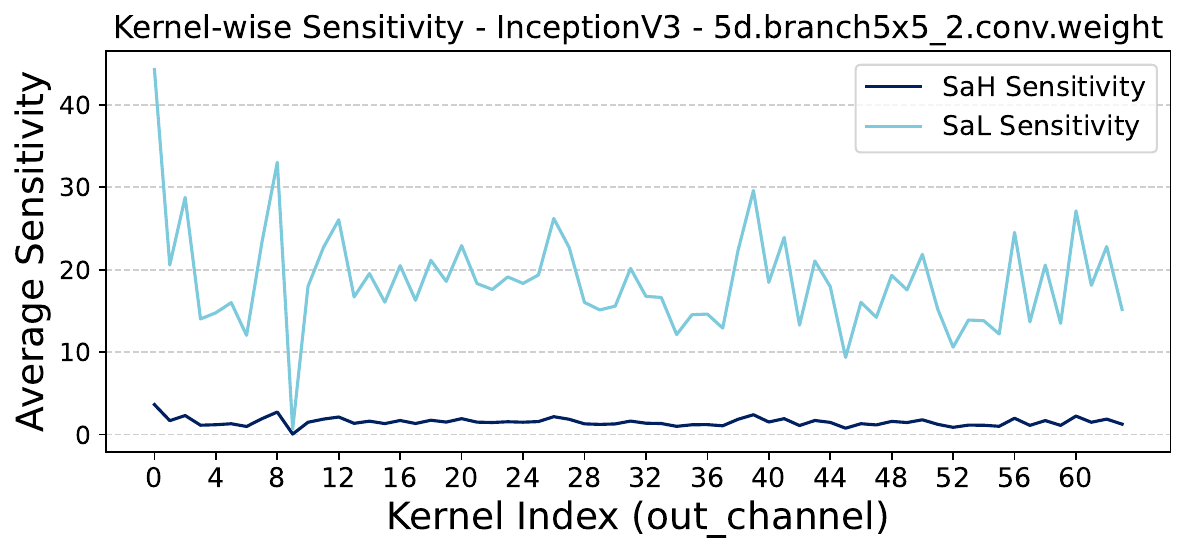}
		\includegraphics[width=0.5\linewidth]{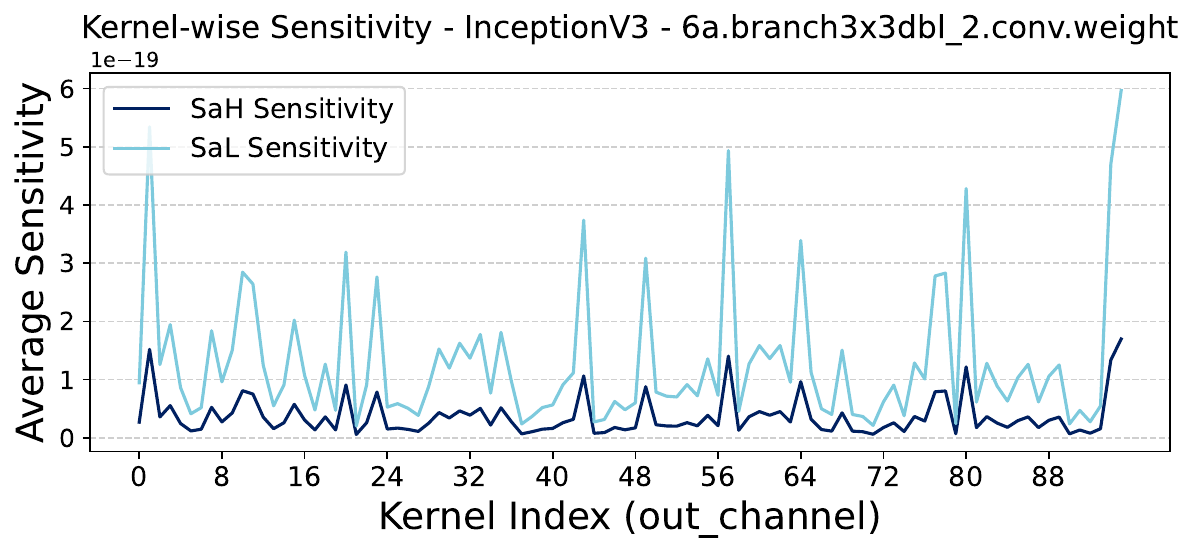}
	}
	
	\subfloat{
		\includegraphics[width=0.5\linewidth]{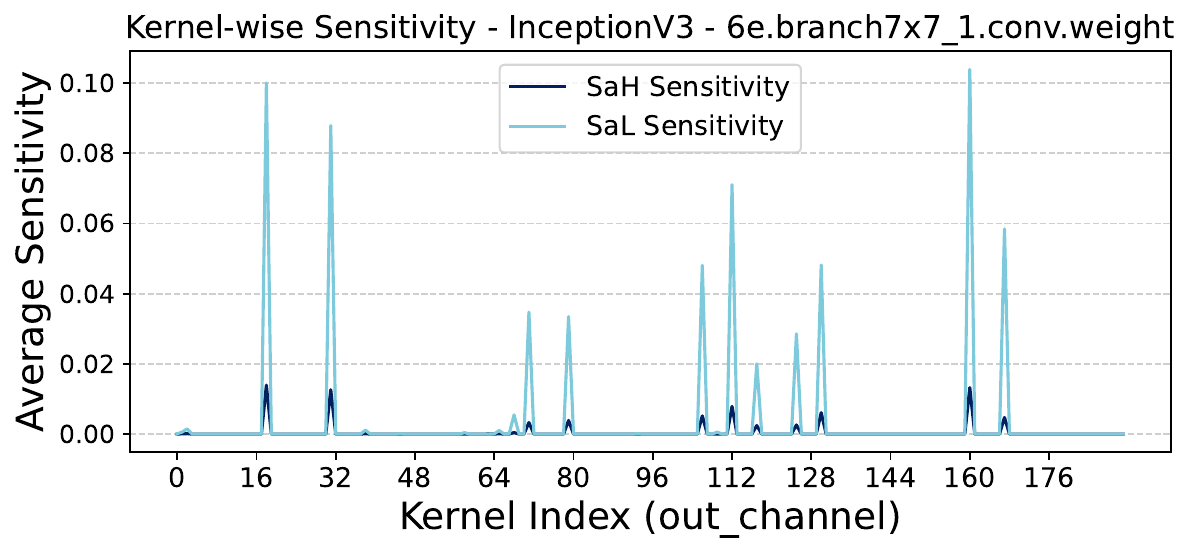}
		\includegraphics[width=0.5\linewidth]{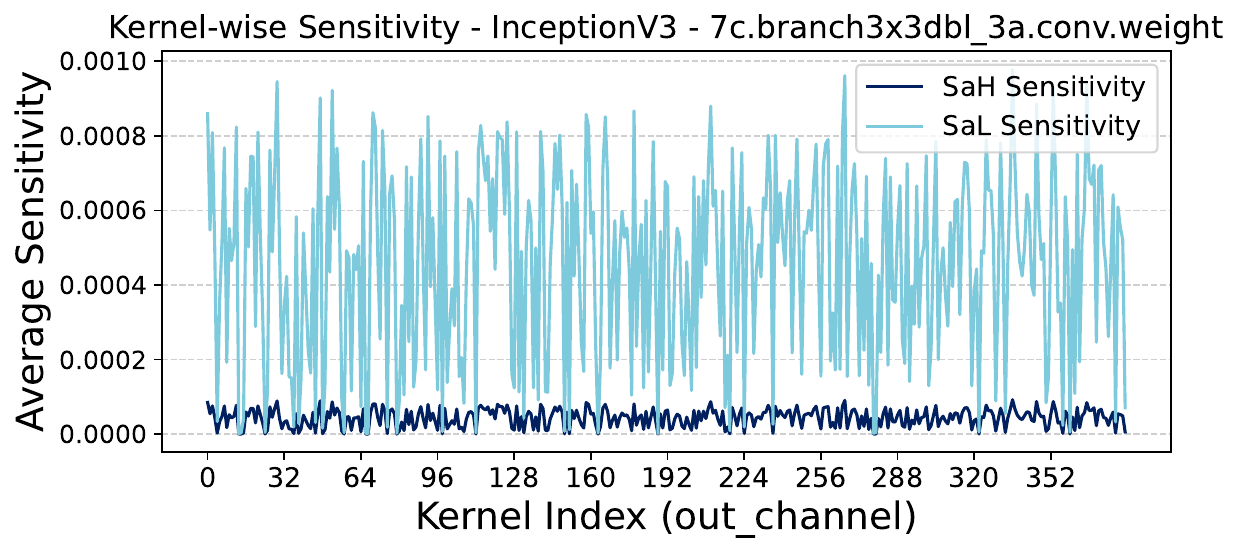}
	}
	\caption{The kernel-wise sensitivity analyzing of CONV layers in \textit{InceptionV3}.}
	\label{Fig17}
\end{figure}

\begin{figure}[!t]
	\centering
	\includegraphics[width=1\linewidth]{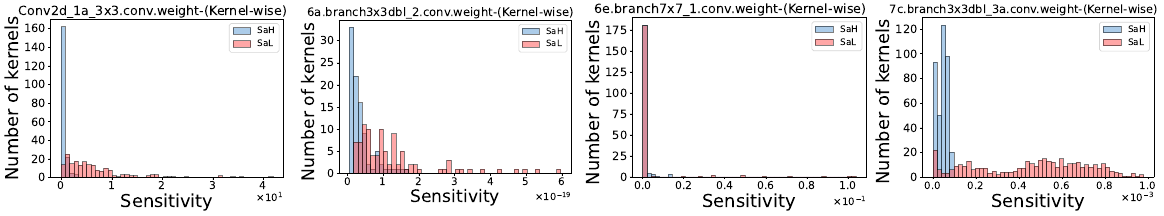}
	\caption{Kernel-wise sensitivity histogram of \textit{InceptionV3}.}
	\label{Fig18}
\end{figure}

\begin{figure}[!t]
	\centering
	\includegraphics[width=1\linewidth]{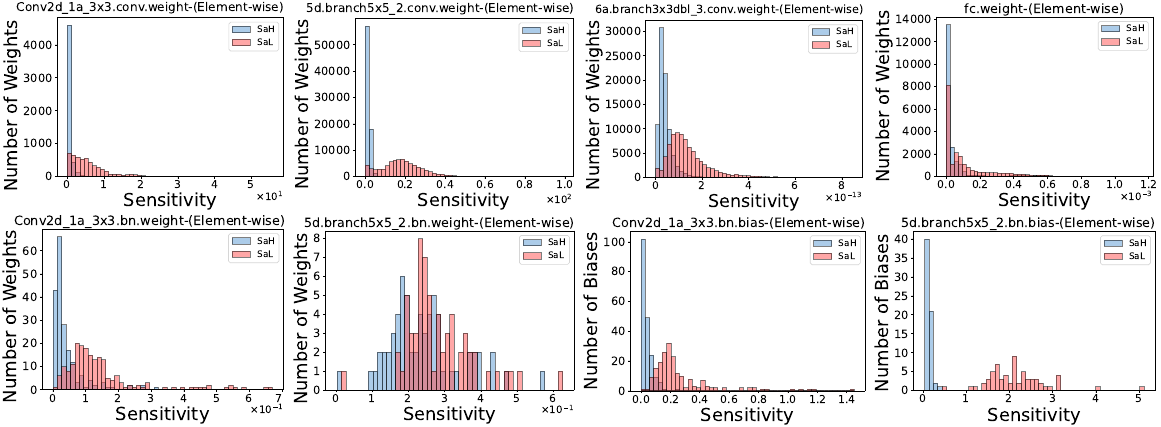}
	\caption{Element-wise sensitivity histogram of \textit{InceptionV3}.}
	\label{Fig19}
\end{figure}

The result of kernel-wise sensitivity values and kernel-wise histograms for some selected CONV layers from different modules of \textit{InceptionV3} are shown in Fig.~\ref{Fig17} and~\ref{Fig18} respectively. As illustrated in Fig~\ref{Fig17}, most kernels in \textit{Conv2d\_1a\_3x3} are highly sensitive to \textit{SaH} and \textit{SaL} faults, and can thus be regarded as critical kernels. In contrast, the sensitivity of the most kernels in module \textit{Inception C}, block \textit{6e} (such as \textit{6e.branch7x7\_1}) is close to zero, ndicating that these kernels are non-critical.
Although some layers, such as \textit{7c\_branch3x3dbl\_3a}, exhibit overall lower sensitivity compared to other layers, the kernels within these layers still show a wide range of sensitivities.As reflected in the histograms, some kernels have low sensitivity, while others are significantly more vulnerable. This heterogeneity indicates that the variations in sensitivity are not uniform across all kernels, suggesting that kernel-wise analysis is necessary rather than relying solely on the layer-wise average. Similarly, the elements of kernels and other elements of FC and BatchNorms are not equal. To better understand the distribution of sensitivity in each weight and biases of CONV layers and BatchNorms, the element-wise histogram of some selected layers is shown in Fig.~\ref{Fig19}. 

\begin{figure}[!t]
	\centering
	\includegraphics[width=0.99\linewidth]{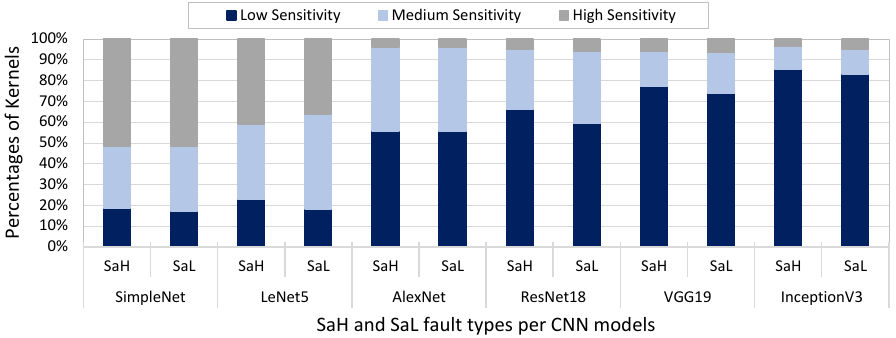}
	\caption{ The percentages of kernels at different \textit{SaH} and \textit{SaL} sensitivity levels across various models.}
	\label{Fig20}
\end{figure}

We categorized the kernel sensitivity based on sensitivity values into three levels: low, medium, and high. These categorizations are based on the range of sensitivity values that vary in each layer. The percentage of all kernels of CONV layers at each sensitivity level is illustrated in Fig.~\ref{Fig20}. As depicted, the number of kernels within each category varies depending on the fault type and CNN models. In \textit{LeNet5} and \textit{SimpleNet}, the proportion of kernels with high sensitivity level is greater than those at lower levels. For instance, \textit{SimpleNet} consists of 22 kernels in total, among which approximately 20\% belong to the low-sensitivity level for both \textit{SaL} and \textit{SaH} faults, 31.40\% correspond to the medium sensitivity level, and 48.60\% belong to the high-sensitivity level. In \textit{LeNet5}, the number of kernels with medium sensitivity under \textit{SaL} faults exceeds tha of \textit{SaH}. However, the number of kernels with low sensitivity in \textit{AlexNet}, \textit{ResNet18}, and \textit{VGG19} is more than in other levels, by 56\%, 76.12\%, and 79.62\% of all kernels, respectively. In \textit{InceptionV3}, 86.51\%, 9.36\% and 4.13\% of kernels belong to low-, medium-, and high sensitivity levels, respectively.

Overall, the observed sensitivity trends are strongly influenced by architectural properties such as layer depth, parameter reuse, connectivity patterns, and the functional role of each layer in feature extraction and classification. In addition, the sensitivity of parameter depends on the distribution of weight values within each layer and the magnitude of weight perturbation caused by \textit{SaH} and \textit{SaL}.

\subsection{Validation of the Proposed Method}

To validate the effectiveness of the sensitivity evaluation and to assess its suitability as a candidate for vulnerability analysis of CNNs in the presence of faults, we conduct a comprehensive fault injection experiment targeting the parameters of CNNs using the ReConFI framework~\cite{Dorostkar2024}. The experimental results are compared with our analytical VS metric considering the accuracy, loss, mean squared error (MSE), and prediction hamming distance of \textit{SimpleNet}, \textit{LeNet5}, and \textit{ResNet18}.

\begin{figure}[!t]
	\centering
	\subfloat[\label{Fig21-a}]{
		\includegraphics[width=0.35\linewidth]{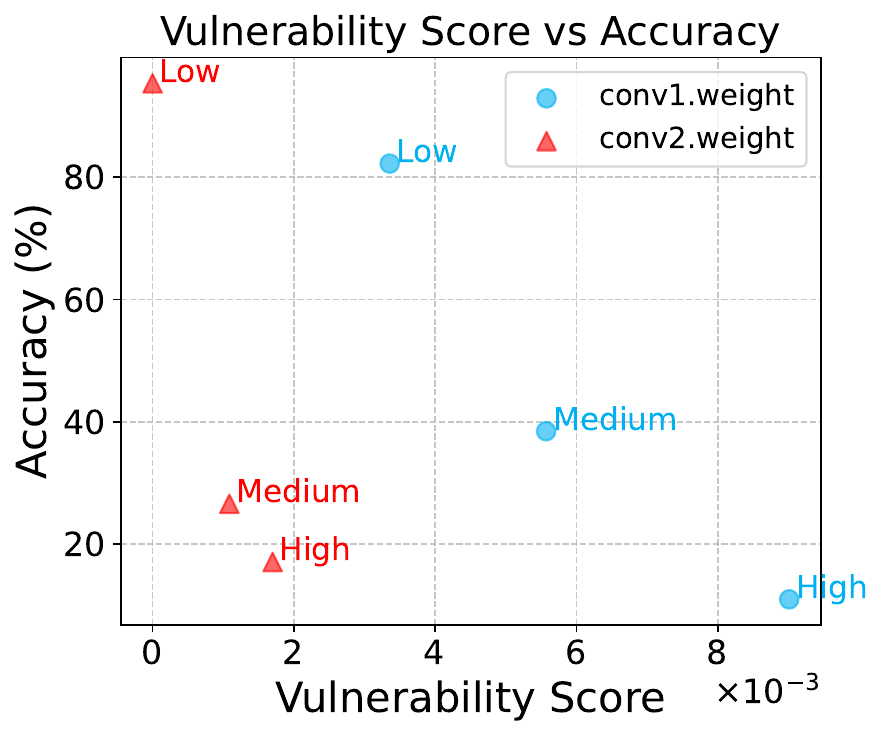} 
	}
	\subfloat[\label{Fig21-b}]{
		\includegraphics[width=0.35\linewidth]{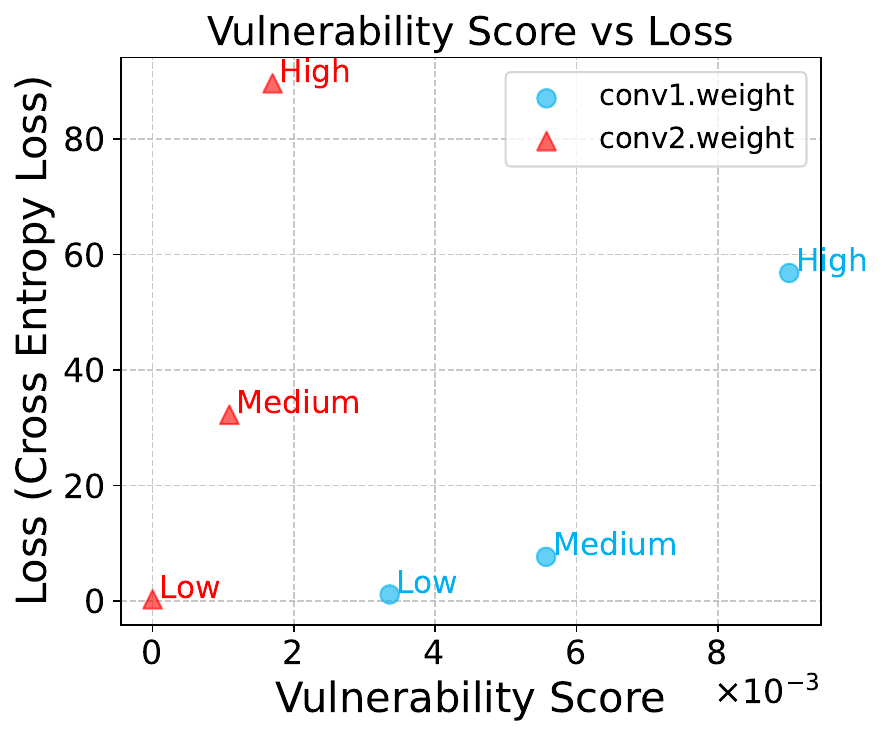}
	}
	
	\subfloat[\label{Fig21-c}]{
		\includegraphics[width=0.35\linewidth]{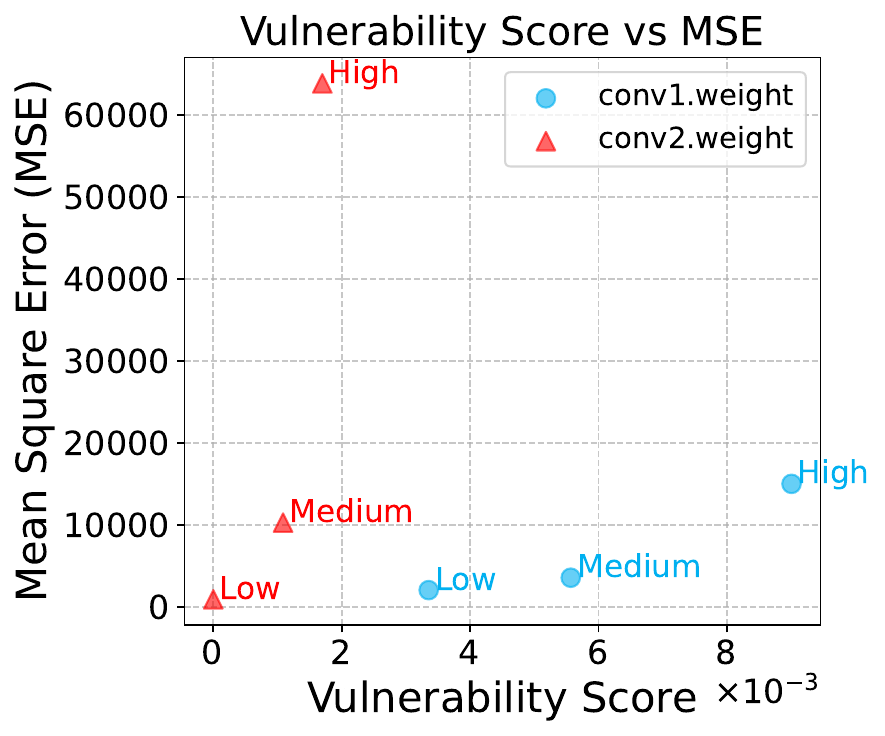}\
	}
	\subfloat[\label{Fig21-d}]{
		\includegraphics[width=0.35\linewidth]{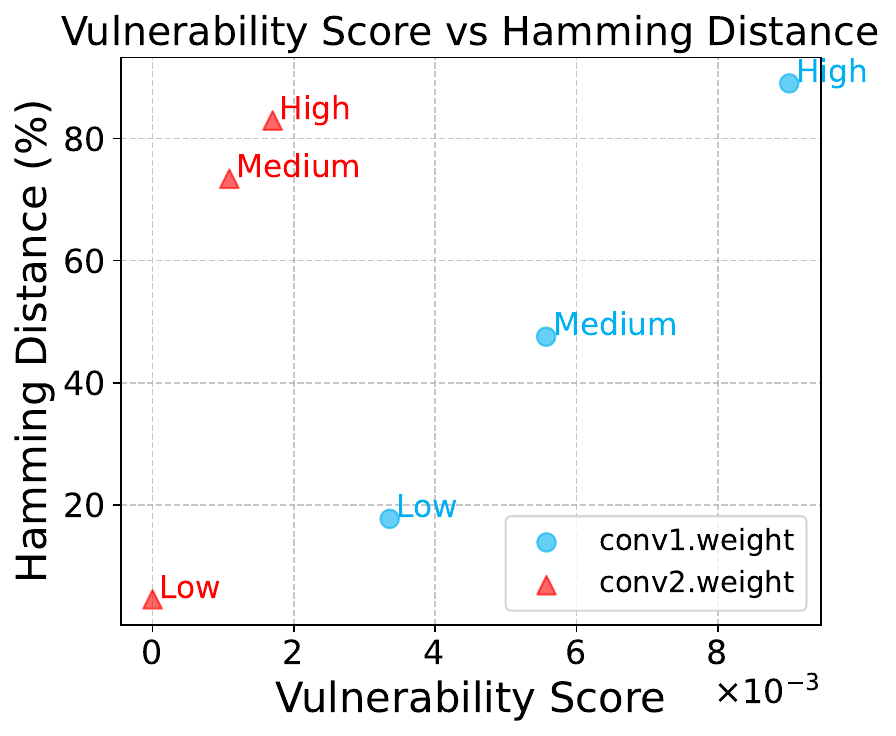}\
	}
	\caption{The comparison of the proposed VS and (a) accuracy, (b) loss and (c) MSE and (d) prediction hamming distance in \textit{SimpleNet}.}
	\label{Fig21}
\end{figure}

\begin{figure}[!t]
	\centering
	\subfloat[\label{Fig22-a}]{
		\includegraphics[width=0.35\linewidth]{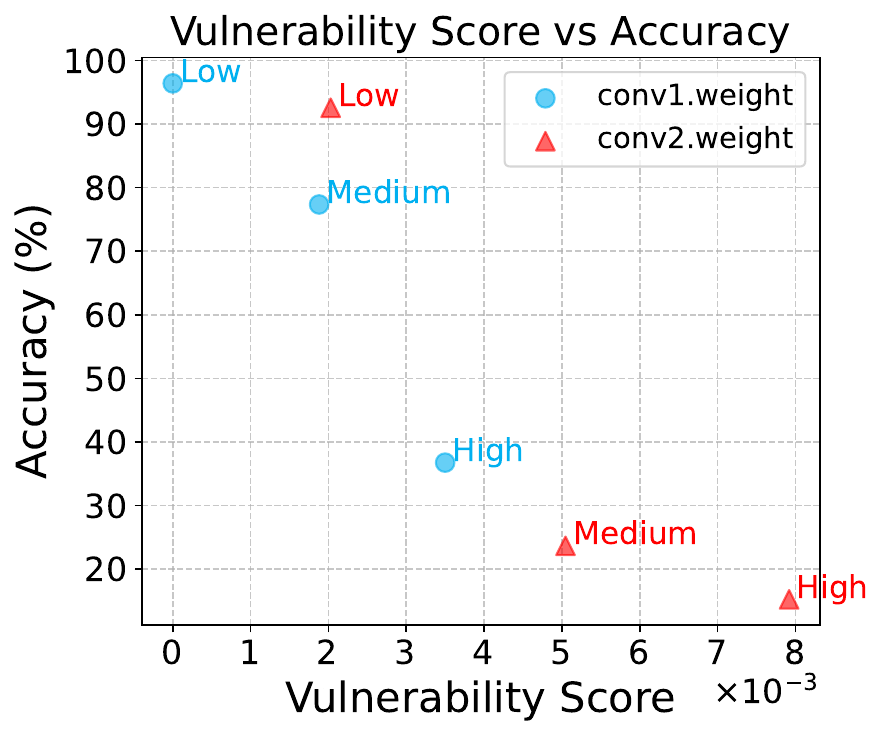} 
	}
	\subfloat[\label{Fig22-b}]{
		\includegraphics[width=0.35\linewidth]{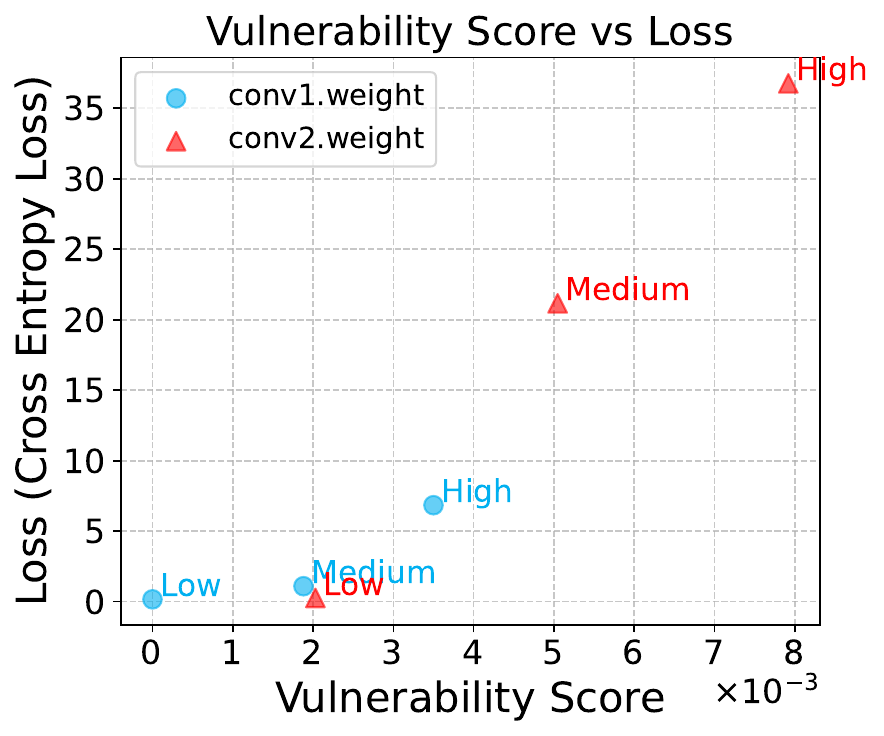}
	}
	
	\subfloat[\label{Fig22-c}]{
		\includegraphics[width=0.35\linewidth]{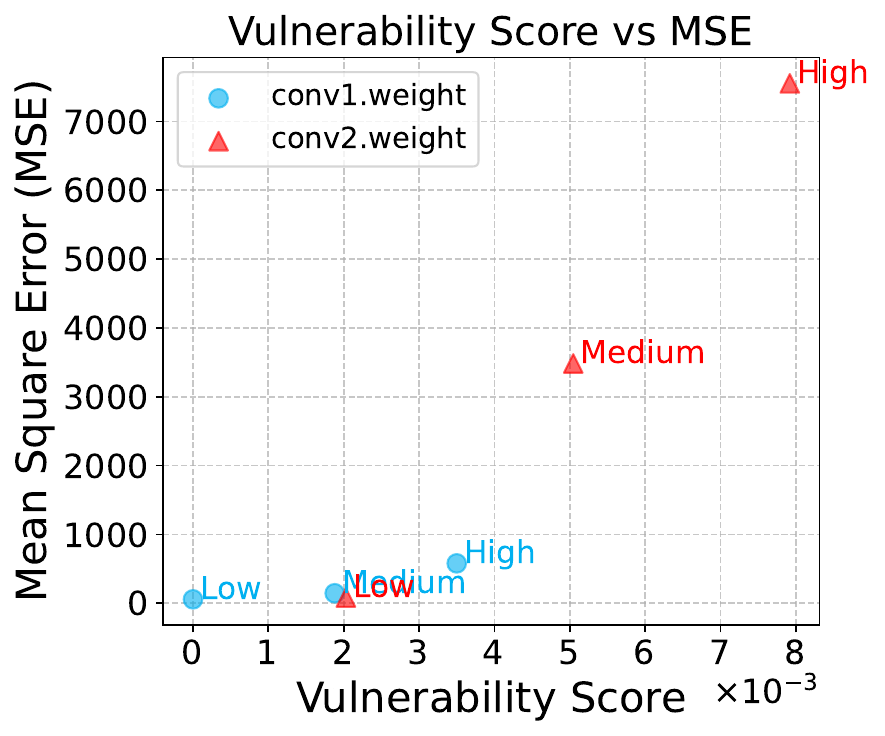}\
	}
	\subfloat[\label{Fig22-d}]{
		\includegraphics[width=0.35\linewidth]{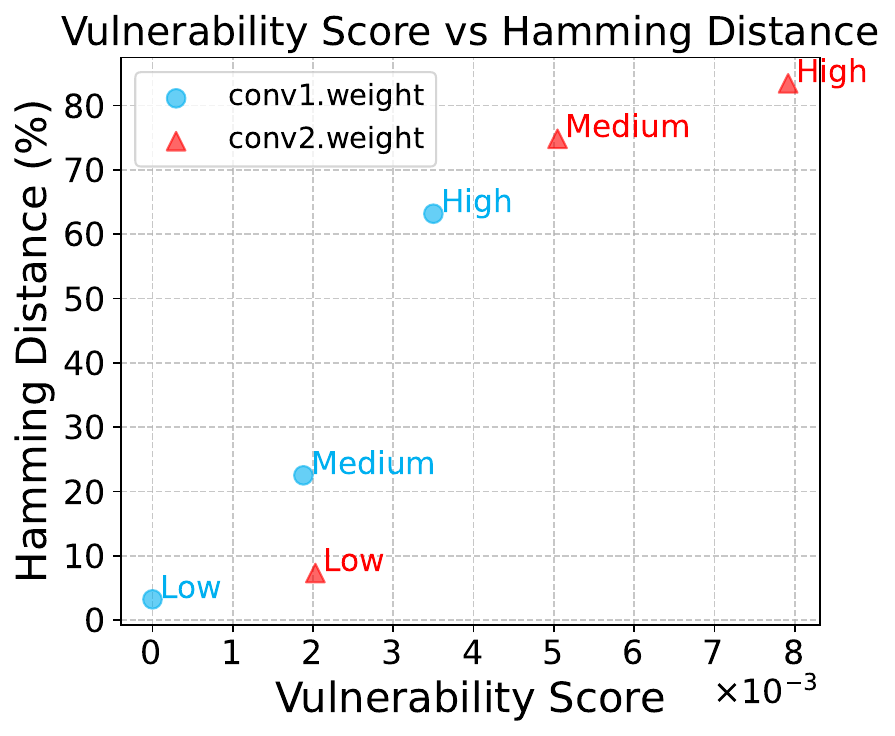}\
	}
	\caption{The comparison of the proposed VS and (a) accuracy, (b) loss and (c) MSE and (d) prediction hamming distance in \textit{LeNet5}.}
	\label{Fig22}
\end{figure}

\begin{figure}[t]
	\centering
	\subfloat[\label{Fig23-a}]{
		\includegraphics[width=0.35\linewidth]{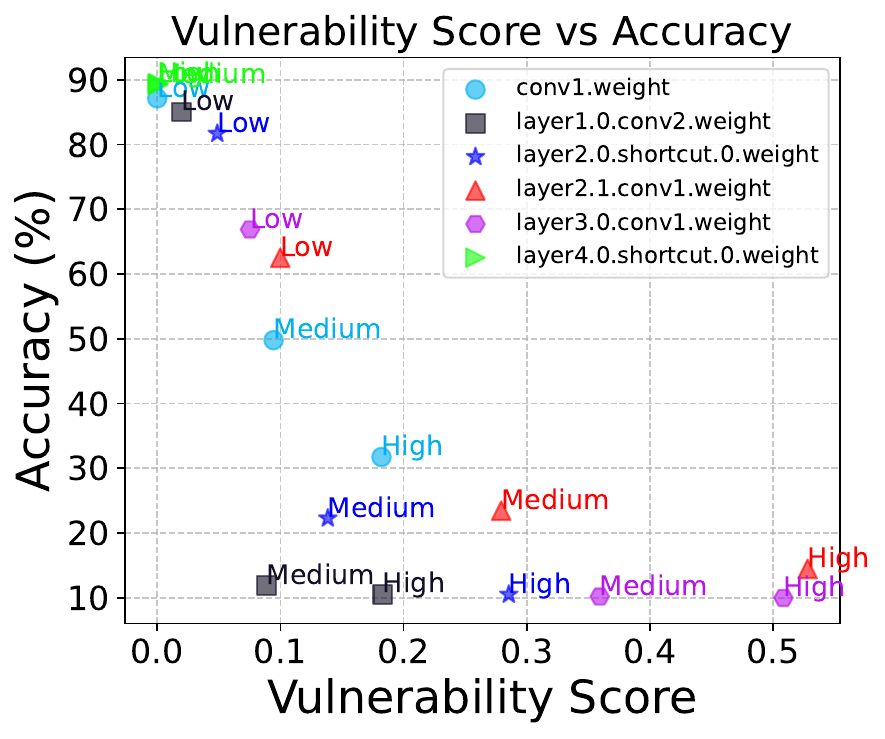} 
	}
	\subfloat[\label{Fig23-b}]{
		\includegraphics[width=0.35\linewidth]{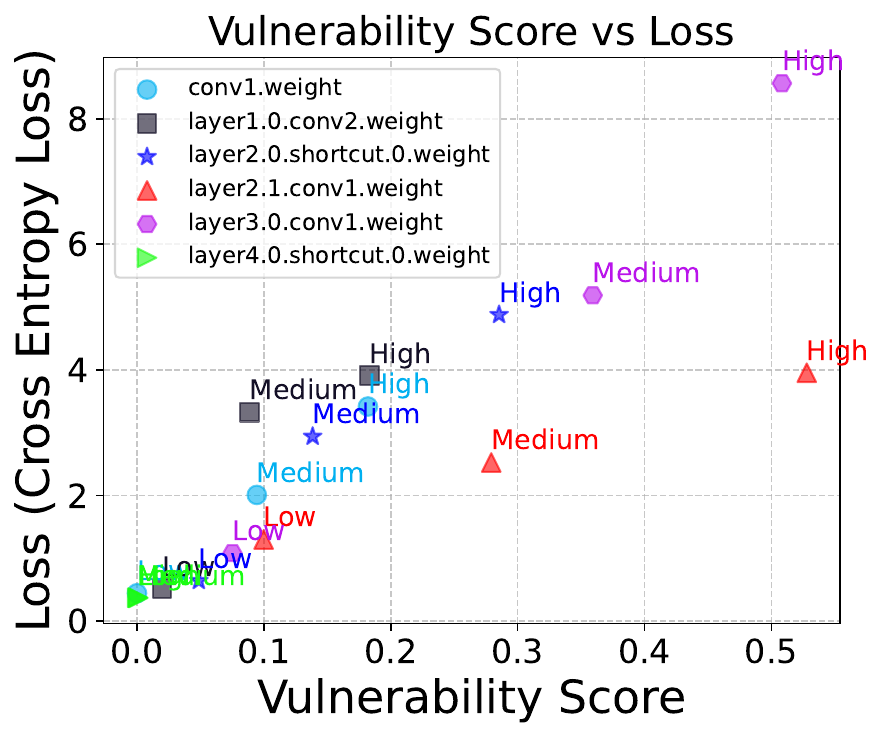}
	}
	
	\subfloat[\label{Fig23-c}]{
		\includegraphics[width=0.35\linewidth]{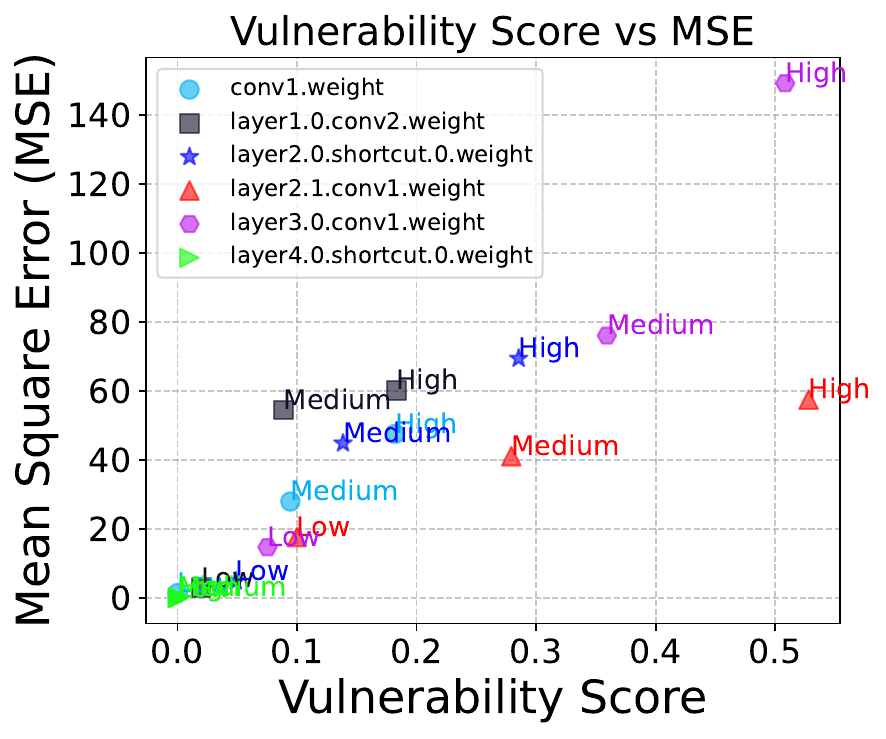}\
	}
	\subfloat[\label{Fig23-d}]{
		\includegraphics[width=0.35\linewidth]{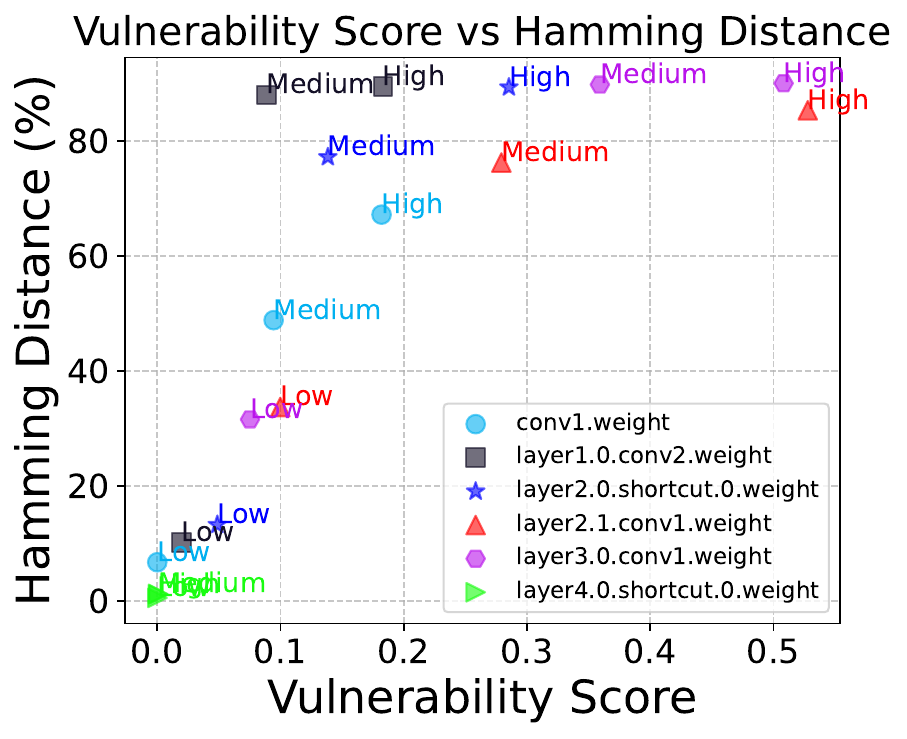}\
	}
	\caption{The comparison of the proposed VS and (a) accuracy, (b) loss and (c) MSE and (d) prediction hamming distance in \textit{ResNet18}.}
	\label{Fig23}
\end{figure}

For this purpose, the kernels of each layer are classified into three categories, low, medium, and high sensitivity, according to the sensitivity range within the layer.
Subsequently, we injected 10\% \textit{SaH} and 2\% \textit{SaL} faults into each kernel. These rates were chosen based on findings from~\cite{Ching-YiChen2015}, where fault occurrences in ReRAM cells were reported to be approximately 9.04\% \textit{SaH} and 1.75\% \textit{SaL}.
The average classification accuracy, loss, MSE, and prediction Hamming distance are then calculated and compare to VS metric. The comparison between VS and these performance metrics across the CONV layers of \textit{LeNet5}, \textit{SimpleNet}, and \textit{ResNet18} are illustrated in Figs.~\ref{Fig21}~to~\ref{Fig23}.

In these charts, each layer is represented by a distinct marker and color with three sensitivity categories. For every category, the average VS and the corresponding parameter values of all kernels are shown in the charts. The \textit{SimpleNet} and \textit{LeNet5} models each contain two CONV layers, and thus three kernel categories are shown for each of the two layers. In contrast, for the \textit{ResNet18} model, five layers are selected out of the total eighteen, resulting in five separate plots, each containing three kernel categories. These charts allow examining the relationship between VS, the chosen parameter, and the sensitivity differences across layers. The value of the VS depends on the sensitivity of each layer to \textit{SaH} and \textit{SaL} faults, as well as on the proportion of each fault type. Since the grouping is performed independently within each layer, the high-sensitivity value of one layer does not necessarily represent the same sensitivity range as the high-sensitivity value of another. For instance, in \textit{SimpleNet}, the VS separation between categories in layer \textit{conv2} is much smaller than that of \textit{conv1}. 

As can be observed, an increase in VS generally leads to a decrease in accuracy and an increase in loss, MSE, and Hamming distance. However, the rate of degradation or increment differs significantly across layers. For example, as shown in Fig.~\ref{Fig21-a}, in layer conv1 of the \textit{SimpleNet} model, the reduction is gradual across sensitivity categories; whereas for \textit{conv2}, even medium sensitivity kernels cause a dramatic accuracy drop. This trend is also observable in loss, MSE, and Hamming distance. 

As shown in Fig.~\ref{Fig22}, the \textit{conv2} layer exhibits a higher VS than \textit{conv1} in \textit{LeNet5}. Therefore, the accuracy degradation and loss increment are more pronounced in this layer. Both the MSE and Hamming distance increase as the VS increases. Faults in the low-sensitivity kernels of \textit{conv1} have negligible impact on performance metrics. In contrast, the accuracy in medium- and high-sensitivity kernels in this layer decrease by 13.7\% and 62.51\%, respectively. The predicted Hamming distance in low sensitivity kernels is about 0.2\%, indicating that faults in this category change 0.2\% of results compared to non-faulty model. This value is increased to 14.25\% and 63.15\% in medium- and high-sensitivity kernels of \textit{conv1}. 

In \textit{ResNet18}, most of the performance degradation is due to faults in the medium- and high-sensitivity categories. As shown in Fig.~\ref{Fig23-a}, the layer \textit{layer4.0.shortcut.0.weight} exhibits the lowest sensitivity (close to zero), and the corresponding fault injection results indicate that faults in this layer have minimal impact on performance metrics. In contrast, \textit{layer3.0.conv1.weight} demonstrates a higher sensitivity value compared to other layers, and faults in this layer result in a more pronounced degradation of accuracy. On average, fault injection into low-, medium-, and high-sensitivity kernels decreases accuracy by 10.60\%, 54.95\% and 61.64\%, respectively. Similarly, the predicted Hamming distances increase by 16.00\%, 63.54\% and 70.40\% across low-, medium-, and high-sensitivity kernels, respectively. 

To evaluate the statistical robustness of the proposed VS, we performed a correlation analysis between VS and the performance metrics. Since the relationship between VS and the evaluation metrics is not strictly linear, Spearman's rank correlation coefficient ($\rho$) was employed to capture monotonic dependencies without assuming linearity. The analysis was performed both across all kernels (global correlation) of each network and separately for kernels within individual layers (layer-wise), and the results are reported in Table~\ref{Table2}. The statistical significance of the observed correlations was evaluated using the corresponding $p$-values.
	
As reported in Table~\ref{Table2}, VS demonstrates strong and statistically significant correlations with multiple performance metrics in networks such as \textit{LeNet5} and \textit{ResNet18}. For \textit{LeNet5}, the global correlations range from $\rho = 0.71$ to $0.85$ ($p < 10^{-3}$), while for \textit{ResNet18} they remain consistently significant with $\rho \approx 0.66$--$0.69$ ($p < 10^{-3}$). These results indicate that kernels with higher VS values tend to produce larger accuracy degradation and higher error metrics when subjected to faults, confirming that VS reliably captures the relative vulnerability of network components. Layer-wise analysis further confirms this trend. For example, in several layers of \textit{ResNet18}, such as \textit{conv1.weight} and \textit{layer3.0.conv1}, VS shows strong monotonic relationships with the fault-impact metrics ($\rho$ up to $0.89$, $p < 0.01$). Although the global correlation in \textit{SimpleNet} is relatively weak, the analysis reveals that VS accurately identifies the most sensitive layer. In particular, for the \textit{conv2.weight} layer, extremely strong correlations are observed ($\rho = 0.86$ for accuracy drop, $\rho = 0.96$ for Hamming distance, and $\rho = 1.00$ for MSE, all with $p < 0.01$). This indicates that while structural sensitivity varies across layers, VS effectively highlights the critical layers and kernels.
	
Overall, the statistically significant correlations observed across multiple models and metrics demonstrate that the proposed VS provides a stable and reliable ranking-based indicator for identifying vulnerable kernels, particularly in deeper neural network architectures.

\begin{table}[t]
	\centering
	\caption{Spearman's rank correlation ($\rho$) between the proposed VS 
		and performance metrics.}
	\label{Table2}
		\begin{tabular*}{\textwidth}{
			@{\extracolsep{\fill}}
			>{\raggedright\arraybackslash}p{1.1cm}
			>{\raggedright\arraybackslash}p{3cm}
			>{\raggedright\arraybackslash}p{2cm}
			>{\raggedright\arraybackslash}p{1cm}
			>{\raggedright\arraybackslash}p{1cm}
			>{\raggedright\arraybackslash}p{2.6cm}
		}
		\hline
		Model & Layer & Accuracy Drop & Loss & MSE & Hamming Distance \\
		\hline
		\multirow{3}{*}{LeNet} 
		& All kernels            & 0.76$^{\ast\ast\ast}$ &  0.85$^{\ast\ast\ast}$ &  0.81$^{\ast\ast\ast}$ &  0.71$^{\ast\ast}$ \\
		& conv1.weight           & 0.94$^{\ast\ast}$ &  0.94$^{\ast\ast}$ &  0.89$^\ast$ &  0.94$^{\ast\ast}$ \\
		& conv2.weight           & 0.43      &  0.62      &  0.47      &  0.22      \\
		\hline
		\multirow{3}{*}{SimpleNet} 
		& All kernels            & 0.12      & 0.19      & 0.20      & 0.09      \\
		& conv1.weight           & 0.31      &  0.31      &  0.36      &  0.40      \\
		& conv2.weight           & 0.86$^\ast$ &  0.86$^\ast$ &  1.00$^{\ast\ast\ast}$ &  0.96$^{\ast\ast}$ \\
		\hline
		\multirow{6}{*}{ResNet18} 
		& All kernels            & 0.68$^{\ast\ast\ast}$ &  0.66$^{\ast\ast\ast}$ &  0.69$^{\ast\ast\ast}$ &  0.69$^{\ast\ast\ast}$ \\
		& conv1.weight           & 0.87$^{\ast\ast}$ &  0.83$^\ast$ &  0.83$^\ast$ &  0.89$^{\ast\ast}$ \\
		& layer1.0.conv2.weight  & 0.70$^\ast$ &  0.27      &  0.32      &  0.67$^\ast$ \\
		& layer2.0.shortcut.0.weight & 0.60$^\ast$ &  0.54$^\ast$ &  0.56$^\ast$ &  0.59$^\ast$ \\
		& layer2.1.conv1.weight  & 0.50$^\ast$ &  0.49$^\ast$ &  0.52$^\ast$ &  0.50$^\ast$ \\
		& layer3.0.conv1.weight  & 0.55$^{\ast\ast}$ &  0.50$^\ast$ &  0.54$^{\ast\ast}$ &  0.58$^{\ast\ast}$ \\
		\hline
	\end{tabular*}
	
	\vspace{0.5em}
	\footnotesize
	$^\ast$~$p<0.05$, 
	$^{\ast\ast}$~$p<0.01$, 
	$^{\ast\ast\ast}$~$p<0.001$.
\end{table}

\begin{figure}[!t]
	\centering
	\subfloat[\label{Fig24-a}]{
		\includegraphics[width=0.35\linewidth]{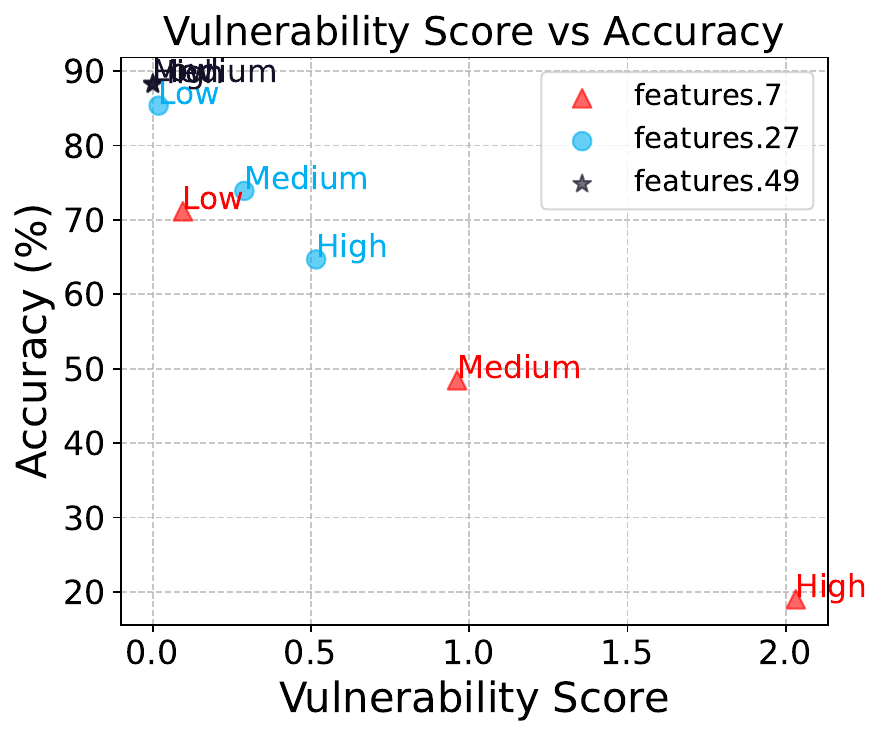} 
	}
	\subfloat[\label{Fig24-b}]{
		\includegraphics[width=0.35\linewidth]{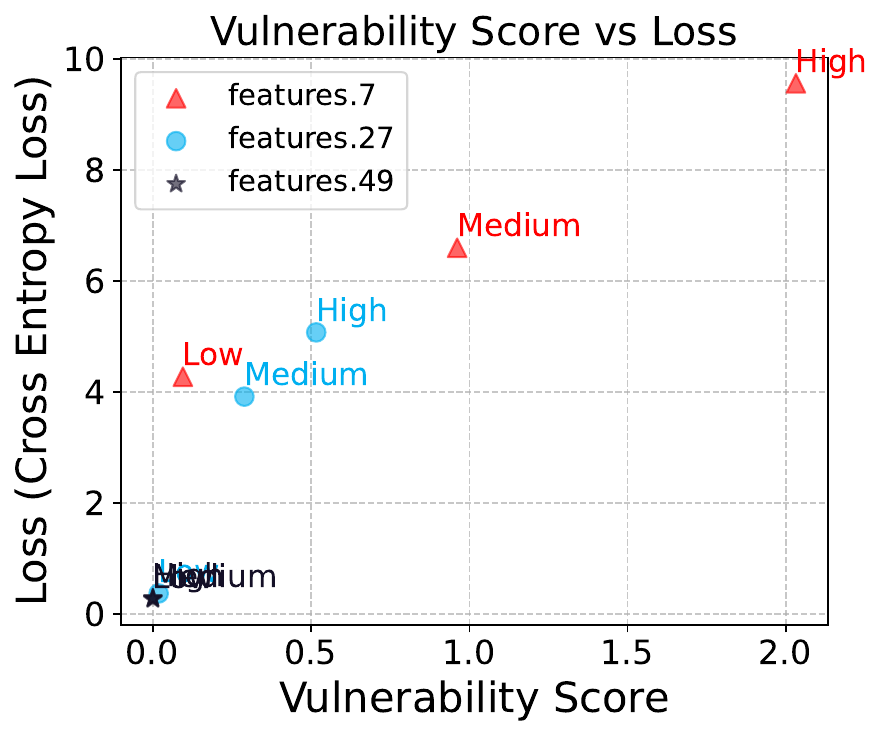}
	}
	
	\subfloat[\label{Fig24-c}]{
		\includegraphics[width=0.35\linewidth]{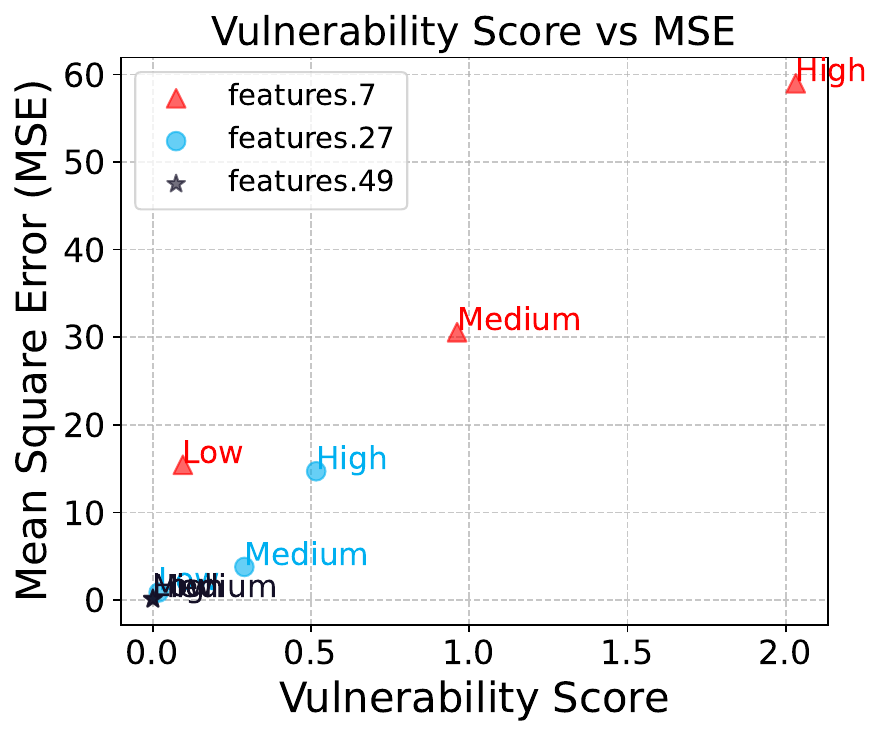}\
	}
	\subfloat[\label{Fig24-d}]{
		\includegraphics[width=0.35\linewidth]{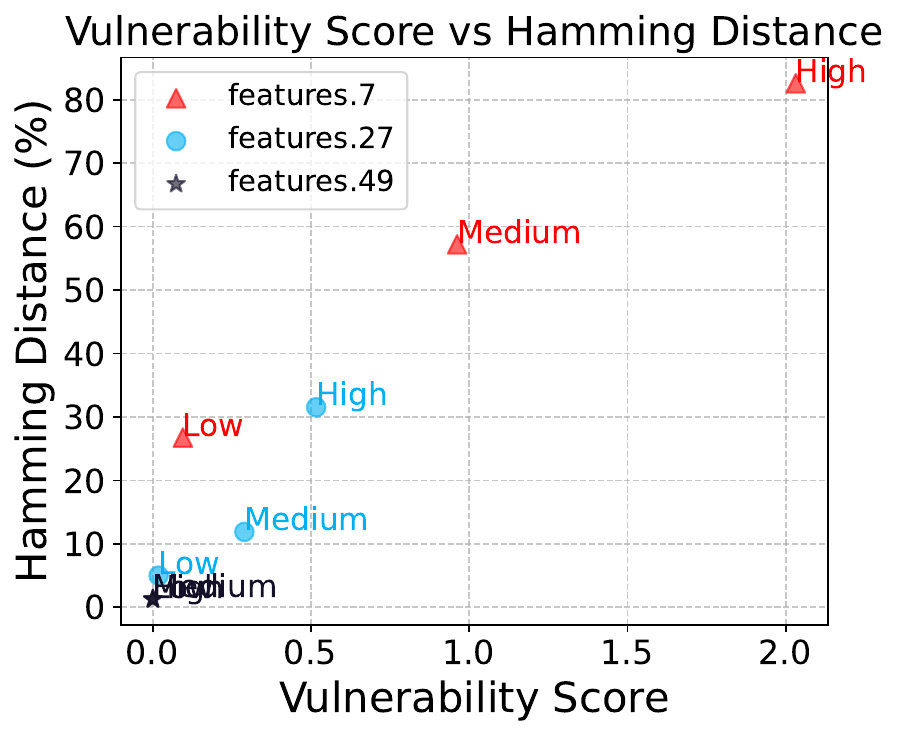}\
	}
	\caption{The comparison of the proposed VS and (a) accuracy, (b) loss and (c) MSE and (d) prediction hamming distance in \textit{VGG19}.}
	\label{Fig24}
\end{figure}

To further evaluate the predictive capability of the proposed vulnerability metric on large-scale CNNs, we performed selective fault injection on layers of the \textit{VGG19}. Since exhaustive fault injection across all kernels is computationally prohibitive for large models, three layers were selected from different stages of the architecture. For each selected layer, one high-sensitivity kernel (top-ranked), one medium-sensitivity kernel, and one low-sensitivity kernel (bottom-ranked) were chosen. The comparison between VS metric and performance metric of the selected layers of \textit{VGG19} is show in Fig.~\ref{Fig24}. As shown in Fig.~\ref{Fig24}, the kernels with higher VS in layers \textit{features.7} and \textit{features.27} consistently produce larger accuracy drops compared to medium- and low-sensitivity kernels within the same layer, while in layer \textit{features.49}, which has a low VS, the accuracy drop is negligible. This trend is observed across early, middle, and late layers, providing additional evidence that the proposed analytical metric remains predictive even for large-scale architectures.

As shown in Table~\ref{Table3}, exhaustive fault injection is prohibitively expensive for large-scale networks, since each parameter requires a separate inference pass.
A runtime comparison demonstrates that exhaustive fault injection is prohibitively expensive for large-scale networks, as each parameter requires one inference pass. For example, a single fault injection takes about 0.453 sec on \textit{LeNet5} and about 0.972 sec on \textit{SimpleNet}, running on an \textit{Intel Core i7, 3.2 GHz} processor with a \textit{NVIDIA GeForce GTX 2080Ti} GPU. Therefore, analyzing all parameters of \textit{LeNet5} (44,426 parameters) and \textit{SimpleNet} (431,080 parameters) would take approximately 5.6 hours and 4.9 days, respectively. In contrast, sensitivity analysis derives the vulnerability of all parameters with only one forward and backward pass to calculate first- and second-order gradients, requiring 1.51 sec for \textit{LeNet5} and 4.696 sec for \textit{SimpleNet}. Consequently, exhaustive fault-injection analysis is on the order of $10^4-10^5$ times slower than gradient-based sensitivity analysis (about $8.9\times10^4$ times slower for \textit{SimpleNet} and about $1.3\times10^4$ times slower for \textit{LeNet5}), which makes sensitivity analysis more computationally efficient for whole-network vulnerability assessment. Table~\ref{Table2} shows the time required for sensitivity analysis. However, fault injection remains valuable for targeted, fine-grained fault studies, but it is prohibitively expensive when applied exhaustively to large-scale CNNs.

\begin{table}[!t]
	\centering
	\caption{Time required for gradient-based sensitivity analysis.}
	\label{Table3}
	\begin{tabular*}{\textwidth}{
			@{\extracolsep{\fill}}
			>{\raggedright\arraybackslash}p{1cm}
			>{\centering\arraybackslash}p{1.2cm}
			>{\centering\arraybackslash}p{1.2cm}
			>{\centering\arraybackslash}p{1.5cm}
			>{\centering\arraybackslash}p{1.5cm}
			>{\centering\arraybackslash}p{1.5cm}
			>{\centering\arraybackslash}p{2cm}
		}
		\toprule
		CNN model   & LeNet5 & SimpleNet & AlexNet & VGG19 & ResNet19 & InceptionV3 \\
		\midrule
		Time (sec)  & 1.510 & 4.696 & 1069.844 $\approx$19 min & 2066.676 $\approx$35 min & 3107.888 $\approx$53 min & 142056.613 $\approx$39 h, 27 min \\
		\bottomrule
	\end{tabular*}
\end{table}

\section{Conclusion}\label{concl}

In this work, we presented a fast, scalable, and analytical framework for analyzing and evaluating the fault vulnerability of CNNs deployed on ReRAM-based PIM accelerators. Unlike traditional fault injection, which is prohibitively time-consuming for large-scale models, our method provides an efficient alternative by approximating fault sensitivity through gradient-based first- and second-order analysis. By quantifying vulnerability across multiple levels of granularity, element-wise, kernel-wise, and layer-wise, our approach enables a hierarchical ranking of CNN components according to their susceptibility to \textit{SaH} and \textit{SaL} faults. Experimental validation against fault injection simulations demonstrates that the proposed vulnerability score (VS) provides a reliable and accurate estimation of performance degradation within only a few minutes, instead of days or weeks. The results show that as the VS increases, the loss, MSE, and Hamming distance also increase, while the classification accuracy decreases. We comprehensively analyze the sensitivity of CNN learning parameters across different topologies and network scales. Our investigation reveals that the vulnerability of CNNs to permanent faults strongly depends on multiple factors, including the type and magnitude of the fault, the underlying network topology, the number and configuration of layers, their relative distance from the output, and the role of specific learning parameters such as weights and biases. These results demonstrate the potential of our framework as a practical tool for designing fault-tolerant ReRAM-based accelerators, particularly in edge and safety-critical applications where reliability is essential.

\bibliography{mybibfile-others}

\end{document}